\documentclass{article}     

\usepackage{amsmath, amsfonts,amsthm,amssymb,bm}
\usepackage{dsfont}

\usepackage[english]{babel}

\usepackage{latexsym,amssymb,amsfonts,graphicx}
\usepackage{amsmath,dsfont}
\usepackage{verbatim}
\usepackage{mathrsfs}
\usepackage{bm}
\usepackage{color}
\usepackage{epsfig}

\usepackage{url}

\usepackage{bm}
\usepackage{natbib}

\def\esssup_#1{\underset{#1}{\Xi}}
\def\essinf_#1{\underset{#1}{\mathrm{ess\,inf\, }}}
\def\argmax_#1{\underset{#1}{\mathrm{arg\,max\, }}}
\def\argmin_#1{\underset{#1}{\mathrm{arg\,min\, }}}

\newcommand{\Px}{ \mathbb{P} }

\newcommand{\N}{ \mathds{N} }
\newcommand{\Ex}{ \mathbb{E} }

\newcommand{\A}{{\cal A}}

\newcommand{\Gx}{\mathbb{G}}

\newcommand{\D}{\mathrm{d}}
\newcommand{\idc}{\mathbf{1}}
\newcommand{\too}{-\!\!\!\!\!\to}

\newcommand{\F}{\mathcal{F}}
\newcommand{\G}{\mathcal{G}}
\newcommand{\R}{\mathds{R}}

\newtheorem{theorem}{Theorem}[section]
\newtheorem{definition}{Definition}[section]

\newtheorem{proposition}[theorem]{Proposition}
\newtheorem{remark}[theorem]{Remark}
\newtheorem{lemma}[theorem]{Lemma}

\allowdisplaybreaks[4]

\definecolor{Red}{rgb}{0.00, 0.00, 0.00}
\newcommand{\Red}{\color{Red}}
\definecolor{DRed}{rgb}{0.5, 0.00, 0.00}

\definecolor{Blue}{rgb}{0.00, 0.00, 1.00}

\definecolor{Green}{rgb}{0.0, 0.4, 0.0}

\small

\title{
Optimal Investment under Information Driven Contagious Distress
}
\author{
Lijun Bo \thanks{Email: lijunbo@ustc.edu.cn,  School of Mathematical Sciences, University of Science and Technology of China, and Wu Wen Tsun Key
Laboratory of Mathematics, Chinese Academy of Science, Hefei, Anhui Province 230026, China.} \and
Agostino Capponi \thanks{E-mail: ac3827@columbia.edu, Department of Industrial Engineering and Operations Research, Columbia University, New York, NY 10027, USA. }}

\begin{document}
\maketitle
\begin{abstract}
We introduce a dynamic optimization framework to analyze optimal portfolio allocations within an information driven contagious distress model. The investor allocates his wealth across several stocks whose growth rates and distress intensities are driven by a hidden Markov chain, and also influenced by the distress state of the economy. We show that the optimal investment strategies depend on the gradient of value functions, recursively linked to each other via the distress states. We establish uniform bounds for the solutions to a sequence of approximation problems, show their convergence to the unique Sobolev solution of the recursive system of Hamilton-Jacobi-Bellman partial differential equations (HJB PDEs), and prove a verification theorem. We provide a numerical study to illustrate the sensitivity of the strategies to contagious distress, stock volatilities and risk aversion.
\vspace{0.3 cm}


\vspace{0.3 cm}


\end{abstract}

\section{Introduction}
The phenomenon of contagious distress has received a lot of attention in the finance community. Crisis episodes like the great recession have shown that unexpected shocks and distress events can seriously affect the financial status of other market participants due to direct or indirect linkages. One stream of literature has focused on contagious distress stemming from direct causal relationships between obligors. These types of relations have been shown to be empirically relevant for some sectors such as commercial banks, where the distress likelihood of an entity is likely to increase if some of its major borrowers or counterparties default, see also \cite{FreyBackhaus04} and \cite{Yu}. There are, however, other effects arising from {\it information contagion}, where news about financial distress may lead investors to update the valuations of securities held in their portfolios. These information effects arise when investors have incomplete information about the creditworthiness of other obligors, and the distress risk depends on a number of (possibly correlated) market factors which none of the market participants can directly observe. 

We introduce a novel framework for deciding on optimal portfolio allocations within an information driven contagious distress model. To the best of our knowledge, ours is the first study to analyze this form of contagion in a portfolio optimization context. A few studies have incorporated defaultable securities into the portfolio optimization framework.
 \cite{BielJang} derive optimal investment strategies for an investor with constant relative risk aversion (CRRA), allocating her wealth among a defaultable bond, risk-free bank account and a stock. \cite{CapFig}, develop a portfolio framework where investors can trade a stock and a defaultable security in a Markov modulated framework with observable regimes. \cite{BoWang} consider an investor maximizing utility from consumption, and investing her wealth in a defaultable perpetual bond.
Few other works have considered contagion risk in a portfolio optimization or hedging context. {\cite{KSAll} discuss contagion effects on defaultable bond prices. 
 \cite{bielecki08} develop a framework for hedging defaultable claims using credit default swaps.}
 {\cite{BoCappMF}} introduce a credit portfolio framework in which default times are correlated via an interacting intensity model. This results in a causal form of contagion which ignores information effects. 

{\Red Our paper is also related to a branch of literature studying partially observed control problems. \cite{Davisincom} consider the problem of maximizing expected terminal utility in a model where the investor only observes asset prices, but not the return processes. \cite{Sass} consider a related problem and assume that the instantaneous rates of return are driven by a hidden continuous time finite state Markov chain. \cite{NagaiPeng} study a partially observed optimal portfolio problem with a risk-sensitive investor. \cite{FreyAbdWun} analyze the impact of expert options on optimal investment under incomplete information.  \cite{NagaiRung2} and \cite{NagaiRung1} consider risk averse investors, respectively of logarithmic and power type, who invest in stocks whose price processes are modulated by a hidden finite state continuous time Markov chain, and only observed at random times.}

We consider a portfolio model consisting of {stocks} which can enter into a state of distress. The growth rates and distress intensities are driven by a hidden Markov chain and are influenced by the distress states of other stocks in the portfolio. This introduces correlation among stock prices, given that the hidden regime affects their future comovements. Hence, our model may be interpreted as a frailty correlation model in the same spirit as \cite{DuffieJF}, with the frailty factor being the unobserved Markov chain. The latter may be interpreted as a model for the macro-economy, where market regimes correspond to inflation/recession states, or high/low distress risk (see \cite{Longstaff} for empirical evidence). We consider the partially observed utility maximization problem of a power investor who only observes stock prices and distress events, but not the modulating Markov chain. Using existing results from \cite{CappLopezhidd} (see also \cite{NagaiRung}), such an optimization problem can be reduced to an equivalent fully observed risk-sensitive control problem.

We next summarize our main contributions. We develop a rigorous analysis of contagious distress, arising through the dependence of the optimal investment strategies on the gradient of value functions associated with the distress states of the portfolio. 
We show that each value function may be recovered as the unique Sobolev solution of the corresponding recursive HJB PDE. This is achieved by means of the following steps. {We first consider a Stampacchia{'s} approximation problem to deal with the presence of two quadratic gradient terms and an exponential nonlinearity coming from information contagion effects. Then, we prove that the approximation problem admits a sequence of uniformly bounded regular Sobolev solutions in a suitably chosen Sobolev space. We then show that the above approximating sequence of solutions converges to the unique Sobolev solution of the formal recursive HJB PDE in some appropriate Sobolev space.}  {Lastly, we prove a verification theorem, in which the unique Sobolev solution obtained earlier is shown to coincide with the value function of the risk-sensitive control problem.} 

We develop a numerical study assess the sensitivity of the optimal investment strategies to contagious distress, stock volatilities and risk aversion of the investor. We consider a minimal market model with two stocks and two hidden regimes, so to exemplify the typical behavior of the investor. We find that the investor increases his holdings in the stock if the filter probability of being in the ``most'' profitable regime, i.e. the one associated with higher growth rate of the stock, increases. As the distress risk of a stock gets higher, he decreases his holdings in the riskier stock and invests the saving proceeds in the safer stock. {\Red Extrapolating these results to the case of multiple regimes, we expect the investor's strategy to be sensitive to the ratio of the filter probabilities. For instance, in the case of three regimes, we expect the investor to increase his stock holdings as the filter probability of being in the most profitable regime increases. An increase, yet of more moderate size, is also expected if such a probability increases, relative to that of being in the regime of medium profitability.} 
The distress of a stock leads the investor to reduce the optimal holdings in other stocks because of the contagion effect. Moreover, we find that the optimal investment decisions are most sensitive to the risk aversion level of the investor if all stocks are in the safe state. As the volatility of a stock increases, the investor shifts his wealth to the less volatile stock if the filter probability of staying in the high distress regime is not too high, and to the money market account otherwise.

Contagious distress leads to a recursive dependence structure between the HJB-PDEs. An explicit solution can only be obtained for the PDE associated with the state where all stocks are distressed. Hence, the analysis of HJB PDEs associated with states in which two or more stocks are distressed must rely on properties of solutions to PDEs associated with states in which a higher number of stocks are distressed. Our recursive system of HJB PDEs (see Eq.~\eqref{eq:HJBeqn1fix} below) presents technical challenges coming from the simultaneous presence of (i) two terms with quadratic gradient growth, and (ii) an unbounded non-Lipschitz continuous exponential term which is nonlinear in the solution. {\Red Because of this complex structure, the existence of classical solutions cannot be established using standard arguments. In the context of a liquidity control problem, \cite{Pham} characterize
the value function as the unique viscosity solution to a system of integro-ordinary differential equations. The regularization technique discussed by \cite{Pham} (see also \cite{FeGaGo})
considers viscosity solutions to the HJB equation and then prove the $C^{1,2}$ regularity. This is not applicable to our risk-sensitive control framework because the utility function does not possess the homogeneity property.
Despite PDEs with quadratic gradient growth have been thoroughly analyzed by \cite{Kob00}, the uniqueness of the viscosity solution is guaranteed under growth conditions which fail to
be satisfied in our case due to the unbounded nonlinearity, see also Remark \ref{rem:weak-why}. These considerations lead us to consider a different type of solution, namely the Sobolev
solution, for our recursive system of HJB PDEs.}  {\Red Prior literature has studied existence and uniqueness of Sobolev solutions to HJB equations, see
for instance \cite{Cerrai}, \cite{Gozzi96}, \cite{SIAM1} and \cite{Masiero14}. 
    \cite{Cerrai} (see Section 9.4 therein) considers the same HJB equation (see Eq.~(9.4.1) therein) as \cite{Gozzi96}, but only discusses mild solutions. \cite{Masiero14} study
    mild solutions to a class of semilinear Kolmogorov equations where the nonlinearity has superquadratic growth with respect to the derivative of the solution. 
    Their analysis requires the global Lipschitz continuity of the nonlinearity, which fails to be satisfied in our case because of (ii). For the same reason, the regularity method proposed by \cite{Gozzi96} is not directly applicable to our setup as an a priori $C^2$ estimate for the norm of local solutions cannot be obtained.
    Our proposed approach is to establish an a priori estimate in a Sobolev space, and then prove the existence of {approximating solutions} using fixed point methods.
}


    {The rest of the paper is organized as follows. Section \ref{sec:model} introduces the market model. Section \ref{sec:utilmax} derives the recursive system of HJB PDEs for our control problem}.
Section \ref{sec:default-contagion} develops a rigorous mathematical analysis of the optimal strategies and HJB equations. Section \ref{sec:verif} proves the verification theorem. Section~\ref{sec:numerical} presents a numerical analysis. Some proofs of technical results are delegated to an Appendix.

\section{The Market Model}\label{sec:model}
Section \ref{sec:notations} introduces notations and definitions which will be extensively used throughout the paper. Section \ref{markmod} introduces the securities at disposal of the investor.

\subsection{Notation and Definitions}\label{sec:notations} Bold letters are used to denote real vectors and matrices. We use ${\bm 0}=(0,\ldots,0)$ to denote a vector consisting of all zero entries, and ${\bm 1}$ to denote a row vector with all entries equal to $1$.
For a real vector or matrix ${\bm v}$, we use ${\bm v}^{\top}$ to denote its transpose. We use $|\bm{x}|$ to denote the {$L^2$-norm} of the real vector $\bm {x}$. For a given real matrix $\bm{a}$, ${\rm Tr}[\bm a]$ denotes its trace. For a given binary $N$-dimensional vector ${\bf z} =(z_1,\ldots,z_N)\in {\cal S}:=\{0,1\}^N$ such that $z_i = 0$, we use
\begin{equation}
{\bf z}^i:=(z_1,\ldots,z_{i-1},1-z_i,z_{i+1},\ldots,z_n)
\label{eq:flip}
\end{equation}
to denote the vector obtained by flipping the $i$-th component of the vector from zero to one. For a given real vector ${\bm v}$, and binary vector ${\bf z}$, we denote by
\begin{equation}
{\bm v}_{\bf z} := ((1-z_1)v_1,\ldots,(1-z_N)v_N)^{\top}
\label{eq:vz}
\end{equation}
the vector obtained by multiplying ${\bm v}$ with ${\bf 1}-{\bf z}$ componentwise. For a given real $K \times N$ dimensional matrix ${\bm B}=(B_{k,i};\ (k,i)\in\{1,\ldots,K\}\times\{1,\ldots,N\})$, we denote by
\begin{equation}
{\bm B}_{\bf z}:= \big( B_{k,i} (1-z_i);\ (k,i) \in \{1,\ldots,K) \times \{1,\ldots,N\}\big).
\label{eq:Bz}
\end{equation}
For a given vector $\bm{v}$, we use $diag(\bm{v})$ to denote a diagonal matrix whose $(i,i)$-th entry is $v_i$.

Let $j_1, j_2 ,\ldots,j_n$ be distinct indices in $\{1,\ldots,N\}$. We use ${\bf0}^{j_1,\ldots,j_n}$ to denote the $N$-dimensional vector whose entries are zero except for entries $j_1,\ldots,j_n$ which are set to one.
We use the notations ${\rm H}$ and ${\rm W}$ to denote the Sobolev spaces. We use $\bm{W}$ and $\bm{H}$ to denote, respectively, the multidimensional Brownian motion and default indicator vector.

\subsection{The Portfolio Securities}\label{markmod}
We consider a frictionless financial market  consisting of $N+1$ securities: a risk-free bank account, and $N$ stocks.
We next specify the dynamics of each of the market securities:

\noindent\textbf{Risk-free bank account.} The dynamics of the risk-free bank account $B=(B(t);\ t\geq0)$ is given by $\D B(t) = r B(t) \D t$, $B(0)=1$, where $r$ is the instantaneous interest rate.

\noindent\textbf{Stock prices.} We use doubly stochastic intensity models to model the distress times of the stocks. Concretely, let $\tau_i$ be the distress time of the $i$-th stock, defined as
\begin{eqnarray}\label{eq:i-defult-time}
\tau_i := \inf\left\{t>0;\ \int_0^t h_i(u,{\bm X}(u),{\bm H}(u)) \D u \geq \Theta_i\right\},\ \ \ \ \ i=1,\ldots,N,
\end{eqnarray}
where $(\Theta_i;\ i=1,\ldots,N)$ are mutually independent unit mean exponential random variables, independent of ${\bm X}$ and ${\bm W}$. We follow the convention $\inf\emptyset=+\infty$. {The vector ${\bm H}(t)=(H_1(t),\ldots,H_N(t))$, $H_i(t):={\bf1}_{\tau_i\leq t}$ for $i=1,\ldots,N$, indicates the occurrence of the distress event of the $i$-th stock. We also refer to \cite{CappFrei} for mathematical details on the construction of such a multi-name default risk model.}

We define 
$\mathcal{H}_t = \bigvee_{i=1}^N {\cal H}_t^i$, where $\mathcal{H}^{i}_t = \sigma(H_i(u);\ u \leq t)$. Further, we set $\mathcal{G}_t^{i} = \mathcal{F}_t^{i} \vee {\cal H}_t^i$, and  $\mathcal{G}_t = \bigvee_{i=1}^N \mathcal{G}^i_t \vee {\F}_t^{{\bm X}}$. We take the right continuous versions of both $\mathcal{G}_t^{i}$, $i=1,\ldots,N$, and $\mathcal{G}_t$.
It can be proven that $(h_i(t,{\bm X}(t),{\bm H(t)});\ t\geq0)$ is the $(\Px,\mathcal{G}_t^{i})$-\emph{hazard rate} of $\tau_i$ {(see \cite{bielecki01}, Section 6.5 for details)}. That is
\begin{equation}\label{MrtRprDftPrcP}
	\Xi^{\bm X}_i(t) := H_i(t) - \int_0^{t\wedge\tau_i} \bar{H}_i({u}) h_i(u,\bm{X}(u),\bm H(u)) \D u
\end{equation}
is a $(\Px,\mathcal{G}_t^{i})$-martingale.
Hence for $i \in \{1,\ldots,N\}$, ${\bm h}_i(t,{\bf z}) = (h_i(t,{\bm e}_1,{\bf z}),\dots,h_i(t,{\bm e}_K,{\bf z}))^{\top}\in(0,\infty)^{K}$ denotes the values that the hazard rate
process of the $i$-th stock can take in the different regimes at time $t$. The price dynamics of the $i$-th stock is given by\footnote{{The focus of our study is on the impact of information-driven contagion on optimal investment decisions. 
For this reason, we consider a diagonal volatility matrix, i.e. no dependence between market risks of stocks is imposed. In presence of two sources of dependencies (default contagion and volatility), it would be difficult to separate the impact of contagion from market risk on the optimal strategies. From a mathematical perspective,
 replacing the noise term $\vartheta_i \D W_i(t)$
with the more general form $\sum_{j=1}^N\vartheta_{ij}\D W_j(t)$, where ${\bm\Sigma}=(\vartheta_{ij})$ is non-diagonal and satisfying invertibility conditions, would require straightforward adjustments to the analysis.} }
\begin{eqnarray}\label{eq:price-dynamics-i-stock1}
\frac{\D P_i(t)}{P_{i}(t)} = \left( b_i(t,{\bm X}(t),\bm H(t)) + h_i(t,{\bm X}(t),\bm H(t)) \right) \D t + \vartheta_i(\bm H(t)) \D W_i(t),\ \ \ \ \ P_i(0) = P^{\circ,i},
\end{eqnarray}
where the volatility $\vartheta_i({\bf z})>0$ for $i=1,\ldots,N$, $P^{\circ,i} \in \mathds{R}^+$, $b_i: \mathbb{R}_{+}\times\{{\bm e}_{1},\dots,{\bm e}_{N}\}\times{\cal S}\to\mathbb{R}_{+}$ is a deterministic function, and $\vartheta_i:{\cal S}\to\R_+$ is the volatility of the $i$-th stock. Hence, both volatility and instantaneous return of the $i$-th stock depend on the distress states of other stocks in the portfolio. This is in line with empirical evidence as market volatility usually spikes at times when firms experience financial distress.

\begin{remark}
Differently from the drift, the volatility of the stock depends on the distress states of the other stocks in portfolio, but is independent of the regime in place. This is justified by two main considerations.
If $\vartheta_i$, $i=1,\ldots,N$, were depending on ${\bm X}$, the Markov chain would be observable. The problem would then reduce to a fully observed, as opposed to a partially observed, stochastic control problem. We next explain why this would be the case. Consider a diffusion process $V(t) = V(0) + \int_0^t \kappa(s) \D s + \int_0^t \vartheta(s) \D W_s$. 
Let $t=t_0<t_1<\cdots<t_n=T$ denote the trading times over the time interval $[t,T]$. It is known that, as $\sup_{k=1,\ldots,n}(t_k-t_{k-1})\too0$, the realized variance converges in probability to the quadratic variation of $V(t)$, i.e.,
\begin{align*}
\sum_{k=1}^n\big(V(t_i)-V(t_{i-1})\big)^2\too \int_t^T \vartheta^2(s)\D s.
\end{align*}
This approximation is commonly adopted in practice.
Suppose that the volatility process depends on the Markov chain ${\bm X}$. This means that $\vartheta_i(t) = \left( \vartheta_i(t,{\bm e}_1,{\bf z}),\ldots,\vartheta_i(t,{\bm e}_K,{\bf z}) \right)$, where all vector components are distinct. Set $V(t) = {\log P_i(t)}$, and apply the above result. Since we observe returns $(\log\frac{P_i(t_k)}{P_i(t_{k-1})};\ k=1,\ldots,n)$ on a high-frequency scale ($\sup_{m=1,\ldots,n}(t_m-t_{m-1})\too0$), i.e. we know $\sum_{i=1}^n\big(\log\frac{P_i(t_k)}{P_i(t_{k-1})}\big)^2$, then we would know $\int_t^T \vartheta_i^2(s,{\bm X}(s),{\bm H}(s)) \D s$ for all $0<t<T$. This means that we can deduce $\vartheta_i(t,{\bm X}(t),{\bm H}(t))$ if we take $T$ sufficiently close to $t$. But then, being the values $\vartheta_i(t,{\bm e}_k,{\bf z})$'s distinct for $k=1,\ldots,K$, we would be able to recover the state ${\bm X}$. Hence, our problem would no longer be partially observed but rather fully observed. In addition to the theoretical argument described above, there is also a practical reason for why we have opted for a regime independent volatility process. Regime independent volatilities have been shown to provide a good calibration fit to data; see, for instance, \cite{Liechty} for a regime switching model of this type calibrated to data from the New York Merchantile stock exchange.
\end{remark}

{\Red
\begin{remark}
We outline a calibration procedure for the proposed investment model.
While drift and volatilities can be readily estimated using historical time series of equity data, estimates of historical {distress} intensities are more difficult to obtain given that financial distress occurs rarely in practice. Previous literature has considered interacting intensity models of the form $h_i(t,{{\bm H}(t)}) = \hat{h}_i(t,\sum_{j=1}^N H_j(t))$, and calibrated them to tranched credit products. An example of such a model is the so-called convex counterparty risk model proposed in \cite{FreyBackhaus04}, which assumes a distress intensity of the form
$$
\hat{h}_i(t,m) = \omega_{i,0} + \frac{\omega_1}{\omega_2} \left(e^{\omega_2 \frac{(m-o(t))^+}{m}} -1 \right),\qquad {m=0,1,\ldots,N},
$$
where $\omega_{i,0} > 0$, $\omega_1 \geq 0$, and $\omega_2 > 0$. The parameter $o(t)$ measures the expected number of firms entered into a distress state by time $t$. The parameter $\omega_1$ gives the slope of $\hat{h}_i(t,m)$ at $o(t)$ and intuitively models the strength of contagious distress under ``normal'' conditions. The interaction parameter $\omega_2$ controls the tendency of the model to generate contagion and produce {distress propagation}. \cite{FreyBackhaus04} calibrate this model to single-name credit derivatives and five-year tranche spreads on the iTraxx Europe. In the above specification of the distress intensity, we can interpret the parameter $\omega_{i,0}$ as the idiosyncratic component measuring the credit quality of the $i$-th stock. Next, using empirical estimates of distress risk premia, defined as the ratio of risk-neutral to historical distress intensities, such as those provided by \cite{Driessen}, 
 it is possible to produce historical distress intensities from the above estimates $\hat{h}_i(t,m)$'s of risk-neutral distress intensities.
It then remains to be determined the impact of the hidden Markov chain on distress risk. For this, we can use the numerical estimates provided by \cite{Longstaff}, who employ a three-state homogenous regime-switching model to
explain variations in the realized distress intensities of the U.S. corporate bond market over the course of 150 years. They find the existence of three regimes, ``low'', ``middle'' and ``high'', which constitute a proxy for a good, medium and bad macroeconomic environment. They also produce estimates for the regime switching probabilities, which can be readily used as a proxy for the transition rates of our hidden Markov chain. 
\end{remark}
}
We also impose the following mild technical assumption throughout the paper:
\begin{itemize}
  \item[({\bf A1})] For $(i,k,{\bf z}) \in \{1,\dots,N\}\times\{1,\ldots,K\}\times{\cal S}$, $b_i(\cdot,{\bm e}_k,{\bf z})\in C^1(\R_+)$, and the distress intensity of the $i$-th stock $h_i(\cdot,{\bm e}_k,{\bf z}) \in C^1(\R_+)$.
\end{itemize}
{The continuity of $b_i(\cdot,{\bm e}_k,{\bf z})$ and $h_i(\cdot,{\bm e}_k,{\bf z})$ is the only required condition for the proofs of Lemma \ref{lem:lower-bound-xi} and Lemma~\ref{lem:mtMt} in Section~\ref{sec:default-contagion}, and guarantees that the change of measure developed in Section~\ref{sec:filterprob} is well defined. However, the proof of Lemma \ref{lem:sing2-b} in the next section also requires that the derivatives of $b_i(\cdot,{\bm e}_k,{\bf z})$ and $h_i(\cdot,{\bm e}_k,{\bf z})$ are continuous, i.e. ({\bf A1}). These lemmas will play a key role in establishing the main results and in the proof of the verification theorem. We impose the assumption ({\bf A1}) henceforth.
}

We define the market filtration $\mathcal{G}_t^I := \mathcal{F}_t  \vee \mathcal{H}_t$
to be the subfiltration of {$\G_t$} capturing market available information, which includes stock security prices and occurrence of distress events. We take the right continuous version of $\mathcal{G}_t^I$. We may also write $\mathcal{G}_t = \mathcal{F}_t^{{\bm X}} \vee \mathcal{G}_t^I$. 

\section{Optimal Portfolio Problem}\label{sec:utilmax}
We consider a power investor who maximizes his expected utility from terminal wealth by dynamically allocating his wealth into a risk free bank account and $N$ stocks. We denote by $\nu_B(t)$ the number of shares of the risk-free bank account held by the investor at {time} $t$, and by $\nu_{i}(t)$ the shares of the $i$-th stock ($i=1,\ldots,N$) held by him at {time} $t$. We denote by $V^{{\bm \nu}}(t)$ the wealth of the portfolio process ${\bm \nu}=(\nu_B,\nu_{i};\ i=1,\ldots,N)$ at time $t$. We consider strategies where no investment is made on a stock after it entered the distress state. When this happens, the investor liquidates the position at the current market value, and allocates the available wealth to the remaining securities in the portfolio. The wealth process is given by
\begin{eqnarray*}
V^{{\bm \nu}}(t)=\nu_B(t) B(t) + \sum_{i=1}^N \nu_i(t) P_i(t).
\end{eqnarray*}
Under the self-financing condition, we obtain its dynamics
\[
\D V^{\bm \nu}(t) = \nu_B(t) \D B(t) + \sum_{i=1}^N \nu_i(t) \D P_i(t).
\]
For $t\geq0$, define the fractions of wealth invested into the securities
\begin{eqnarray}\label{eq:fraction}
\pi_B(t) := \frac{\nu_B(t) B(t)}{V^{\bm \nu}(t)},\ \ \ \ \pi_i(t):=\frac{\nu_i(t) P_i(t)}{V^{\bm \nu}(t)}, \ \ \ \ i=1,\ldots,N,
\end{eqnarray}
if $V^{\bm \nu}(t) > 0$, and we set $\pi_B(t) = \pi_i(t) = 0$ whenever $V^{\bm \nu}(t) = 0$. Hence, $\pi_B(t)$ is the proportion of wealth invested in the money market account, while $\pi_i(t)$ is the proportion of wealth invested in the $i$-th stock at time $t$. Next, we define the class of admissible strategies.
{
\begin{definition} \label{def:admissi}
The class of admissible strategies $\tilde{{\cal U}}(t,T)$, $0 \leq t \leq T$, is a class of $\mathbb{G}^I$-adapted self-financing trading strategies given by
${\bm \pi}(s):=(\pi_1(s),\ldots, \pi_N(s))^{\top}$, $s\geq t$,
where $\pi_{i}(s)$ denotes the number of shares of the $i$-th risky stock before it enters into a distressed state, i.e. $\pi_i(s) = (1-H_i(s)) \pi_i(s)$ for $s\geq t$,
such that there exists a sequence of stopping times $t= \varsigma_0 < \varsigma_1 < \cdots \varsigma_{l} = T$, ${l} > 1$, so that
\begin{equation}
\Ex^{\hat{\Px}} \left[e^{\int_{\varsigma_j}^{\varsigma_{j+1}}|{\bm \pi}(s)|^2\D s} \right] <+\infty, \qquad \qquad {j} =0,\ldots,{l}-1.
\label{cond:admiss}
\end{equation}
\end{definition}
}
For a given admissible trading strategy $\bar{{\bm \pi}}(t)=(\pi_B(t),{\bm \pi}(t))$, ${\bm \pi}(t)=(\pi_1(t),\ldots,\pi_N(t))^{\top}$, the dynamics of the resulting wealth process may be written in terms of $\bar{{\bm \pi}}$  as:
\begin{eqnarray}\label{eq:sde-Vpi}
\frac{\D V^{\bar{{\bm \pi}}}(t)}{V^{\bar{{\bm \pi}}}(t-)} = \pi_B(t) \frac{\D B(t)}{B(t)} + \sum_{i=1}^N \pi_i(t) \frac{\D P_i(t)}{P_{i}(t)}.
\end{eqnarray}
Set $V^{{\bm \pi}}(0) = v>0$, where $v>0$ is the initial budget. Using the dynamics \eqref{eq:price-dynamics-i-stock1} along with the identity $\pi_B(t)=1-\sum_{i=1}^N\pi_i(t)$, it follows that the wealth process only depends on ${\bm \pi}(t)$ and is given by
\begin{eqnarray}\label{eq:sde-Vpi1}
\frac{\D V^{{\bm \pi}}(t)}{V^{{\bm \pi}}(t)} = r\D t + \sum_{i=1}^N \pi_i(t)(b_i(t,\bm{X}(t),\bm H(t)) + h_i(t,\bm{X}(t),\bm H(t)) -r)\D t +\sum_{i=1}^N \vartheta_i(\bm H(t))\pi_i(t) \D W_i(t).
\end{eqnarray}
Then the solution to Eq.~\eqref{eq:sde-Vpi1} is given by
\begin{eqnarray}\label{eq:solution-sde-Vpi}
 V^{{\bm \pi}}(t) = v \exp\left(\int_0^t\phi(s,\bm{X}(s),{\bm H}(s),{\bm \pi}(s))\D s  + \int_0^t{\bm \pi}(s)^{\top} {\bm \Sigma} \D {\bm W}(s) - \frac{1}{2}\int_0^t{{\bm \pi}}(s)^{\top} {\bm \Sigma} {\bm \Sigma}^{\top}{{\bm \pi}}(s)\D s \right),
\end{eqnarray}
where ${\bm \Sigma}=diag((\vartheta_1({\bf z}),\ldots,\vartheta_N({\bf z}))$ and $\phi(t,{\bm e}_j,{\bf z},{{\bm \pi}}):= r + \sum_{i=1}^N\pi_i(b_i(t,{\bm e}_j,{\bf z}) + h_i(t,{\bm e}_j,{\bf z})-r)$.
The objective of the power investor is to choose an admissible trading strategy ${\bm \pi}(t)$ so to maximize his expected terminal utility
\begin{equation}
\Ex[U(V_T^{\bm\pi})], \qquad U(v) = \frac{v^{\gamma}}{\gamma},\quad v>0,
\label{eq:partialobs}
\end{equation}
where $\gamma \in (0,1)$ is a fixed constant. The problem given in~\eqref{eq:partialobs} is a maximization problem with partial information, since
the economic factors $\bm{X}$ are not directly observable, and the strategies can only be based on past information of stock prices. We transform it to a
fully observed control problem in three steps. First, we give the filter probabilities in section \ref{sec:filterprob}. We then use them to define the state of a fully observed risk sensitive control problem in section \ref{sec:risksens}. We derive the recursive system of HJB PDEs associated with it in section \ref{sec:HJB}.

Before proceeding further, we introduce some notation and terminology.
We denote by $\mathcal{E}(I)$ the stochastic exponential of a stochastic integral process $I(t)=\int_0^t {\bm a}(s)^{\top} \D {\bm C}(s)$, where ${\bm C}=({\bm C}(t);\ t\geq0)$ is an $\mathbb{R}^{N}$-valued {continuous It\^o} process, and ${\bm a}(t)=(a_1(t),\ldots,a_N(t))^{\top}$ is $\G_t$-predictable. It holds that
\begin{equation}
\mathcal{E}_t(I) = {\exp\left(\int_0^t {\bm a}(s)^{\top} \D {\bm C}(s) - \frac{1}{2} \int_0^t {\bm a}(s)^{\top}  {{\bm a}(s) \D [{\bm C},{\bm C}]_s}\right),}
\label{eq:stocexpcont}
\end{equation}
where $[{\bm C},{\bm C}]_s := ([C_i, C_i]_s; \ i = 1,\ldots,N)$, and $[C_i,C_i]_s$ denotes the quadratic variation of the It\^{o} process ${C}_i$ at time $s$. Let $Z(t) = \sum_{i=1}^N\int_0^t \kappa_i(s) \D \Xi^{\bm X}_i(s)$, where $\Xi^{\bm X}_i(s)$ is of the form defined in \eqref{MrtRprDftPrcP}, and the vector ${\bm \kappa}(s)=(\kappa_1(s),\ldots,\kappa_N(s))$ is $\G_s$-predictable, with $\kappa_i(s) > -1$ for all $s\geq0$ and $i=1,\ldots,N$. Then
\begin{equation}
\mathcal{E}_t(Z) = {\exp\left[\sum_{i=1}^N\left(\int_0^t \log(1+\kappa_i(s)) \D H_i(s) - \int_0^{t \wedge \tau_i} \kappa_i(s) h_i(s,\bm{X}(s),{\bm H}(s)) \D s\right)\right].}
\label{eq:stocexpdisc}
\end{equation}
It is well known (see \cite{bielecki01}, section 4.3) that $R(t) := \mathcal{E}_t({I}) \mathcal{E}_t(Z)$ follows the SDE:
\begin{equation}
R(t) = 1 + \int_{{(0,t]}} R(s-) \left({\bm a}(s)^{\top} \D {\bm C}(s) + \sum_{i=1}^N\kappa_i(s) \D \Xi^{\bm X}_i(s) \right),\ \ \ \ t\geq0.
\label{eq:stocexpdyn}
\end{equation}

\subsection{Filter Probabilities} \label{sec:filterprob}
Denote the filter probabilities of the Markov chain $\bm{X}$ by
\begin{eqnarray}\label{eq:filter-p}
p_k(t) := \Px\left(\bm{X}(t)={\bm e}_k|\G_t^I\right),\ \ \ \ k\in\{1,\ldots,K\},
\end{eqnarray}
where we recall that $K$ is the number of states of $\bm{X}$. Further, let $\mathcal{P}$ be the class of probability vectors. Define the function $\hat{g}:D_1 \times \mathcal{P} \times D_2\too\R$ by
\begin{eqnarray}\label{eq:tildeg}
\hat{g}(y,{\bm p},\upsilon) := \sum_{k=1}^{K}g(y,{\bm e}_k,\upsilon) p_k,
\end{eqnarray}
where $g:D_1\times\{{\bm e}_1,\ldots,{\bm e}_K\}\times D_2 \too \R$, and $D_1, D_2$ are arbitrary, possibly empty, domains.
Consider the log-price process $Y_i(t):=\log(P_i(t))$. Then ${\bm Y}(t)=(Y_1(t),\ldots,Y_N(t))^{\top}$ satisfies the SDE:
\begin{align}\label{eq:pre-default-logprice}
\D {\bm Y}(t) &= {\bm \mu}(t,\bm{X}(t),{\bm H}(t))\D t + {\bm \Sigma}({\bm H}(t)) \D {{\bm W}}(t)\nonumber\\
&= {\bm \Sigma}({\bm H}(t)) \left({\bm \Sigma}^{-1} {\bm \mu}(t,\bm{X}(t),{\bm H}(t))\D t + \D {\bm W}(t)\right)=: {\bm \Sigma}\D\hat{{\bm W}}(t),
\end{align}
where the $N$-dimensional column vector
\begin{eqnarray}\label{eq:Y-coefficients}
{\bm \mu}(t,{\bm e}_k,{\bf z}) := \left[b_1(t,{\bm e}_k,{\bf z}) + h_1(t,{\bm e}_k,{\bf z}) -{\vartheta_1^2({\bf z})}/{2},\ldots,b_N(t,{\bm e}_k,{\bf z}) + h_N(t,{\bm e}_k,{\bf z})-{\vartheta_N^2({\bf z})}/{2}\right]^{\top}.
\end{eqnarray}
We have the following system of SDEs satisfied by the vector ${\bm p}(t)=(p_1(t),\ldots,p_K(t))^{\top}$ of filter probabilities:
\begin{proposition}[\cite{FreySchmidt}, Proposition 3.6]\label{prop:normzlized-sde}
The vector ${\bm p}(t)$ of filter probabilities satisfies
\begin{align}\label{eq:normzlized-sde}
\D p_k(t) &= \sum_{\ell=1}^K \varpi_{\ell,k}(t)p_{\ell}(t)\D t + p_k(t)\left({\bm \mu}(t,{\bm e}_k,{\bm H}(t))-\hat{{\bm \mu}}(t,{\bm p}(t),{\bm H}(t))\right)^{\top}({\bm \Sigma}{\bm \Sigma}^{\top})^{-1}
\left(\D {\bm Y}(t)-\hat{{\bm \mu}}(t,{\bm p}(t),{\bm H}(t))\D t\right)\nonumber\\
&\quad+p_k(t-)\sum_{i=1}^N\left(\frac{h_i(t,{\bm e}_k,{\bm H}(t-))}{\hat{h}_i(t,{\bm p}(t-),{\bm H}(t-))}-1\right)\D \Xi_i(t).
\end{align}
Above, the $\Gx^I$-martingale $\Xi_i=(\Xi_i(t);\ t\geq0)$ is defined, for $i=1,\ldots,N$, as
\begin{eqnarray}\label{eq:GIdefault-mart}
\Xi_i(t) := H_i(t) - \int_0^{t\wedge\tau_i}\hat{h}_i(s,{\bm p}(s),{\bm H}(s))\D s,\ \ \ \ \ t\geq0.
\end{eqnarray}
\end{proposition}

Next, we introduce the following change of probability measure:
\begin{eqnarray}\label{eq:P->hatP}
\frac{\D\hat{\Px}}{\D\Px}\Big|_{\G_t} = \Psi(t),
\end{eqnarray}
where $\Psi(t)$ is a density process which satisfies the following SDE:
\begin{eqnarray}\label{eq:density1}
\Psi(t) = 1 + \int_0^{t}\Psi(s-)\left(-({\bm \Sigma}^{-1}{{\bm \mu}}(s,\bm{X}(s),{\bm H}(s)))^{\top}\D {\bm W}(s) + \sum_{i=1}^N\frac{1-h_i(s,{\bm X}(s-),{\bm H}(s-))}{h_i(s,{\bm X}(s-),{\bm H}(s-))}\D \Xi_i^{\bm X}(s)\right),
\end{eqnarray}
with $\Xi^X_i(t)=H_i(t)-\int_0^{t\wedge\tau_i}h_i(s,\bm{X}(s),{\bm H}(s))\D s$. More precisely, we can write the above density as
\begin{eqnarray}\label{eq:stoch-exp1}
\Psi(t) ={\cal E}_t\left(\int_0^{\cdot}-({\bm \Sigma}^{-1}{\bm \mu}(s,\bm{X}(s),{\bm H}(s)))^{\top}\D {{\bm W}}(s)\right)
{\cal E}_t\left(\sum_{i=1}^N\int_0^{\cdot}\frac{1-h_i(s,{\bm X}(s-),{\bm H}(s-))}{h_i(s,{\bm X}(s-),{\bm H}(s-))}\D \Xi^{\bm X}_i(s)\right).
\end{eqnarray}
We next show that $\hat\Px$ is well-defined, by verifying that $\mathbb{E}\left[\Psi(T)\right]=1$.  To this purpose, we use a general version of Novikov's condition, proven in \cite{ProtterShimbo} (see Theorem 9 therein), stating that the stochastic exponential $\mathcal{E}(M)$ of a locally square integrable martingale $M$ is a martingale on $[0,T]$ if
\begin{equation}\label{GeneralNovikov}
	\Ex\left[e^{\frac{1}{2}\left<M^{c},M^{c}\right>_{T}+\left<M^{d},M^{d}\right>_{T}}\right]<\infty,
\end{equation}
where $M^{c}$ and $M^{d}$ are the continuous and purely discontinuous martingale parts of $M$. Here, $\left<M^{c},M^{c}\right>_{T}$ and $\left<M^{d},M^{d}\right>_{T}$ denote the conditional quadratic variations for $M^c$ and $M^d$ at time $T$ respectively (see \cite{Protter}, Page 70). From Eq.~\eqref{eq:density1}, $\Psi(t)=\mathcal{E}_{t}(M)$ where
\[
	M(t)= -\int_{0}^{t} {\bm \mu}(s,\bm{X}(s),{\bm H}(s))^{\top} {\bm \Sigma}^{-1} \D {\bm W}(s) + \sum_{i=1}^N\int_{0}^{t}\frac{1-h_i(s,{\bm X}(s-),{\bm H}(s-))}{h_i(s,{\bm X}(s-),{\bm H}(s-))}\D \Xi^{\bm X}_i(s).
\]
Hence, we have
\begin{align*}
\left<M^{c},M^{c}\right>_{T}&=\int_{0}^{T}{\bm \mu}(s,\bm{X}(s),{\bm H}(s))^{\top}({\bm \Sigma}^{\top} {\bm \Sigma})^{-1} {\bm \mu}(s,\bm{X}(s),{\bm H}(s))\D s,\\
	\left<M^{d},M^{d}\right>_{T}&= \sum_{i=1}^N\int_0^T\left[\frac{1-h_i(s,{\bm X}(s),{\bm H}(s))}{h_i(s,{\bm X}(s),{\bm H}(s))}\right]^2h_i(s,{\bm X}(s),{\bm H}(s))\bar{H}_i(s)\D s.
\end{align*}
Clearly, $\left<M^{c},M^{c}\right>_{T}$ is bounded in view of Assumption ({\bf A1}). It remains to prove that $\left<M^{d},M^{d}\right>_{T}$ is also bounded. Indeed, we have
\begin{align*}
\left<M^{d},M^{d}\right>_{T}&=\sum_{i=1}^N\int_0^T\frac{(1-h_i(s,{\bm X}(s),{\bm H}(s)))^2}{h_i(s,{\bm X}(s),{\bm H}(s))}\bar{H}_i(s)\D s
\nonumber\\
&\leq \sum_{i=1}^N\int_0^T\left(\frac{1}{h_i(s,\bm{X}(s),{\bm H}(s))}+h_i(s,\bm{X}(s),{\bm H}(s))\right)\D s.
\end{align*}
{Notice that by ({\bf A1}), for all $(i,k,{\bf z})\in\{1,\ldots,N\}\times\{1,\ldots,K\}\times{\cal S}$, the function $h_i(t,{\bm e}_k,{\bf z})$ is continuous w.r.t. time $t$, and strictly positive. Hence, $1/h_i(t,{\bm e}_k,{\bf z})$ is also continuous and strictly positive.} Thus
\[
\max_{(i,k,{\bf z})\in\{1,\ldots,N\}\times\{1,\ldots,K\}\times{\cal S}}\left\{\sup_{0\leq t\leq T}\left(h_i^{-1}(t,{\bm e}_k,{\bf z})+h_i(t,{\bm e}_k,{\bf z})\right)\right\}<C,
\]
for some constant $C>0$. Thus, we conclude that $\left<M^{d},M^{d}\right>_{T}$ is also bounded.

Hence, it follows that under the new probability measure $\hat{\Px}$, the process $\hat{{\bm W}}$ defined by~\eqref{eq:pre-default-logprice} is a $\Gx^I$-Brownian motion and $\hat{\Xi}_i(t):=H_i(t)-\int_0^{t\wedge\tau_i}\D u$ is a $\Gx^I$-martingale for each $i=1,\ldots,N$. Thus, under the new probability measure $\hat{\Px}$, the system of SDE \eqref{eq:normzlized-sde} is given by
\begin{align}\label{eq:normzlized-sde-HatP}
\D p_k(t) &= \sum_{\ell=1}^K \varpi_{\ell,k}(t)p_{\ell}(t)\D t + p_k(t)\left({\bm \mu}(t,{\bm e}_k,{\bm H}(t))-\hat{{\bm \mu}}(t,{\bm p}(t),{\bm H}(t))\right)^{\top}({\bm \Sigma}{\bm \Sigma}^{\top})^{-1}
\left({\bm \Sigma}\D\hat{{\bm W}}(t)-\hat{{\bm \mu}}(t,{\bm p}(t),{\bm H}(t))\D t\right)\nonumber\\
&\quad+p_k(t-)\sum_{i=1}^N\left(\frac{h_i(t,{\bm e}_k,{\bm H}(t-))}{\hat{h}_i(t,{\bm p}(t-),{\bm H}(t-))}-1\right) \left( \D \hat{\Xi}_i(t) + \bar{H}_i(t)(1-\hat{h}_i(t,{\bm p}(t),{\bm H}(t)))\D t\right).
\end{align}

We conclude this section by recalling a useful lemma which will be used in the forthcoming sections.

\begin{lemma}[\cite{CappLopezhidd}, Lemma B.1]\label{lemma:TryPrvRmnPs}
For any $T>0$ and each $k\in\{1,\dots,K\}$, it holds that
\[
{\Px}\big(p_k(t)\in(0,1),\ \ \textrm{for\ all }\ t\in [0,T)\big)=1.
\]
\end{lemma}

\subsection{Risk Sensitive Control Problem} \label{sec:risksens}
This section shows how the partially observed control problem may be reduced to a fully observed control problem, whose state is given by the filter probabilities,. This reduction is done following similar arguments to \cite{CappLopezhidd}. Recall that we are considering a power investor, whose wealth process is given by \eqref{eq:solution-sde-Vpi} under $\Px$. Then, we have
\begin{eqnarray*}
(V^{{\bm \pi}}(T))^{\gamma} 
= v^\gamma\exp\left(-\gamma\int_0^T\eta(s,\bm{X}(s),{\bm H}(s),{{\bm \pi}}(s))\D s+\gamma\int_0^T{{\bm \pi}}(s)^{\top}{\bm \Sigma} \D {{\bm W}}(s)-\frac{\gamma^2}{2}\int_0^T{{\bm \pi}}(s)^{\top}{\bm \Sigma}{\bm \Sigma}^{\top}{\bm \pi}(s)\D s\right),
\end{eqnarray*}
where the function
\begin{eqnarray}\label{eq:eta}
\eta(t,{\bm e}_j,{\bf z},{\bm \pi}) := -r + \sum_{i=1}^N\pi_i(r-b_i(t,{\bm e}_{j},{\bf z}) - h_i(t,{\bm e}_{j},{\bf z}))
+ \frac{1-\gamma}{2}{{\bm \pi}}^{\top}{\bm \Sigma}{\bm \Sigma}^{\top}{{\bm \pi}}.
\end{eqnarray}
Moreover, under the probability measure $\hat{\Px}$, we have that
\begin{eqnarray}
\Ex\left[U(V^{{{\bm \pi}}}(T) )\right] = \frac{v^\gamma}{\gamma}\Ex^{\hat{\Px}}\left[L^{{{\bm \pi}}}(T)\right],
\label{eq:gammaeq}
\end{eqnarray}
where the process $L^{{{\bm \pi}}}(t)$ is defined as
\begin{align}\label{eq:Lt}
L^{{{\bm \pi}}}(t) &:= {\cal E}_t\left(\int_0^{\cdot}{\bm q}(s,\bm{X}(s),{\bm H}(s),{{\bm \pi}}(s))^{\top}{\bm \Sigma}\D\hat{{\bm W}}(s)\right){\cal E}_t\left(\sum_{i=1}^N\int_0^{\cdot}(h_i(s,\bm{X}(s-),{\bm H}(s-))-1)\D\hat{\Xi}_i(s)\right)\nonumber\\
&\quad\times \exp\left(-\gamma\int_0^t\eta(s,\bm{X}(s),{\bm H}(s),{{\bm \pi}}(s))\D s\right),
\end{align}
and the function
\begin{align}
{\bm q}(t,{\bm e}_{j},{\bf z},{{\bm \pi}}) := ({\bm \Sigma}{\bm \Sigma}^{\top})^{-1}{{\bm \mu}}(t,{\bm e}_{j},{\bf z}) + \gamma{{\bm \pi}}.
\label{eq:qfunc}
\end{align}
Define the process
\begin{align}\label{eq:hatL}
\hat{L}^{{{\bm \pi}}}(t) &:= {\cal E}_t\left(\int_0^{\cdot}\hat{\bm q}(s,{\bm p}(s),{\bm H}(s),{{\bm \pi}}(s))^{\top}{\bm \Sigma}\D\hat{\bm W}(s)\right){\cal E}_t\left(\sum_{i=1}^N\int_0^{\cdot}(\hat{h}_i(s,{\bm p}(s-),{\bm H}(s-))-1)\D\hat{\Xi}_i(s)\right)\nonumber\\
&\quad\times \exp\left(-\gamma\int_0^t\hat{\eta}(s,{\bm p}_{s},{\bm H}(s),{{\bm \pi}}(s))\D s\right).
\end{align}
Based on the above established result that {$(\Psi(t);\ 0\leq t\leq T)$ given by \eqref{eq:stoch-exp1} is a $(\Px,\Gx)$}-martingale, and using Proposition 3.3 in \cite{CappLopezhidd}, we obtain the following
\begin{proposition}\label{prop:equivform}
Let ${\bm \pi}$ belong to the class of admissible strategies, i.e. ${\bm \pi} \in \tilde{{\cal U}}(0,T)$, where $\tilde{{\cal U}}(t,T)$ has been introduced in Definition~\ref{def:admissi}. Then the objective functional has the representation
\begin{eqnarray*}
J_T(v,{{\bm \pi}}):=\Ex\left[U(V^{{{\bm \pi}}}(T))\right] = \frac{v^{\gamma}}{\gamma}\Ex^{\hat{\Px}}\left[\hat{L}^{{{\bm \pi}}}(T)\right].
\end{eqnarray*}
\end{proposition}

Since ${\bm p}(t)=(p_1(t),\ldots,p_K(t))^{\top}$ is a degenerate diffusion (because $\sum_{k=1}^K p_k(t)=1$), we consider the projected filter process
$\tilde{{\bm p}}(t):=(p_1(t),\ldots,p_{K-1}(t))^{\top}$. By Lemma \ref{lemma:TryPrvRmnPs}, we have that $\tilde{\bm p}(t)\in\Delta_{K-1}$ for all $t\geq0$, $\tilde{\Px}$-a.s., where the $K-1$ dimensional simplex is defined by
\begin{eqnarray}
\Delta_{K-1}:=\left\{{\bm \lambda}:=(\lambda_1,\ldots,\lambda_{K-1})\in(0,1)^{K-1};\ \sum_{k=1}^{K-1}\lambda_k<1\right\}.
\label{eq:simplexdef}
\end{eqnarray}
Further, we define the function $\tilde{g}:D_1\times{\Delta}_{K-1}\times D_2\too\R$ by
\begin{eqnarray}\label{eq:tildeg}
\tilde{g}(y,{\bm \lambda},\upsilon) := g(y,{\bm e}_K,\upsilon) + \sum_{k=1}^{K-1}(g(y,{\bm e}_k,\upsilon)-g(y,{\bm e}_K,\upsilon))\lambda_k,
\end{eqnarray}
where $g:D_1\times\{{\bm e}_1,\ldots,{\bm e}_K\}\times D_2 \too \R$, and $D_1, D_2$ are arbitrary, possibly empty, domains. Notice that it holds that $\tilde{g}(y,\tilde{\bm p}(t),v)=\hat{g}(y,{\bm p}(t),v)$. To proceed further and develop the reduction to the risk sensitive control problem, we first introduce the following change of probability measure:
\begin{align}
\frac{\D\tilde{\Px}}{\D\hat{\Px}}\Big|_{\G_t^I}&= \zeta(t)\nonumber\\
&:={\cal E}_t\left(\int_0^{\cdot}\hat{\bm q}(s,{\bm p}(s),{\bm H}(s),{\bm \pi}(s))^{\top}{\bm \Sigma}\D\hat{{\bm W}}(s)\right){\cal E}_t\left(\sum_{i=1}^N\int_0^{\cdot}(\hat{h}_i(s,{\bm p}(s-),{\bm H}(s-))-1)\D\hat{\Xi}_i(s)\right).
\label{eq:dPxdtPx}
\end{align}
Under the above probability measure $\tilde{\Px}$, we have that the processes:
\begin{align}
\tilde{{\bm W}}(t)&:=\hat{{\bm W}}(t) - \int_0^t{\bm \Sigma}^{\top}\hat{\bm q}(s,{\bm p}(s),{\bm H}(s),{\bm \pi}(s))\D s,\ \ {\rm and}\ \ \
\tilde{\Xi}_i(t):=H_i(t)-\int_0^{t\wedge\tau_i}\hat{h}_i(s,{\bm p}(s),{\bm H}(s))\D s
\label{eq:ptildeBM}
\end{align}
are respectively a $\Gx^I$-Brownian motion, and a $\Gx^I$-martingale for each $i=1,\ldots,N$.

Notice that differently from $\hat{\Px}$, the measure $\tilde{\Px}$ depends on the investment strategy ${\bm \pi}$. Nevertheless, we can still show that $\tilde{\Px}$ is well-defined, by verifying that $\mathbb{E}^{\hat{\Px}} \left[\zeta(T)\right]=1$.
For this, we use a general version of Novikov's condition, proven in \cite{ProtterShimbo} (see Lemma 13 therein), stating that the stochastic exponential $\mathcal{E}(\bar{M})$ of a locally square integrable martingale $\bar{M}$ is a martingale on $[0,T]$ if there exists a sequence of stopping times satisfying $0=\varsigma_0<\varsigma_1<\cdots<\varsigma_l=T$ with $ l>1$ so that
\begin{equation}\label{GeneralNovikov3}
	\Ex^{\hat{\Px}}\left[e^{\left<\bar{M},\bar{M}\right>_{\varsigma_{j+1}}-\left<\bar{M},\bar{M}\right>_{\varsigma_{j}}}\right]<\infty, \qquad \qquad j=1,\ldots,{l}-1. 
\end{equation}
Here
\[
	\bar{M}(t)= \int_0^{t}\hat{\bm q}(s,{\bm p}(s),{\bm H}(s),{{\bm \pi}}(s))^{\top}{\bm \Sigma}\D\hat{{\bm W}}(s) + \sum_{i=1}^N\int_0^{t}(\hat{h}_i(s,{\bm p}(s-),{\bm H}(s-))-1)\D\hat{\Xi}_i(s).
\]
Hence we have $\left<\bar{M},\bar{M}\right>_t=\left<\bar{M}^{c},\bar{M}^{c}\right>_{t}+\left<\bar{M}^{d},\bar{M}^{d}\right>_{t}$, where for $0\leq t\leq T$,
\begin{align*}
\left<\bar{M}^{c},\bar{M}^{c}\right>_{t}&=\int_{0}^{t}\hat{\bm q}(s,{\bm p}(s),{\bm H}(s),{{\bm \pi}}(s))^{\top}{\bm \Sigma}{\bm \Sigma}^{\top}\hat{\bm q}(s,{\bm p}(s),{\bm H}(s),{{\bm \pi}}(s))\D s,\\
	\left<\bar{M}^{d},\bar{M}^{d}\right>_{t}&= \sum_{i=1}^N\int_0^t(\hat{h}_i(s,{\bm p}(s),{\bm H}(s))-1)^2\bar{H}_i(s)\D s.
\end{align*}
Using $\hat{h}_i(t,{\bm p}(t),{\bf z})=\sum_{k=1}^Kh_i(t,{\bm e}_k,{\bf z})p_k(t)>0$, it follows that
\begin{eqnarray*}
\left<\bar{M}^{d},\bar{M}^{d}\right>_{T}\leq\sum_{i=1}^N\int_0^T\left[1+(\hat{h}_i(s,{\bm p}(s),{\bm H}(s)))^2\right]\D s.
\end{eqnarray*}
By the assumption ({\bf A1}), we can conclude that $\hat{h}_i(t,{\bm p}(t),{\bf z})$ is bounded for all $(t,{\bf z})\in[0,T]\times{\cal S}$, since $p_k(t)\in(0,1)$ for all $k\in\{1,\ldots,K\}$, as proven in
Lemma~\ref{lemma:TryPrvRmnPs}. Thus, we conclude that $\left<\bar{M}^{d},\bar{M}^{d}\right>_{T}$ is bounded. Next, we notice that $\hat{\bm q}(t,{\bm p}(t),{\bf z},{\bm \pi})=\sum_{k=1}^K{\bm q}(t,{\bm e}_k,{\bf z},{\bm \pi})p_k(t)=({\bm \Sigma}{\bm \Sigma}^{\top})^{-1}\hat{\bm \mu}(t,{\bm p}(t),{\bf z}) + \gamma{{\bm \pi}}$. Then, it holds that
\begin{eqnarray*}
&&\hat{{\bm q}}(t,{\bm p}(t),{\bf z},{{\bm \pi}})^{\top}{\bm \Sigma}{\bm \Sigma}^{\top}\hat{\bm q}(t,{\bm p}(t),{\bf z},{{\bm \pi}}) =
\hat{\bm \mu}^{\top}(t,{\bm p}(t),{\bf z})({\bm \Sigma}{\bm \Sigma}^{\top})^{-1}\hat{\bm \mu}(t,{\bm p}(t),{\bf z})+\gamma\hat{\bm \mu}^{\top}(t,{\bm p}(t),{\bf z}){\bm \pi}\nonumber\\
&&\qquad\qquad+\gamma{\bm \pi}^{\top}\hat{\bm \mu}(t,{\bm p}(t),{\bf z})+\gamma^2{\bm \pi}^{\top}{\bm \Sigma}{\bm \Sigma}^{\top}{\bm \pi}=\left|{\bm \Sigma}^{-1}\hat{\bm \mu}(t,{\bm p}(t),{\bf z})+\gamma{\bm \Sigma}^{\top}{\bm \pi}\right|^2.
\end{eqnarray*}
Then, for each admissible strategy ${\bm \pi}\in \tilde{{\cal U}}(t,T)$,
\begin{align*}
\Ex^{\hat{\Px}}\left[e^{\left<\bar{M}^{c},\bar{M}^{c}\right>_{\varsigma_{j+1}}-\left<\bar{M}^{c},\bar{M}^{c}\right>_{\varsigma_{j}}} \right] &= \Ex^{\hat{\Px}}\left[e^{\int_{\varsigma_j}^{\varsigma_{j+1}} \left|{\bm \Sigma}^{-1}\hat{\bm \mu}(s,{\bm p}(s),{\bm H}(s))+\gamma{\bm \Sigma}^{\top}{\bm \pi}(s)\right|^2\D s} \right]\nonumber\\
&\quad\leq e^{C_T} \cdot \Ex^{\hat{\Px}}\left[e^{C_T \int_{\varsigma_j}^{\varsigma_{j+1}}|{\bm \pi}(s)|^2\D s}\right] <+\infty,\ \ j=1,\ldots,l-1,
\end{align*}
where we have used ({\bf A1}) and introduced a positive constant $C_T$ depending on $T>0$. The last inequality follows from the definition of an admissible control set, see Eq.~\eqref{cond:admiss} therein.
Thus, under $\tilde{\Px}$, the system of SDEs \eqref{eq:normzlized-sde-HatP} may be rewritten as
\begin{align}\label{eq:normzlized-sde-TildeP}
\D p_k(t) &= \left(\sum_{\ell=1}^K \varpi_{\ell,k}(t)p_{\ell}(t)+\gamma p_k(t)\left({\bm \mu}(t,{\bm e}_k,{\bm H}(t))-\hat{{\bm \mu}}(t,{\bm p}(t),{\bm H}(t))\right)^{\top}{{\bm \pi}}(t)\right)\D t\nonumber\\
&\quad + p_k(t)\left({\bm \mu}(t,{\bm e}_k,{\bm H}(t))-\hat{{\bm \mu}}(t,{\bm p}(t),{\bm H}(t))\right)^{\top}{\bm \Sigma}^{-1}\D\tilde{\bm W}(t)\\
&\quad+p_k(t-)\sum_{i=1}^N\frac{h_i(t,{\bm e}_k,{\bm H}(t-))-\hat{h}_i(t,{\bm p}(t-),{\bm H}(t-))}{\hat{h}_i(t,{\bm p}(t-),{\bm H}(t-))}\D \tilde{\Xi}_i(t).\nonumber
\end{align}
Using Proposition \ref{prop:equivform} along with equations~\eqref{eq:hatL} and~\eqref{eq:dPxdtPx}, the objective functional may be rewritten as
\begin{align}\label{eq:problem-tildeP}
J_T(v,{{\bm \pi}}) &=  \frac{v^{\gamma}}{\gamma}\Ex^{\hat{\Px}}\left[\hat{L}^{{{\bm \pi}}}(T)\right] = \frac{v^{\gamma}}{\gamma}\Ex^{\tilde{\Px}}\left[\zeta^{-1}(T)\hat{L}^{{{\bm \pi}}}(T)\right]  \\\notag
&= \frac{v^{\gamma}}{\gamma}\Ex^{\tilde{\Px}}\left[\exp\left(-\gamma\int_0^T\hat{\eta}(s,{\bm p}(s),{\bm H}(s),{{\bm \pi}}(s))\D s\right)\right] = \frac{v^{\gamma}}{\gamma}\Ex^{\tilde{\Px}}\left[\exp\left(-\gamma\int_0^T\tilde{\eta}(s,\tilde{\bm p}(s),{\bm H}(s),{{\bm \pi}}(s))\D s\right)\right].
\end{align}
Eq.~\eqref{eq:problem-tildeP} indicates that our original problem has been reduced to a risk sensitive control problem, where the criterion to maximize is of exponential type and the risk averse attitude of the investor is taken into account through the parameter $\gamma$. In such a problem, the controlled diffusion process is a filtering process, living in the $K-1$ dimensional simplex.

\subsection{Problem Formulation and HJB Equation}\label{sec:HJB}
This section gives the HJB equation associated with the risk-sensitive control problem. We first write the dynamics of $\tilde{\bm p}(t)$ in a matrix form. To this purpose, define the $N\times (K-1)$-dimensional matrix
\begin{align*}
&{\bm \mu}^{\bot}(t,{\bf z}) := \left[{\bm\mu}(t,{\bm e}_1,{\bf z}),\ldots,{\bm\mu}(t,{\bm e}_{K-1},{\bf z})\right]_{N\times (K-1)}\nonumber\\
&\quad = \left[\begin{array}{ccc}
b_1(t,{\bm e}_1,{\bf z}) + h_1(t,{\bm e}_1,{\bf z}) -{\vartheta_1^2({\bf z})}/{2} & \cdots & b_{1}(t,{\bm e}_{K-1},{\bf z}) + h_{1}(t,{\bm e}_{K-1},{\bf z}) -{\vartheta_{1}^2({\bf z})}/{2} \\
\vdots & \vdots & \vdots \\
b_{N}(t,{\bm e}_{1},{\bf z}) + h_{N}(t,{\bm e}_{1},{\bf z}) -{\vartheta_{N}^2({\bf z})}/{2} & \cdots & b_{N}(t,{\bm e}_{K-1},{\bf z}) + h_{N}(t,{\bm e}_{K-1},{\bf z}) -{\vartheta_{N}^2({\bf z})}/{2} \\
\end{array}\right].
\end{align*}
Further, define ${\bm \beta}_{\varpi}(t,{\bm\lambda})$ to be the $K-1$-dimensional column vector defined as, for $(t,{\bm\lambda})\in\R_+\times\Delta_{K-1}$,
\begin{eqnarray*}
{{\bm \beta}}_{\varpi}(t,{\bm\lambda}):=\left[\varpi_{K,1}(t)+\sum_{k=1}^{K-1}(\varpi_{k,1}(t)-\varpi_{K,1}(t))\lambda_k,\ldots,
\varpi_{K,K-1}(t)+\sum_{k=1}^{K-1}(\varpi_{k,K-1}(t)-\varpi_{K,K-1}(t))\lambda_k\right]^{\top}.
\end{eqnarray*}
Using the relation $\tilde{g}(y,\tilde{\bm p}(t),v)=\hat{g}(y,{\bm p}(t),v)$,
the $\hat{\Px}$-dynamics of the filter probabilities given in Eq.~\eqref{eq:normzlized-sde-HatP}, along with equations~\eqref{eq:ptildeBM} and~\eqref{eq:qfunc}, we may write the $\tilde{\Px}$-dynamics of $\tilde{\bm p}(t)$ as
\begin{eqnarray}\label{eq:sde-tildep}
\D\tilde{\bm p}(t) = {\bm \beta}_{\gamma}(t,\tilde{\bm p}(t),{\bm H}(t),{\bm \pi}(t))\D t + {\bm \sigma}(t,\tilde{\bm p}(t),{\bm H}(t))\D\tilde{\bm W}(t)+
\sum_{i=1}^N{\bf J}_i(t,\tilde{\bm p}(t-),{\bm H}(t-))\D\tilde{\Xi}_i(t),
\end{eqnarray}
where the coefficients
\begin{align}\label{eq:coeffdef}
\nonumber {\bm \sigma}(t,{\bm \lambda},{\bf z}) &:= diag(\bm \lambda)\left[{\bm \mu}^{\bot}(t,{\bf z})^{\top}-{\bf1}_{K-1}\tilde{{\bm \mu}}(t,{\bm\lambda},{\bf z})^{\top}\right]{\bm \Sigma}^{-1},\nonumber\\
{\bm \beta}_\gamma(t,{\bm\lambda},,{\bf z},{{\bm \pi}})&:= {{\bm \beta}}_{\varpi}(t,{\bm \lambda}) + \gamma{\bm \sigma}(t,{\bm \lambda},{\bf z}){\bm \Sigma}^{\top}{{\bm \pi}},\\
{\bf J}_i(t,{\bm \lambda},{\bf z}) &:=  diag({\bm \lambda})\frac{1}{\tilde{h}_i(t,{\bm \lambda},{\bf z})}\left[{\bm h}_i^{\bot}(t,{\bf z})-{\bf1}_{K-1}\tilde{h}_i(t,{\bm \lambda},{\bf z})\right],\ \ i=1,\ldots,N.\nonumber
\end{align}
Above ${\bm h}_i^{\bot}(t,{\bf z})=(h_i(t,\bm{e}_1,{\bf z}),\ldots,h_i(t,\bm{e}_{K-1},{\bf z}))^{\top}$ for $i=1,\ldots,N$, and ${\bf1}_{K-1}$ denotes the $(K-1)$-dimensional column vector with all entries equal to $1$. For $T>0$, define the bounded domain $Q_T:=[0,T]\times\Delta_{K-1}$.

The following lemma, proven in the Appendix, will be later used to prove the verification theorem.
\begin{lemma}\label{lem:sing2-b}
Under Assumption ({\bf A1}), there exists a linear increasing function $\varrho(\cdot):\R_+\too\R_+$, $\varrho(0)=0$, so that for distinct $(t,{\bm\lambda}_1)$ and $(s,{\bm\lambda}_2)\in Q_T$ with $T>0$ and ${\bf z}\in{\cal S}$,
\begin{eqnarray}\label{eq:sing2-b}
\left|{\bm \sigma} {\bm \sigma}^{\top}(t,{\bm\lambda}_1,{\bf z})-{\bm \sigma}{\bm \sigma}^{\top}(s,{\bm\lambda}_2,{\bf z})\right|\leq \varrho(\left|(t,{\bm\lambda}_1)-(s,{\bm\lambda}_2)\right|).
\end{eqnarray}
\end{lemma}

The generator of the bivariate Markov process $(\tilde{\bm p}(t),{\bm H}(t);\ t\geq0)$ is given in the following lemma, proven in the Appendix.
\begin{lemma}\label{eq:generatorp-H}
The generator of the bivariate process $(\tilde{\bm p}(t),{\bm H}(t);\ t\geq0)$ is of the following form: for $(t,{\bm \lambda},{\bf z})\in Q_T\times{\cal S}$,
\begin{align}\label{eq:generator}
\A^{{\bm\pi}} f(t,{\bm \lambda},{\bf z}) &=  \frac{\partial f}{\partial t}(t,{\bm\lambda},{\bf z}) + \bm \nabla f(t,{\bm\lambda},{\bf z}) {\bm \beta}_\gamma(t,{\bm \lambda},{\bf z},{\bm \pi})+\frac{1}{2}{\rm Tr}\left[{\bm \sigma}{\bm \sigma}^{\top}D^2f\right](t,{{\bm \lambda}},{\bf z})\nonumber\\
&\quad+\sum_{i=1}^N\left[f\left(t, \frac{1}{\tilde{h}_i(t,{\bm \lambda},{\bf z})}\left[{\bm \lambda}\cdot{\bm h}_i^{\bot}(t,{\bf z})\right],{\bf z}^{i}\right)-f(t,{\bm \lambda},{\bf z})\right](1-z_i)\tilde{h}_i(t,{\bm \lambda},{\bf z}),
\end{align}
where the function $f$ is $C^1$ w.r.t. $t$, and $C^2$ w.r.t. ${\bm \lambda}$.
\end{lemma}
Above, the gradient operator $\bm \nabla f$ is intended to be a row vector. Next, for a generic $0 \leq t \leq T$ such that $\tilde{\bm p}(t)= {\bm \lambda}\in\Delta_{K-1}$ and ${\bm H}(t) = {\bf z} \in{\cal S}$, we define
\[
G(t,{\bm \lambda},{\bf z},{\bm \pi}):=\Ex^{\tilde{\Px}}\left[e^{-\gamma\int_t^T\tilde{\eta}(s,\tilde{\bm p}(s),,{\bm H}(s),{\bm \pi}(s))\D s} \big| \tilde{\bm p}(t) = {\bm \lambda}, {\bm H}(t) = {\bf z} \right],
\]
where the function
\begin{eqnarray}\label{eq:tilde-eta}
\tilde{\eta}(t,{\bm \lambda},{\bf z},{\bm \pi})=\eta(t,{\bm e}_K,{\bf z},{\bm \pi}) + \sum_{k=1}^{K-1}(\eta(t,{\bm e}_k,{\bf z},{\bm \pi})-\eta(t,{\bm e}_K,{\bf z},{\bm \pi}))\lambda_k.
\end{eqnarray}
Further, define the value function
\begin{eqnarray}\label{eq:value-fcn}
w(t,{\bm \lambda},{\bf z}) := \sup_{{\bm \pi}\in \tilde{\cal U}(t,T)}\log\left(G(t,{\bm \lambda},{\bf z},{\bm \pi})\right).
\end{eqnarray}
The following equalities show the equivalence between our original problem~\eqref{eq:partialobs} and the problem in Eq.~\eqref{eq:value-fcn}:
\begin{align}
\nonumber \sup_{{\bm \pi} \in \tilde {\cal U}(0,T)} \frac{1}{\gamma} \Ex\left[V^{\gamma}(T)|V(0) = v,{\bm H}(0) = {\bf z} \right]
&= \frac{v^{\gamma}}{\gamma}\sup_{{\bm \pi} \in \tilde{\mathcal{U}}(0,T)} \Ex^{\tilde{\Px}} \left[e^{-\gamma \int_0^T {\tilde{\eta}(s,\tilde{\bm p}(s),{\bm H}(s),{\bm \pi}(s))}\D s} | \tilde{\bm p}(0) = {\bm \lambda},{\bm H}(0) = {\bf z} \right]\\
&=\frac{v^{\gamma}}{\gamma} e^{w(0,{\bm \lambda},{\bf z})}.
\label{eq:connect}
\end{align}
The first equality follows from~\eqref{eq:problem-tildeP}, and the last equality follows directly from Eq.~\eqref{eq:value-fcn}.

Next, we specialize the class of admissible strategies to be Markovian.
\begin{definition}\label{eq:def32}
The class of $\mathbb{G}^I$-adapted feedback trading strategies, denoted by ${\cal U}_t={\cal U}(t,T;{\bm \lambda},{\bf z})$, $ 0 \leq t \leq T$, is given by
$$
{\bm \pi}(s):=(\pi_1(s),\ldots, \pi_N(s))^{\top}:=(\pi_1(s,\tilde{\bm p}(s{}),{\bm H}({s{}})), \ldots, \pi_N(s,\tilde{\bm p}(s{}),{\bm H}({s{}}))^{\top},\ \ \ \ s\geq t,
$$
with ${\bm \pi}(t) \in \tilde{\cal U}(t,T)$, and $(\tilde{\bm p}(t),{\bm H}(t)) = ({\bm \lambda},{\bf z}) \in \Delta_{K-1} \times{\cal S}$.
\end{definition}
We now proceed to derive the formal HJB equation based on heuristic arguments first.
{In light of the dynamic programming principle (see e.g. Theorem 3.1 of \cite{OkSu})}, we have that for $s\in(t,T]$,
\begin{equation}
	w(t,{\bm \lambda},{\bf z}) = \sup_{{\bm \pi} \in \mathcal{U}_t} \log \Ex^{{\tilde{\Px}}}\left[e^{w(s,\tilde{\bm p}(s),{\bm H}(s)) - \gamma \int_t^s \tilde{\eta}(u,\tilde{\bm p}(u),{\bm H}(u),{\bm \pi}(u))\D u} \big| \tilde{\bm p}(t) = {\bm \lambda}, {\bm H}(t) = {\bf z}  \right].
\label{eq:vf}
\end{equation}
Next, define $\varepsilon({s}, {\bm \lambda},{\bf z}) := e^{w({s},{\bm \lambda},{\bf z})}$. Using the generator~\eqref{eq:generator},
we obtain the following recursive HJB system of PDEs after straightforward algebraic manipulations:
\begin{align}
 &0 = \sup_{\bm \pi\in\R^N} \varepsilon(t,{\bm \lambda},{\bf z}) \Bigg\{\frac{\partial w}{\partial t}(t,{\bm \lambda},{\bf z}) + \bm \nabla w(t,{\bm \lambda},{\bf z})\,  {\beta}_{\gamma}(t,{\bm \lambda},{\bf z},{\bm\pi}_{\bf z}) + \frac{1}{2} {\rm Tr}[{\bm \sigma} {\bm \sigma}^{\top} D^2 w](t,{\bm \lambda},{\bf z}) \label{eq:HJBalm} \\
\nonumber &\quad+ \frac{1}{2} \left[(\bm \nabla w){\bm \sigma}{\bm \sigma}^{\top}(\bm \nabla w)^{\top}\right](t,{\bm \lambda},{\bf z}) +  \sum_{i=1}^N(1-z_i)\tilde{h}_i(t,{\bm \lambda},{\bf z})\left[e^{w\left(t, \frac{{\bm \lambda}\cdot{\bm h}_{i}^{\bot}(t,{\bf z})}{\tilde{h}_i(t,{\bm \lambda},{\bf z})},{\bf z}^{i}\right)-w(t,{\bm \lambda},{\bf z})}-1\right]  - \gamma \tilde{\eta}(t,{\bm \lambda},{\bf z},{\bm\pi}_{{\bf z}})\Bigg\}.
\end{align}
Noting that
\begin{eqnarray}\label{eq:tilde-eta}
\tilde{\eta}(t,{\bm \lambda},{\bf z},{{\bm \pi}})=-r + \sum_{i=1}^N \pi_i(r-\tilde{b}_i(t,{\bm \lambda},{\bf z}) - \tilde{h}_i(t,{\bm \lambda},{\bf z}))+\frac{1-\gamma}{2}{{\bm \pi}}^{\top}{\bm \Sigma}{\bm \Sigma}^{\top}{{\bm \pi}},
\end{eqnarray}
we may rewrite Eq.~\eqref{eq:HJBalm} as
\begin{align}\label{eq:hjEqn}
&\frac{\partial w}{\partial t}(t,{\bm \lambda},{\bf z}) + \frac{1}{2}{\rm Tr}\left[{\bm \sigma}{\bm \sigma}^{\top}D^2w\right](t,{\bm \lambda},{\bf z}) + \frac{1}{2}\left[(\bm \nabla w){\bm \sigma}{\bm \sigma}^{\top}(\bm \nabla w)^{\top}\right](t,{\bm \lambda},{\bf z}) +\gamma r \nonumber\\
&\ + \sum_{i=1}^N(1-z_i)\tilde{h}_i(t,{\bm \lambda},{\bf z})\left[e^{w\left(t, \frac{{\bm \lambda}\cdot{\bm h}_{i}^{\bot}(t,{\bf z})}{\tilde{h}_i(t,{\bm \lambda},{\bf z})},{\bf z}^{i}\right)-w(t,{\bm \lambda},{\bf z})}-1\right]
+\sup_{{\bm \pi}\in\R^N}\Phi\left({\bm \nabla} w(t,{\bm \lambda},{\bf z});t,{\bm \lambda},{\bf z},{{\bm \pi}}\right)=0,
\end{align}
with terminal condition $w(T,{\bm \lambda},{\bf z}) = 0$ for all $({\bm \lambda},{\bf z})\in\Delta_{K-1}\times{\cal S}$. Above, the function
\begin{equation}\label{eq:Phi}
\Phi({\bm \nabla} w;t,{{\bm \lambda}},{\bf z},{{\bm \pi}}):=({\bm \nabla} w)\left[{{\bm \beta}}_{\varpi}(t,{\bm \lambda})+\gamma{\bm \sigma}(t,{\bm\lambda},{\bf z}){\bm \Sigma}^{\top}{\bm \pi}_{\bf z}\right]-\gamma{\bm \pi}_{\bf z}^{\top}{\bm \Gamma}(t,{\bm \lambda},{\bf z})-\frac{\gamma(1-\gamma)}{2}{\bm \pi}_{\bf z}^{\top}{\bm \Sigma}^{\top}{\bm \Sigma}{\bm \pi}_{\bf z},
\end{equation}
where
\begin{eqnarray}\label{eq:Gammat-p}
{\bm \Gamma}(t,{\bm \lambda},{\bf z}):=\left[r-\tilde{b}_1(t,{{\bm \lambda}},{\bf z}) - \tilde{h}_1(t,{{\bm \lambda}},{\bf z}),\ldots,r-\tilde{b}_N(t, {{\bm \lambda}},{\bf z}) - \tilde{h}_N(t,{{\bm \lambda}},{\bf z})\right]^{\top},
\end{eqnarray}
and $\tilde{b}_i(t,{\bm \lambda},{\bf z})=b_i(t,{\bm e}_K,{\bf z})+\sum_{k=1}^{K-1}(b_i(t,{\bm e}_k,{\bf z})-b_i(t,{\bm e}_K,{\bf z}))\lambda_k$,
while $\tilde{h}_i(t,{\bm \lambda},{\bf z})=h_i(t,{\bm e}_K,{\bf z})+\sum_{k=1}^{K-1}(h_i(t,{\bm e}_k,{\bf z})-h_i(t,{\bm e}_K,{\bf z}))\lambda_k$, for $i=1,\ldots,N$.
Using the first-order condition, the optimal feedback strategy is given by
${{\bm \pi}}_{\bf z}^*(t,{\bm\lambda}) = \frac{1}{1-\gamma}\left[({\bm \Sigma}^{\top}{\bm \Sigma})^{-1}\left({\bm \Sigma}{\bm \sigma}(t,{\bm \lambda},{\bf z})^{\top}
(\bm \nabla w)^{\top}(t,{\bm \lambda},{\bf z})-{\bm \Gamma}(t,{\bm \lambda},{\bf z})\right)\right]_{\bf z}$.
Hence, using \eqref{eq:Phi}, we obtain
\begin{align}\label{eq:Phistar}
\Phi^*({\bm \nabla} w;t,{\bm \lambda},{\bf z})&:=\Phi({\bm \nabla} w;t,{\bm \lambda},{\bf z},{{\bm \pi}}_{\bf z}^*(t,{\bm\lambda}))=(\bm \nabla w)\left[{{\bm \beta}}_{\varpi}(t,{\bm \lambda})-\frac{\gamma}{1-\gamma}{\bm \sigma}_{\bf z}(t,{\bm \lambda}){\bm \Sigma}^{-1}{\bm \Gamma}(t,{\bm \lambda},{\bf z})\right]\\
&+\frac{\gamma}{2(1-\gamma)}\left[(\bm \nabla w)({\bm \sigma}_{\bf z}{\bm \sigma}_{\bf z}^{\top})(t,{\bm \lambda},{\bf z})(\bm \nabla w)^{\top}\right]
+\frac{\gamma}{2(1-\gamma)}{\bm \Gamma}_{\bf z}^{\top}(t,{\bm \lambda})({\bm \Sigma}^{\top}{\bm \Sigma})^{-1}{\bm \Gamma}_{\bf z}(t,{\bm \lambda}).\nonumber
\end{align}
For ${\bf z}\in{\cal S}$, we recall that ${\bm \sigma}_{\bf z}$ is obtained from the matrix ${\bm \sigma}$ as specified in Eq.~\eqref{eq:Bz}, while
${\bm \Gamma}_{\bf z}$ is derived from ${\bm \Gamma}$ as described in Eq.~\eqref{eq:vz}. Above, we have used the simplified notations ${\bm\sigma}:=\sigma(t,{\bm\lambda},{\bf z})$, ${\bm\sigma}_{\bf z}(t,{\bm \lambda}):={\bm\sigma}_{\bf z}(t,{\bm\lambda},{\bf z})$ and ${\bm\Gamma}_{\bf z}(t,{\bm \lambda}):={\bm\Gamma}_{\bf z}(t,{\bm\lambda},{\bf z})$. Plugging the expression~\eqref{eq:Phistar} into \eqref{eq:hjEqn}, we obtain the final form of the HJB equation, which will be analyzed in this paper: on $(t,{\bm\lambda},{\bf z})\in[0,T)\times\Delta_{K-1}\times{\cal S}$,
\begin{align}\label{eq:HJBeqn1fix}
&\frac{\partial w}{\partial t}(t,{\bm \lambda},{\bf z}) + \frac{1}{2}{\rm Tr}\left[{\bm \sigma}{\bm \sigma}^{\top}D^2 w\right](t,{\bm \lambda},{\bf z})+ \frac{1}{2}\left[(\bm \nabla w){\bm \sigma}{\bm \sigma}^{\top}(\bm \nabla w)^{\top}\right](t,{\bm \lambda},{\bf z})\nonumber\\
 &\qquad+ \frac{\gamma}{2(1-\gamma)}\left[(\bm \nabla w){\bm \sigma}_{\bf z}{\bm \sigma}_{\bf z}^{\top}(\bm \nabla w)^{\top}\right](t,{\bm \lambda},{\bf z})+(\bm \nabla w)(t,{\bm \lambda},{\bf z}){\bm \theta}(t,{\bm \lambda},{\bf z})\nonumber\\
 &\qquad+\sum_{i=1}^N(1-z_i)\tilde{h}_i(t,{\bm \lambda},{\bf z})e^{w\left(t, \frac{{\bm \lambda}\cdot{\bm h}_{i}^{\bot}(t,{\bf z})}{\tilde{h}_i(t,{\bm \lambda},{\bf z})},{\bf z}^{i}\right)-w(t,{\bm \lambda},{\bf z})}+{\rho}(t,{\bm \lambda},{\bf z})=0,
\end{align}
with terminal condition $w(T,{\bm \lambda},{\bf z})=0$ for all $({\bm \lambda},{\bf z})\in\Delta_{K-1}\times{\cal S}$, where the coefficients
\begin{align}
{\bm \theta}(t,{\bm \lambda},{\bf z}) &:={{\bm \beta}}_{\varpi}(t,{\bm \lambda})-\frac{\gamma}{1-\gamma}{\bm \sigma}_{\bf z}(t,{\bm \lambda}){\bm \Sigma}^{-1}{\bm \Gamma}(t,{\bm \lambda},{\bf z}),\\\notag
{\rho}(t,{\bm \lambda},{\bf z}) &:= \gamma r - \sum_{i=1}^N(1-z_i)\tilde{h}_i(t,{\bm \lambda},{\bf z})+\frac{\gamma}{2(1-\gamma)}{\bm \Gamma}_{\bf z}^{\top}(t,{\bm \lambda})({\bm \Sigma}^{\top}{\bm \Sigma})^{-1}{\bm \Gamma}_{\bf z}(t,{\bm \lambda}).
\label{eq:thrhodef}
\end{align}

{
\begin{remark}
We analyze how information driven contagion affects the optimal strategy of an investor. 
We refer to this as information driven contagion because, even if the {distress} intensity of name $i$ does not change in reaction to the {distress} event of name $j$, the investor still indirectly accounts for financial distress of name $j$ when deciding on his optimal strategy in the $i$-th stock. He does so by revising his beliefs of the Markov chain being in a specific state according to the mechanism described next.
First, notice that the optimal feedback strategy in the distress state ${\bf z}$ depends on the gradient of the solution to the HJB equation in the same state ${\bf z}$.
The impact of information driven contagion is captured by the following term in the HJB PDE~\eqref{eq:HJBeqn1fix}:
\[
\sum_{i=1}^N(1-z_i)\tilde{h}_i(t,{\bm \lambda},{\bf z})e^{w\left(t, \frac{{\bm \lambda}\cdot{h}_{i}^{\bot}(t,{\bf z})}{\tilde{h}_i(t,{\bm \lambda},{\bf z})},{\bf z}^{i}\right)-w(t,{\bm \lambda},{\bf z})}.
\]
Consider first the situation when all $N$ stocks are in a distress state, i.e., the state is ${\bf z}={\bf1}$. Then, the above term becomes zero. This is consistent with intuition given that no contagion would be present if all stocks have already experienced distress.
Next, consider the situation when ${\bf z} = (\underbrace{0,0,\ldots 0}_{m \; terms}, \underbrace{1,1,\ldots,1}_{N-m \; terms})$,
i.e. the first $m$ stocks are alive and the remaining $N-m$ are in a distress state. The above term will then become
\[
\sum_{i=1}^m \tilde{h}_i(t,{\bm \lambda},{\bf z})e^{w\left(t, \frac{{\bm \lambda}\cdot{h}_{i}^{\bot}(t,{\bf z})}{\tilde{h}_i(t,{\bm \lambda},{\bf z})},{\bf z}^{i}\right)-w(t,{\bm \lambda},{\bf z})}.
\]
Hence, the solution $w$ to the PDE in Eq.~\eqref{eq:HJBeqn1fix} depends on the set $\left \{w\left(t, \frac{{\bm \lambda}\cdot{\bm h}_{i}^{\bot}(t,{\bf z})}{\tilde{h}_i(t,{\bm \lambda},{\bf z})},{\bf z}^{i}\right)\right \}_{i=1}^m$. Such a set consists of solutions to $m$ PDEs of the form~\eqref{eq:HJBeqn1fix}, but associated with
(1) the {distress} state ${\bf z}^i$, $i=1,\ldots,m$, in which the $i$-th stock is distressed, and (2) the revised filter probabilities
$\frac{{\bm \lambda}\cdot{h}_{i}^{\bot}(t,{\bf z})}{\tilde{h}_i(t,{\bm \lambda},{\bf z})}$ obtained from ${\bm \lambda}$ by adjusting for the probability that the Markov chain is in
regime $k$ when stock $i$ is distressed, $k=1,\ldots,K$. Such a probability is given by $\frac{h_i(t,{\bm e}_k,{\bf z})}{\tilde{h}_i(t,{\bm \lambda},{\bf z})}$.
As a consequence, the investor will consider the optimal expected utilities achievable in any state reached when additional stocks enter distress (accordingly revising the probability of the hidden chain being in a specific regime based on the additional distress information) when determining his strategy. This observation will also guide the analysis of Eq.~\eqref{eq:HJBeqn1fix} in the following sections.
\end{remark}
}

{\Red

\begin{remark}
In a fully heterogenous portfolio, one would need to solve $2^N$ PDEs to compute $w(t,{\bm \lambda},{\bf 0})$. Because of the contagion effects highlighted in the previous remark, the solution to the PDE associated with the state in which all names are alive recursively depends on the solutions to PDEs associated with all possible distress states of the portfolio. In practice, however, it is common to divide the portfolio of names into groups having identical characteristics and hence exchangeable. These groups would correspond to firms with identical credit rating or belonging to the same industry, which are thus expected to have similar credit quality, equity returns and volatility. Similar decompositions have been proposed in the credit risk literature. \cite{FreyBackhaus04} propose a mean-field model with homogenous groups and use it to price credit derivatives. \cite{Tolotti} split the portfolio into two groups of obligors with homogeneous characteristics and estimate losses arising in large credit portfolios.

Next, we discuss the reduction in computational complexity obtained under the grouping assumption. Consider first a fully homogenous portfolio where all stocks have identical characteristics, i.e. they have the same {distress} intensity ($h_{i}(t,{\bm e}_k,{\bf z}) = h(t,{\bm e}_k,{\bf z})$ for all {$i$}), expected instantaneous return ($b_{i}(t,{\bm e}_k,{\bf z})=b(t,{\bm e}_k,{\bf z})$ for all {$i$}), and volatility
($\vartheta_{i}({\bf z})=\vartheta({\bf z})$ {for all $i$}). Under this homogeneity assumption, the complexity would be linear in $N$ given that $w(t,{\bm \lambda},{\bf z}) = w(t,{\bm \lambda},{\bf \tilde{z}})$, if
$\sum_{j=1}^N z_j = \sum_{j=1}^N \tilde{z}_j$, i.e. the solution only depends on the number of distressed stocks and not on their identities. Next, consider the situation in which stocks are partitioned into two groups with identical characteristics. More concretely, assume that $N = 2n$. Let
$h_{i}(t,{\bm e}_k,{\bf z}) = h(t,{\bm e}_k,{\bf z})$ {for $i=1,\ldots,n$}, and $h_{j}(t,{\bm e}_k,{\bf z}) = \tilde{h}(t,{\bm e}_k,{\bf z})$ {for $j=n+1,\ldots,{N}$}. Additionally, assume that
$b_{i}(t,{\bm e}_k,{\bf z}) = b(t,{\bm e}_k,{\bf z})$ {for $i=1,\ldots,n$}, $b_{j}(t,{\bm e}_k,{\bf z}) = \tilde{b}(t,{\bm e}_k,{\bf z})$ {for $j=n+1,\ldots,{N}$}, and
$\vartheta_{i}({\bf z}) = \vartheta({\bf z})$ for $i=1,\ldots,n$,  $\vartheta_{j}({\bf z}) = \tilde{\vartheta}({\bf z})$ for $j=n+1,\ldots,{N}$. Under these assumptions, the required number of PDEs to solve may be estimated as follows. {One PDE solution needs to be computed when ${\bf z}$ has all zero entries. Two PDE solutions need to be computed when ${\bf z}$ has $N-1$ zero entries} ($w(t,{\bm \lambda},{\bf 0}^i) =w(t,{\bm \lambda},{\bf 0}^j)$, if $1 \leq i,j \leq  n$, and $w(t,{\bm \lambda},{\bf 0}^l) =w(t,{\bm \lambda},{\bf 0}^m)$, for $n+1 \leq l, m \leq N$) {or has only one zero entry
($w(t,{\bm \lambda},{\bf 0}^{j_1,\ldots,j_{N-1}}) = {u}(t,{\bm \lambda})$ if {$j_i = i$ for all $i=1,\ldots,n$,} i.e. all names in the first group are in a distress state and, moreover, $w(t,{\bm \lambda},{\bf 0}^{j_1,\ldots,j_{N-1}}) = \tilde{u}(t,{\bm \lambda})$ if $j_i=n+i$ for all $i=1,\ldots,n$, i.e. all names in the second group are in a distress state)}. Iterating this argument we obtain that the total complexity, measured in terms of the number of PDE solutions required, is $2{\sum_{i=1}^ni}+{n+1}=(N+2)^2/4$, i.e quadratic in $N$ as opposed to exponential in $N$. 
\end{remark}

}

\section{Analysis of HJB Equations}\label{sec:default-contagion}
The objective of this section is to analyze the solution of the recursive HJB system of PDEs given in Eq.~\eqref{eq:HJBalm}. For notional convenience, we use
$w_{j_1,\ldots,j_n}(t,{\bm \lambda}):=w(t,{\bm \lambda},{\bm 0}^{j_1,\ldots,j_n})$ to denote the solution of the HJB equation associated
with the distress state ${\bf z}={\bm 0}^{j_1,\ldots,j_n}$, and similarly $h_{i;j_1,\ldots,j_n}(t,{\bm\lambda}):=h_i(t,{\bm\lambda},{\bm 0}^{j_1,\ldots,j_n})$ for $i\notin\{j_1,\ldots,j_n\}$. 
We separately consider the following cases: (1) $n = N$, (2) $n = N-1$, and (3) $0 \leq n < N-1$, in each of the following subsections.

\subsection{The Case $n=N$}

When all stocks are distressed, ${\bf z}={\bm 0}^{j_1,\ldots,j_N}={\bm 1}$. Hence, ${{\bm \pi}}^*(t)={\bm 0}$ and $\pi_B(t)=1$, i.e. the investor allocates his entire wealth
to the money market account. The HJB equation satisfied by $w_{\bm 1}(t,{\bm \lambda})$ is given by
\begin{eqnarray}\label{eq:HJBeqn1-N1fix}
&&\frac{\partial w_{\bm 1}}{\partial t}(t,{\bm \lambda}) +
\frac{1}{2}{\rm Tr}\left[{\bm \sigma}{\bm \sigma}^{\top}D^2 w_{\bm 1} \right](t,{\bm \lambda})
+\frac{1}{2}\left[(\bm \nabla w_{\bm 1}){\bm \sigma}{\bm \sigma}^{\top}(\bm \nabla w_1)^{\top}\right](t,{\bm \lambda})\nonumber\\
&&\qquad+(\bm \nabla w_{\bm 1})(t,{\bm \lambda}){{\bm \beta}}_{\varpi}(t,{\bm \lambda})+\gamma r=0,\ \ \ \ (t,{\bm \lambda})\in [0,T)\times\Delta_{K-1},
\end{eqnarray}
with terminal condition $w_{\bm 1}(T,{\bm \lambda}) = 0$ for all ${\bm \lambda}\in\Delta_{K-1}$.
Next, we consider the transformation $\psi(t,{\bm \lambda}) := e^{w_{\bm 1}(t,{\bm \lambda})}$.
It can be seen that $w_{\bm 1}(t,{\bm \lambda})$ solves Eq.~\eqref{eq:HJBeqn1-N1fix} if and only if
$\psi(t,{\bm \lambda})$ solves the following PDE:
\begin{eqnarray}\label{eq:HJBeqn1-N1fixtr}
&&\frac{\partial \psi}{\partial t}(t,{\bm \lambda}) +
\frac{1}{2}{\rm Tr}\left[{\bm \sigma}{\bm \sigma}^{\top}D^2 \psi\right](t,{\bm \lambda})+
 (\bm \nabla \psi)(t,{\bm \lambda}){\bm \beta}_{\varpi}(t,{\bm \lambda})+ \psi(t,{\bm \lambda})\gamma r=0,
\end{eqnarray}
with terminal condition $\psi(T,{\bm \lambda})=1$ for all ${\bm \lambda}\in\Delta_{K-1}$. A direct application of the Feynman-Kac formula yields
$\psi(t,{\bm \lambda}) = e^{\gamma r(T-t)}$ for $0\leq t\leq T$, and hence the solution is $w_{\bm 1}(t,{\bm \lambda})=\gamma r(T-t)$.

\subsection{The Case $n=N-1$}
When all names, except for one, are distressed, ${\bf z}={\bm 0}^{j_1,\ldots,j_{N-1}}$. The only non-distressed stock is $j_N$, while stocks $j_1,\ldots,j_{N-1}$ are distressed. Hence, $z_{j_N}=0$, ${j_N}=\{1,\ldots,N\}\setminus\{j_1,\ldots,j_{N-1}\}$. The HJB equation~\eqref{eq:HJBeqn1fix}  takes the form
 \begin{eqnarray}\label{eq:HJBeqn-j1-jN-1}
&&\frac{\partial w_{j_1,\ldots,j_{N-1}}}{\partial t}(t,{\bm \lambda}) + \frac{1}{2}{\rm Tr}\left[{\bm \sigma}{\bm \sigma}^{\top}D^2 w_{j_1,\ldots,j_{N-1}}\right](t,{\bm \lambda})+ \frac{1}{2}\left[(\bm \nabla w_{j_1,\ldots,j_{N-1}}){\bm \sigma}{\bm \sigma}^{\top}(\bm \nabla w_{j_1,\ldots,j_{N-1}})^{\top}\right](t,{\bm \lambda})\nonumber\\
 &&\qquad+ \frac{\gamma}{2(1-\gamma)}\left[(\bm \nabla w_{j_1,\ldots,j_{N-1}}){\bm \sigma}_{\bf z}{\bm \sigma}_{\bf z}^{\top}(\bm \nabla w_{j_1,\ldots,j_{N-1}})^{\top}\right](t,{\bm \lambda})+(\bm \nabla w_{j_1,\ldots,j_{N-1}})(t,{\bm \lambda}){\bm \theta}_{j_1,\ldots,j_{N-1}}(t,{\bm \lambda})\nonumber\\
 &&\qquad+\rho_{j_1,\ldots,j_{N-1}}(t,{\bm \lambda})+\tilde{h}_{j_N;j_1,\ldots,j_{N-1}}(t,{\bm \lambda})e^{w_{\bm 1}\left(t, \frac{{\bm \lambda}\cdot{\bm h}_{j_N;j_1,\ldots,j_{N-1}}^{\bot}(t)}{\tilde{\bm h}_{j_N;j_1,\ldots,j_{N-1}}(t,{\bm \lambda})}\right)-w_{j_1,\ldots,j_{N-1}}(t,{\bm \lambda})}=0,
\end{eqnarray}
with terminal condition $w_{j_1,\ldots,j_{N-1}}(T,{\bm \lambda})=0$ for all ${\bm \lambda}\in\Delta_{K-1}$. For $1\leq n\leq N$, we have used the notation
\begin{eqnarray}\label{eq:notation-theta-rho}
{\bm \theta}_{j_1,\ldots,j_{n}}(t,{\bm \lambda})={\bm \theta}(t,{\bm \lambda},{\bm 0}^{j_1,\ldots,j_n}),\ \ \ \ {\rm and}\ \ \ \
{\rho}_{j_1,\ldots,j_{n}}(t,{\bm \lambda})={\rho}(t,{\bm \lambda},{\bm 0}^{j_1,\ldots,j_n}).
\end{eqnarray}
From Eq.~\eqref{eq:HJBeqn-j1-jN-1}, we can observe immediately that the solution $w_{j_1,\ldots,j_{N-1}}(t,{\bm \lambda})$ depends on the solution $w_{\bm 1}(t,{\bm \lambda})$ of the HJB equation in the state of full distress. Using that $w_{\bm 1}(t,{\bm \lambda})=\gamma r(T-t)$, Eq.~\eqref{eq:HJBeqn-j1-jN-1} may then be reduced to
\begin{eqnarray}\label{eq:HJBeqn-j1-jN-1-2}
&&\frac{\partial w_{j_1,\ldots,j_{N-1}}}{\partial t}(t,{\bm \lambda}) + \frac{1}{2}{\rm Tr}\left[{\bm \sigma}{\bm \sigma}^{\top}D^2 w_{j_1,\ldots,j_{N-1}}\right](t,{\bm \lambda})+ \frac{1}{2}\left[(\bm \nabla w_{j_1,\ldots,j_{N-1}}){\bm \sigma}{\bm \sigma}^{\top}(\bm \nabla w_{j_1,\ldots,j_{N-1}})^{\top}\right](t,{\bm \lambda})\nonumber\\
 &&\qquad+ \frac{\gamma}{2(1-\gamma)}\left[(\bm \nabla w_{j_1,\ldots,j_{N-1}}){\bm \sigma}_{\bf z}{\bm \sigma}_{\bf z}^{\top}(\bm \nabla w_{j_1,\ldots,j_{N-1}})^{\top}\right](t,{\bm \lambda})+(\bm \nabla w_{j_1,\ldots,j_{N-1}})(t,{\bm \lambda}){\bm \theta}_{j_1,\ldots,j_{N-1}}(t,{\bm \lambda})\nonumber\\
 &&\qquad+\xi_{j_1,\ldots,j_{N-1}}(t,{\bm \lambda},w_{j_1,\ldots,j_{N-1}}(t,{\bm \lambda}))=0,\ \ \ \ (t,{\bm \lambda})\in [0,T)\times\Delta_{K-1},
\end{eqnarray}
where we have defined the nonlinear term:
\begin{eqnarray}\label{eq:betaJ1-N-1}
\xi_{j_1,\ldots,j_{N-1}}(t,{\bm \lambda},{v}):=\rho_{j_1,\ldots,j_{N-1}}(t,{\bm \lambda})+\tilde{h}_{j_N;j_1,\ldots,j_{N-1}}(t,{\bm \lambda})e^{\gamma r (T-t)-{v}},\ \ \ (t,{\bm \lambda},v)\in Q_T\times\R.
\end{eqnarray}

\begin{remark}\label{rem:weak-why}
As for Eq.~\eqref{eq:HJBeqn1-N1fix}, we have that the PDE in~\eqref{eq:HJBeqn-j1-jN-1-2} is semilinear parabolic with quadratic gradient growth. Differently from that equation, however, the additional quadratic gradient term
\begin{eqnarray*}
\frac{\gamma}{2(1-\gamma)}\left[(\bm \nabla w_{j_1,\ldots,j_{N-1}}){\bm \sigma}_{\bf z}{\bm \sigma}_{\bf z}^{\top}(\bm \nabla w_{j_1,\ldots,j_{N-1}})^{\top}\right](t,{\bm \lambda})
\end{eqnarray*}
appears in Eq.~\eqref{eq:HJBeqn-j1-jN-1-2}. This term reflects the influence of the surviving stock $j_N$ on the solution of the PDE and only disappears in the case of a single regime or in the degenerate case when all regimes are identical. Because of this term, the standard argument of applying the Cole-Hopf transform (as for the case ${\bf z}={\bf1}$) to remove the quadratic gradient term, is not applicable here for the reasons explained next. We would like to find $\delta\in\R$ so that $\psi(t,{\bm\lambda}):=e^{\delta w_{j_1,\ldots,j_{N-1}}(t,{\bm\lambda})}$ satisfies a semilinear PDE without quadratic gradient terms.
We have that $\psi$ satisfies
 \begin{eqnarray}\label{eq:hjb-default-jn2}
&&\frac{\partial \psi}{\partial t}(t,{\bm \lambda}) + \frac{1}{2}{\rm Tr}\left[{\bm \sigma}{\bm \sigma}^{\top}D^2\psi\right](t,{\bm \lambda})+ \frac{1}{2}\psi^{-1}(\delta^{-1}-1)\left[(\bm \nabla\psi){\bm \sigma}{\bm \sigma}^{\top}(\bm \nabla\psi)^{\top}\right](t,{\bm \lambda})\nonumber\\
&&\quad + \frac{\gamma}{2(1-\gamma)}\delta^{-1}\psi^{-1}\left[(\bm \nabla\psi){\bm \sigma}_{\bf z}{\bm \sigma}_{\bf z}^{\top}(\bm \nabla\psi)^{\top}\right](t,{\bm \lambda})
+(\bm \nabla\psi)(t,{\bm \lambda}){\bm \theta}_{j_1,\ldots,j_{N-1}}(t,{\bm \lambda})\nonumber\\
&&\quad +\psi^{1-\frac{1}{\delta}}\tilde{h}_{j_N;j_1,\ldots,j_{N-1}}(t,{\bm \lambda})
e^{\gamma r(T-t)}+\psi\delta\rho_{j_1,\ldots,j_{N-1}}(t,{\bm \lambda})=0.
\end{eqnarray}
Let us analyze the structure of Eq.~\eqref{eq:hjb-default-jn2}. We need to choose $\delta$ so that both quadratic gradient terms in Eq.~\eqref{eq:hjb-default-jn2} vanish. This means that the equations $\delta^{-1}-1=0$ and $\delta^{-1}=0$ should be simultaneously satisfied. Clearly, such a choice of $\delta$ does not exist (notice that $\delta=1$ in the case ${\bf z}={\bf1}$).
An alternative method is to consider viscosity solutions of the HJB equation and then prove that these have $C^{1,2}$ regularity as in the recent work of \cite{Pham} and \cite{FeGaGo}. Therein, they consider one default-free stock and an infinite-horizon framework to study the regularization of their viscosity solution, see \cite{Pham}. 
Since we consider a finite-time risk-sensitive criterion, the objective functional is the exponential utility of the integrated filter probabilities, see Eq.~\eqref{eq:hatL} and Proposition~\ref{prop:equivform}. Such an utility function does not possess the homogeneity property used in \cite{Pham}.

{A classical reference for PDEs with quadratic gradient growth is \cite{Kob00}. Theorem 3.2 therein gives the uniqueness of the viscosity solution under rather standard assumptions (see (H4) and (H5) therein). Although assumption (H4) holds in our case, assumption (H5), consisting itself of four fundamental quadratic growth conditions (a)-(d), fails to hold on the function
\begin{align*}
F_{\bf z}(t,{\bm\lambda},u,{\bm\sigma}(t,{\bm\lambda},{\bf z})p,{\bm\sigma}_{\bf z}(t,{\bm\lambda})p)&:=\frac{1}{2}p\left[({\bm \sigma}{\bm \sigma}^{\top})(t,{\bm \lambda},{\bf z})\right]p^{\top}+\frac{\gamma}{2(1-\gamma)}p\left[({\bm \sigma}_{\bf z}{\bm \sigma}_{\bf z}^{\top})(t,{\bm \lambda})\right]p^{\top}\nonumber\\
&\quad+\rho_{j_1,\ldots,j_{N-1}}(t,{\bm \lambda})+\tilde{h}_{j_N;j_1,\ldots,j_{N-1}}(t,{\bm \lambda})e^{\gamma r(T-t)}e^{-u}.
\end{align*}
Indeed, the above function $F_{\bf z}$ does not even satisfy the growth condition (a): {despite} we can find a constant $C>0$ such that
\begin{align*}
&\left|F_{\bf z}(t,{\bm\lambda},u,{\bm\sigma}(t,{\bm\lambda},{\bf z})p,{\bm\sigma}_{\bf z}(t,{\bm\lambda})p)\right|\leq C(1+|{\bm\sigma}(t,{\bm\lambda},{\bf z})p|^2+
|{\bm\sigma}_{\bf z}(t,{\bm\lambda})p|^2)+Ce^{-u},
\end{align*}
the solution-dependent term $e^{-u}$, $u\in\R$, is unbounded (notice that the solution of our HJB equation may be either positive or negative since we are considering a log-criterion after reformulating the problem as a risk-sensitive control problem). In the recursive case, i.e. when $0\leq n < N-1$, the growth condition (d) also fails to hold. 
Because of these technical challenges, we consider a different type of solution for the recursive system of HJB PDEs, namely the Sobolev solution (see also \cite{SIAM1} for a related application).
}

\end{remark}

\subsection{The General Case $0\leq n \leq N-1$}
In this section, we prove existence and uniqueness of a Sobolev solution in the case of $0\leq n \leq N-1$ distressed stocks in the portfolio. The analysis of the HJB equation \eqref{eq:HJBeqn-j1-jN-1-2} in the case $n=N-1$ thus follows as a special case. The HJB equation satisfied by $w_{j_1,\ldots,j_n}(t,{\bm \lambda})$ depends on the solution of the HJB equation associated with a larger set of distressed stocks, including the distress of stock $i$. The latter is given by $w_{j_1,\ldots,j_n,i}(t,{\bm \lambda})$, $i\in\{1,2,\ldots,N\} \setminus \{j_1,j_2,\ldots,j_n\}$. We proceed inductively, and assume that the Sobolev solution (the exact definition will be given shortly) $w_{j_1,\ldots,j_n,i}(t,{\bm \lambda})$ of the HJB equation corresponding in the distress state ${\bf z}={\bf 0}^{j_1,\ldots,j_n,i}$ has been obtained. We next show the existence of a Sobolev solution to the HJB equation in the distress state ${\bf z}={\bm 0}^{j_1,\ldots,j_n}$. The master HJB equation~\eqref{eq:HJBeqn1fix} takes the form
    \begin{eqnarray}\label{eq:hjb-default-jn}
&&\frac{\partial w_{j_1,\ldots,j_n}}{\partial t}(t,{\bm \lambda}) + \frac{1}{2}{\rm Tr}\left[{\bm \sigma}{\bm \sigma}^{\top}D^2w_{j_1,\ldots,j_n}\right](t,{\bm \lambda})+ \frac{1}{2}\left[(\bm \nabla w_{j_1,\ldots,j_n}){\bm \sigma}{\bm \sigma}^{\top}(\bm \nabla w_{j_1,\ldots,j_n})^{\top}\right](t,{\bm \lambda})\nonumber\\
&&\qquad + \frac{\gamma}{2(1-\gamma)}\left[(\bm \nabla w_{j_1,\ldots,j_n}){\bm \sigma}_{\bf z}{\bm \sigma}_{\bf z}^{\top}(\bm \nabla w_{j_1,\ldots,j_n})^{\top}\right](t,{\bm \lambda})
+(\bm \nabla w_{j_1,\ldots,j_n})(t,{\bm \lambda}){\bm \theta}_{j_1,\ldots,j_n}(t,{\bm \lambda})\nonumber\\
&&\qquad +\xi_{j_1,\ldots,j_n}(t,{\bm \lambda},w_{j_1,\ldots,j_n}(t,{\bm \lambda}))=0,\ \ \ (t,{\bm \lambda})\in [0,T)\times\Delta_{K-1},
\end{eqnarray}
with terminal condition $w_{j_1,\ldots,j_n}(T,{\bm \lambda})=0$ for all ${\bm \lambda}\in\Delta_{K-1}$. Define the coefficient
\begin{eqnarray}\label{eq:beta-Jn}
\xi_{j_1,\ldots,j_n}(t,{\bm \lambda},v):=\sum_{i\in\{j_{n+1},\ldots,j_N\}}\tilde{h}_{i;j_1,\ldots,j_{n}}(t,{\bm \lambda})
e^{w_{j_1,\ldots,j_n,i}\left(t,\frac{{\bm \lambda}\cdot{\bm h}_{i;j_1,\ldots,j_{n}}^{\bot}(t)}{\tilde{h}_{i;j_1,\ldots,j_{n}}(t,{\bm \lambda})}\right)-v}+\rho_{j_1,\ldots,j_n}(t,{\bm \lambda}),
\end{eqnarray}
where we recall that $\rho_{j_1,\ldots,j_n}$ and ${\bm \theta}_{j_1,\ldots,j_n}$ have been defined in~\eqref{eq:notation-theta-rho}. 
For convenience, we reverse the flow of time setting $t\too T-t$, and rewrite Eq.~\eqref{eq:hjb-default-jn} as
 \begin{align}\label{eq:hjb-default-jn-T-t}
\frac{\partial \bar{u}}{\partial t}(t,{\bm \lambda}) &= \frac{1}{2}{\rm Tr}\left[\bar{\bm \sigma}\bar{\bm \sigma}^{\top}D^2\bar{u}\right](t,{\bm \lambda})+ \frac{1}{2}\left[(\bm \nabla \bar{u})\bar{\bm \sigma}\bar{\bm \sigma}^{\top}(\bm \nabla \bar{u})^{\top}\right](t,{\bm \lambda})\nonumber\\
&\quad+ \frac{\gamma}{2(1-\gamma)}\left[(\bm \nabla \bar{u})\bar{\bm \sigma}_{\bf z}\bar{\bm \sigma}_{\bf z}^{\top}(\bm \nabla \bar{u})^{\top}\right](t,{\bm \lambda})
+(\bm \nabla \bar{u})(t,{\bm \lambda})\bar{\bm \theta}_{j_1,\ldots,j_n}(t,{\bm \lambda})\nonumber\\
&\quad +\bar{\xi}_{j_1,\ldots,j_n}(t,{\bm \lambda},\bar{u}(t,{\bm \lambda})),\ \ \ (t,{\bm \lambda})\in (0,T]\times\Delta_{K-1},
\end{align}
with initial condition $\bar{u}(0,{\bm \lambda})=0$ for all ${\bm \lambda}\in\Delta_{K-1}$. For a given function $f(t,{\bm \lambda})$, we are using the notation $\bar{f}(t,{\bm \lambda}):=f(T-t,{\bm \lambda})$. Before studying the uniqueness of a Sobolev solution to Eq.~\eqref{eq:hjb-default-jn-T-t} via approximation arguments, we recall functional spaces which will be used in the analysis. Let $D\subset\R^{K-1}$ be a bounded domain. Then
 \begin{itemize}
  \item $L^p(D)$, $1\leq p<+\infty$ denotes the set of all measurable functions $f$ in $D$ such as the norm $\|f\|_{L^p(D)}:=(\int_D|f(x)|^p\D x)^{\frac{1}{p}}<+\infty$.
  \item $L^{\infty}(D)$ denotes the set of all bounded measurable functions $f$ in $D$ endowed with the norm $\|f\|_{L^{\infty}(D)}:={\rm ess}\sup_{x\in D}|f(x)|$.
  \item ${\rm W}_p^{l}(D)$, $1\leq p<+\infty$, and $l\in\N$, denotes the Sobolev space. It consists of all functions $f$ belonging to $L^p(D)$ which admit $k$-order weak derivatives $D^{k}f$ with $1\leq k\leq l$ so that $D^{k}f\in L^p(D)$. For a given $f\in {\rm W}_p^{l}(D)$, the norm is defined as $\|f\|_{{\rm W}_p^{l}(D)}:=(\int_D\sum_{0\leq k\leq l}|D^{k}f(x)|^p\D x)^{\frac{1}{p}},$ where $D^{0}f:=f$. The closure of $C_0^{\infty}(D)$ in the space of ${\rm W}_p^{l}(D)$ is denoted by ${\rm W}_{p,0}^{l}(D)$. If $p=2$, we use the notation ${\rm H}^l(D):={\rm W}_2^{l}(D)$ and ${\rm H}_0^l(D):={\rm W}_{2,0}^{l}(D)$.
   \item The Sobolev space ${\rm W}_p^{1,2}(Q_T)$ with $Q_T=[0,T]\times\Delta_{K-1}$, is the set of all functions $f(t,{\bm \lambda}):Q_T\too\R$ belonging to $L^p(Q_T)$ which admit  first-order weak derivative $\partial_tf$ w.r.t. time $t$ and $k$-order weak derivative $D^kf$ w.r.t. ${\bm \lambda}$, for $1\leq k\leq 2$.
       The norm of $f\in {\rm W}_p^{1,2}(Q_T)$ is defined as $$\|f\|_{{\rm W}_p^{1,2}(Q_T)}:=\left(\int_{Q_T}|\partial_tf(t,{\bm \lambda})|^p+|Df(t,{\bm \lambda})|^p+|D^2f(t,{\bm \lambda})|^p\D{\bm \lambda}\D t\right)^{\frac{1}{p}}.$$ 
       We will use the space ${\rm W}_p^{1,2}(Q_T)$ in our analysis.
   \end{itemize}
\qquad A function $\bar{u}:[0,T]\too {\rm H}^{1}(\Delta_{K-1})$ is a Sobolev solution of Eq.~\eqref{eq:hjb-default-jn-T-t}
with boundary condition $\bar{u}_Q$ if 
\begin{itemize}
  \item[{\sf(I)}] $\bar{u}-\bar{u}_Q\in L^2([0,T];{\rm H}^{1}(\Delta_{K-1}))$, and $\partial_t\bar{u}\in L^2([0,T];{\rm H}^{-1}(\Delta_{K-1}))$. Here ${\rm H}^{-1}(\Delta_{K-1})$ denotes the dual space of ${\rm H}^{1}(\Delta_{K-1})$.
  \item[{\sf(II)}] For every test function $\phi\in {\rm H}_0^{1}(\Delta_{K-1})$, it holds that
  \begin{eqnarray}\label{eq:HJBeqn-j1-jN-1-re-variation}
&&\int_0^T \left<\partial_t\bar{u},\phi\right>\D t + \frac{1}{2}\int_{Q_T}(\bm \nabla\phi)({\bm \lambda})\left[\bar{\bm \sigma}\bar{\bm \sigma}^{\top}(\bm \nabla\bar{u})^{\top}\right](t,{\bm \lambda})\D{\bm \lambda}\D t\nonumber\\
&&\quad = \frac{1}{2}\int_{Q_T}\left[(\bm \nabla\bar{u})\bar{\bm \sigma}\bar{\bm \sigma}^{\top}(\bm \nabla\bar{u})^{\top}\right](t,{\bm \lambda})\phi(t,{\bm \lambda})\D{\bm \lambda}\D t+\frac{\gamma}{2(1-\gamma)}\int_{Q_T}\left[(\bm \nabla\bar{u})\bar{\bm \sigma}_{\bf z}\bar{\bm \sigma}_{\bf z}^{\top}(\bm \nabla\bar{u})^{\top}\right](t,{\bm \lambda})\phi({\bm \lambda})
\D{\bm \lambda}\D t\nonumber\\
&&\qquad+\int_{Q_T}\left[(\bm \nabla\bar{u})(t,{\bm \lambda})\bar{\bm \theta}_{j_1,\ldots,j_n}(t,{\bm \lambda})+\bar{\xi}_{j_1,\ldots,j_n}(t,{\bm \lambda},\bar{u}(t,{\bm \lambda}))\right]\phi({\bm \lambda})\D{\bm \lambda}\D t\nonumber\\
&&\qquad-\frac{1}{2}\int_{Q_T}{\rm{\bf{div}}}(\bar{\bm \sigma}\bar{\bm \sigma}^{\top})(t,{\bm \lambda})(\bm \nabla\bar{u})^{\top}(t,{\bm \lambda})\phi({\bm \lambda})\D{\bm \lambda}\D t,
\end{eqnarray}
where ${\rm \bf div}(\bar{\bm \sigma}\bar{\bm \sigma}^{\top})(t,{\bm \lambda})$ is a $K-1$-dimensional row vector whose $\ell$-th component is given by
${\rm \bf div}(\bar{\bm \sigma}\bar{\bm \sigma}^{\top})_{\ell}(t,{\bm \lambda})=\sum_{k=1}^{K-1}\frac{\partial(\bar{\bm \sigma}\bar{\bm \sigma}^{\top})_{k,\ell}}{\partial \lambda_k}(t,{\bm \lambda})$ for $\ell=1,\ldots,K-1$.
With slight abuse of notation, $\left<\cdot,\cdot\right>$ represents the dual pair between ${\rm H}^{-1}(\Delta_{K-1})$ and ${\rm H}_0^1(\Delta_{K-1})$.

\item[(III)] $\bar{u}(0,{\bm \lambda})=0$ for all ${\bm \lambda}\in\Delta_{K-1}$, and $\bar{u}(t,{\bm \lambda})=\bar{u}_Q(t,{\bm \lambda})$ for all $(t,{\bm \lambda})\in(0,T]\times\partial\Delta_{K-1}$.
\end{itemize}
We analyze the variational representation \eqref{eq:HJBeqn-j1-jN-1-re-variation}, and show that the Sobolev solution $\bar{u}\in L^{\infty}(Q_T)$. As outlined at the beginning of the section, we proceed inductively and assume the existence of a Sobolev solution $w_{j_1,\ldots,j_n,i}\in L^2([0,T];{\rm H}^{1}(\Delta_{K-1}))\cap L^{\infty}(Q_T)$ in the default state ${\bf z}={\bf0}^{j_1,\ldots,j_n,i}$. Obviously, the base step of the induction is satisfied given $w_{\bm 1}$ has been computed in  closed form, and can be immediately verified that $w_{\bm 1} \in L^2([0,T];{\rm H}^{1}(\Delta_{K-1}))\cap L^{\infty}(Q_T)$. Next, we prove the existence and uniqueness of the solution to Eq.~\eqref{eq:HJBeqn-j1-jN-1-re-variation} in the space $L^2([0,T];{\rm H}^{1}(\Delta_{K-1}))\cap L^{\infty}(Q_T)$, given that $w_{j_1,\ldots,j_n,i}\in L^2([0,T];{\rm H}^{1}(\Delta_{K-1}))\cap L^{\infty}(Q_T)$.

We first provide a bound for the nonlinear term in Eq.~\eqref{eq:HJBeqn-j1-jN-1-re-variation}, which will be later used in our analysis.
\begin{lemma}\label{lem:lower-bound-xi}
{Under Assumption ({\bf A1})}, it holds that
\begin{eqnarray}\label{eq:lower-bound-xi}
\inf_{(t,{\bm \lambda},v)\in Q_T\times\R}\bar{\xi}_{j_1,\ldots,j_n}(t,{\bm \lambda},v)>-\infty,
\end{eqnarray}
where $\bar{\xi}_{j_1,\ldots,j_n}(t,{\bm \lambda},v):=\xi_{j_1,\ldots,j_n}(T-t,{\bm \lambda},v)$, with $\xi_{j_1,\ldots,j_n}(t,{\bm \lambda},v)$ defined by \eqref{eq:beta-Jn}.
\end{lemma}

\noindent{\it Proof.}\quad First, notice that $\tilde{h}_i(t,{\bm \lambda},{\bf z})>0$. This follows from the fact that $h_i(t,{\bm e}_k,{\bf z})>0$ for all pairs $(i,k) \in\{1,\ldots,N\}\times\{1,\ldots,K\}$. By \eqref{eq:beta-Jn}, it follows that
\[
\inf_{(t,{\bm \lambda},v)\in Q_T\times\R}\bar{\xi}_{j_1,\ldots,j_n}(t,{\bm \lambda},v) \geq \inf_{(t,{\bm \lambda})\in Q_T}\rho_{j_1,\ldots,j_n}(T-t,{\bm \lambda}).
\]
On the bounded domain $Q_T$, $\rho_{j_1,\ldots,j_n}(t,{\bm \lambda})$ is bounded by ({\bf A1}). The inequality~\eqref{eq:lower-bound-xi} then follows. \hfill$\Box$

As mentioned above, the appearance of two quadratic gradient terms in Eq.~\eqref{eq:hjb-default-jn-T-t} prevents the use of the exponential transform. Our approach is to deal with the quadratic growth of the gradient by applying Stampacchia{'s} truncation technique, see \cite{Stampacchia} and \cite{DuringJungel} for a related application, and establish an approximation problem to the variational form \eqref{eq:HJBeqn-j1-jN-1-re-variation}. To this purpose, define
\begin{align}\label{eq:min-beta0}
L_{\xi}&:= \inf_{(t,{\bm \lambda},v)\in Q_T\times\R}\bar{\xi}_{j_1,\ldots,j_n}(t,{\bm \lambda},v),\ \ \ \ \ L_{0,\xi}:=\min\left\{0,L_{\xi}\right\},\ \ \ \ \
L_{\xi}(t):=(t+1)L_{0,\xi}, \\
U_{\xi} &:=C_{n,N} e^{B_n-L_{\xi}(T)}+\sup_{(t,{\bm \lambda})\in Q_T}\rho_{j_1,\ldots,j_n}(T-t,{\bm \lambda}),\ \ \ \ U_{0,\xi}:=\max\{0,U_{\xi}\},\ \ \ \ \
U_{\xi}(t):=(t+1)U_{0,\xi},\nonumber
\end{align}
where
\begin{align}
B_n &:=\max_{i\notin\{j_1,\ldots,j_n\}}\left \{\sup_{(t,{\bm \lambda})\in Q_T}|w_{j_1,\ldots,j_n,i}(t,{\bm \lambda})|\right \},\quad
C_{n,N} := \sup_{(t,{\bm \lambda})\in Q_T} \left[\sum_{i\notin\{j_{1},\ldots,j_n\}}\tilde{h}_{i;j_1,\ldots,j_{n}}(t,{\bm \lambda})\right].
\label{eq:BCdef}
\end{align}
We then have the following
\begin{lemma}\label{lem:mtMt}
Under Assumption ({\bf A1}), it holds that, for all $t\geq0$, $L_{\xi}'(t)=L_{0,\xi}$ and
\[
-\infty<L_{\xi}(t)\leq L_{0,\xi}\leq0\leq U_{0,\xi}\leq U_{\xi}(t)<+\infty.
\]
\end{lemma}
\noindent{\it Proof.}\quad Using the definition of $L_{\xi}(t)$, it follows that $L_{\xi}'(t)=L_{0,\xi}$ and the inequality $L_{\xi}(t)\leq L_{0,\xi}\leq0$. Using Lemma \ref{lem:lower-bound-xi}, we obtain $L_{\xi}>-\infty$, and hence $L_{\xi}(t)=(t+1)\min\{0,L_{\xi}\}>-\infty$.
Notice that $B_n$ is finite because $w_{j_1,\ldots,j_n,i}\in L^{\infty}(Q_T)$. In particular, $B_{N-1}=\gamma r T$. Moreover $C_{n,N}$ is finite by Assumption ({\bf A1}). Using the above Lemma \ref{lem:mtMt}, we get $0\leq U_{0,\xi}\leq U_{\xi}(t)<+\infty$.
This completes the proof.
\hfill$\Box$

Next, we use Lemma \ref{lem:mtMt} to prove uniform $L^{\infty}$ bounds of solutions to the {approximating} problem. We will then develop an a priori estimate for solutions of the approximation problem in a suitably chosen functional space, and show that the sequence of approximating solutions converges to the Sobolev solution of our HJB PDE. Finally, we apply a one to one solution transformation technique to establish the uniqueness of the Sobolev solution to Eq.~\eqref{eq:hjb-default-jn-T-t}. 
Before proceeding further, we introduce the following notation:
\begin{equation}
K_{\bm \sigma}:=\inf_{(t,{\bm \lambda},{\bf z})\in Q_T\times{\cal S}}|{\bm \sigma}(t,{\bm \lambda},{\bf z})|^2,\ \ \ \ \ {\rm and}\ \ \ K^{\bm \sigma}:=\sup_{(t,{\bm \lambda},{\bf z})\in Q_T\times{\cal S}}|{\bm \sigma}(t,{\bm \lambda},{\bf z})|^2.
\label{eq:Ksigma}
\end{equation}
Both quantities above are finite from ({\bf A1}). In the following two sections, we will use Assumption ({\bf A1}).

\subsubsection{Uniform Boundedness of Approximating Solutions}

In this section, we establish an approximation problem to \eqref{eq:HJBeqn-j1-jN-1-re-variation} using Stampacchia{'s} truncation method. We then prove the
uniform boundedness in the space $L^{\infty}(Q_T)$ of the solutions of the approximation problem. The challenge in our analysis comes from the exponential form of the nonlinear term in Eq.~\eqref{eq:hjb-default-jn-T-t}. 

As proven in Lemma~\ref{lemma:TryPrvRmnPs}, filter probabilities never reach the boundary of the simplex domain.
Hence, we can set the boundary condition $\bar{u}_Q\equiv0$ without loss of generality.
{\Red
{\begin{remark}
The whole analysis carries through if we prescribe a non-zero boundary function, as long as we set
$L_{0,\xi}:=\min\left\{0,\inf_{(t,{\bm \lambda}) \in Q_T} \bar{u}_Q(t,{\bm \lambda}),L_{\xi}\right\}$, and  $U_{0,\xi}:=\max\{0,\sup_{(t,{\bm \lambda}) \in Q_T} \bar{u}_Q(t,{\bm \lambda}),U_{\xi}\}$.
\end{remark}}
}
Next, we introduce a sequence of truncated solutions corresponding to $\bar{u}$. More precisely, we define
\begin{eqnarray}\label{eq:truncated-solution}
\bar{u}_{L,U}(t,{\bm \lambda}) := \max\left\{L_{\xi}(t),\min\{\bar{u}(t,{\bm \lambda}),U_{\xi}(t)\}\right\},\ \ \ (t,{\bm \lambda})\in Q_T.
\end{eqnarray}
We consider the approximation problem to Eq.~\eqref{eq:HJBeqn-j1-jN-1-re-variation}. More precisely, for $m\in\N$, the approximating solution $\bar{u}^{m}$ is the solution of the following equation:
\begin{align}\label{eq:HJBeqn-j1-jN-1-re-variation-approximation}
\frac{\partial \bar{u}^m}{\partial t}(t,{\bm \lambda}) &= \frac{1}{2}{\rm Tr}\left[\bar{\bm \sigma}\bar{\bm \sigma}^{\top}D^2\bar{u}^{m}\right](t,{\bm \lambda})
+ \frac{1}{2}\frac{\left[(\bm \nabla\bar{u}^{m})\bar{\bm \sigma}\bar{\bm \sigma}^{\top}(\bm \nabla\bar{u}_{L,U}^{m})^{\top}\right](t,{\bm \lambda})}
{1+\frac{1}{m}\left[(\bm \nabla\bar{u}^{m})\bar{\bm \sigma}\bar{\bm \sigma}^{\top}(\bm \nabla\bar{u}^{m})^{\top}\right](t,{\bm \lambda})}\nonumber\\
&\quad+\frac{\gamma}{2(1-\gamma)}\frac{\left[(\bm \nabla\bar{u}^{m})\bar{\bm \sigma}_{\bf z}\bar{\bm \sigma}_{\bf z}^{\top}(\bm \nabla\bar{u}_{L,U}^{m})^{\top}\right](t,{\bm \lambda})}
{1+\frac{1}{m}\left[(\bm \nabla\bar{u}^{m})\bar{\bm \sigma}_{\bf z}\bar{\bm \sigma}_{\bf z}^{\top}(\bm \nabla\bar{u}^{m})^{\top}\right](t,{\bm \lambda})}
+(\bm \nabla\bar{u}^{m})(t,{\bm \lambda})\bar{\bm \theta}_{j_1,\ldots,j_n}(t,{\bm \lambda})\nonumber\\
&\quad+\bar{\xi}_{j_1,\ldots,j_n}(t,{\bm \lambda},\bar{u}^{m}(t,{\bm \lambda})).
\end{align}
We remark that the second and third terms in the approximation problem~\eqref{eq:HJBeqn-j1-jN-1-re-variation-approximation} constitute the bounded approximations for the third and fourth term in Eq.~\eqref{eq:hjb-default-jn}, respectively.
{The existence} of the solution $\bar{u}^{m}\in L^2([0,T];{\rm H}^2(\Delta_{K-1}))\cap {\rm H}^{1}([0,T];L^2(\Delta_{K-1}))$ is guaranteed by Schauder's fixed point
theorem (see section~9.2 of Part III in \cite{Evans}). The next lemma establishes the boundedness property of the approximating solution $(\bar{u}^{m};\ m\in\N)$ in $L^{\infty}(Q_T)$.
\begin{lemma}\label{lem:lower-solution}
For each $0\leq t\leq T$, it holds that $\Delta_{K-1}$-a.s.,
\[
L_{\xi}(t)\leq\bar{u}^{m}(t)\leq U_{\xi}(t),\ \ \ \ \ \forall\ m\in\N,
\]
where the lower bound $L_{\xi}(t)$ and upper bound $U_{\xi}(t)$ are defined in \eqref{eq:min-beta0}.
\end{lemma}

\noindent{\it Proof.}\quad We first establish the lower bound, and then the upper bound.\\
{\it Proof of $L_{\xi}(t)\leq\bar{u}^{m}(t)$, $\Delta_{K-1}$-a.s.}. We choose $\phi(t,{\bm \lambda})=\phi^{-}(\bar{u}^{m}(t,{\bm \lambda})):=\max \left(L_{\xi}(t) -\bar{u}^{m}(t,{\bm \lambda}),0 \right)$ as the test function in (II) in the definition of {Sobolev} solution.
 We next plug this test function into the solution corresponding to~\eqref{eq:HJBeqn-j1-jN-1-re-variation-approximation}. We first compute
\begin{align*}
\int_0^T \left<\partial_t\bar{u}^{m}(t),{\phi}^{-}(\bar{u}^{m}(t)\right>\D t &= \int_0^T \left<\partial_t(\bar{u}^{m}(t)-L_{\xi}(t)),{\phi}^{-}(\bar{u}^{m}(t))\right>\D t +  \int_0^T \left<L_{0,\xi},{\phi}^{-}(\bar{u}^{m}(t))\right>\D t\nonumber\\
&=-\frac{1}{2}\int_{\Delta_{K-1}} \left[\left(\phi^{-}(\bar{u}^{m}(T,{\bm \lambda}))\right)^2-\left(\phi^{-}(\bar{u}^{m}(0,{\bm \lambda}))\right)^2\right]\D{\bm \lambda}
+\int_0^T\left<L_{0,\xi},{\phi}^{-}(\bar{u}^{m}(t))\right>\D t\nonumber\\
&=-\frac{1}{2}\int_{\Delta_{K-1}} \left(\phi^{-}(\bar{u}^{m}(T,{\bm \lambda}))\right)^2\D{\bm \lambda}
+\int_0^T\left<L_{0,\xi},{\phi}^{-}(\bar{u}^{m}(t))\right>\D t,
\end{align*}
where the first equality is obtained using that $L'_{\xi}(t)=L_{0,\xi}$. The second equality follows from integration by parts after using the definition of inner product and exchanging the orders of integration. For the third equality, we used the initial condition $\bar{u}^{m}(0,{\bm \lambda})=0$ for all ${\bm \lambda}\in\Delta_{K-1}$, and $\phi^{-}(0)=0$, since
$L_{\xi}(t)\leq0$.

Next, we show that $[\bm \nabla \bar{u}_{L,U}^{m}\phi^{-}(\bar{u}^{m})](t,{\bm \lambda})={\bf0}$. Using the definitions of $\bar{u}_{L,U}^{m}$, and $\phi^{-}(\bar{u}^m)$,
\begin{eqnarray}
\nonumber [\bm \nabla \bar{u}_{L,U}^{m}\phi^{-}(\bar{u}^{m})](t,{\bm \lambda}) &=& \bm \nabla \left(\bar{u}^{m}(t,{\bm\lambda}) \idc_{L_{\xi}(t) \leq \bar{u}^{m}(t,{\bm\lambda}) \leq U_{\xi}(t)} + U_{\xi}(t) \idc_{ \bar{u}^{m}(t,{\bm \lambda}) \geq U_{\xi}(t)} + L_{\xi}(t) \idc_{ \bar{u}^{m}(t,{\bm \lambda}) < L_{\xi}(t)} \right)   \\
\nonumber & & \times (L_{\xi}(t) - \bar{u}^{m}(t,{\bm \lambda})) \idc_{L_{\xi}(t) \geq \bar{u}^{m}(t,{\bm \lambda})} \\
\nonumber &=& \bm \nabla \bar{u}^{m}(t,{\bm \lambda})\idc_{L_{\xi}(t) \leq \bar{u}^{m}(t,{\bm \lambda}) \leq U_{\xi}(t)} \idc_{L_{\xi}(t) \geq \bar{u}^{m}(t,{\bm \lambda})} (L_{\xi}(t) - \bar{u}^{m}(t,{\bm \lambda})) \\
\nonumber &=& \bm \nabla \bar{u}^{m}(t,{\bm \lambda}) \idc_{L_{\xi}(t) = \bar{u}^{m}(t,{\bm \lambda}) } (L_{\xi}(t) - \bar{u}^{m}(t,{\bm \lambda}))
= {\bf0},
\label{eq:nablacomp}
\end{eqnarray}
where the second step follows from that both $L_{\xi}(t)$ and $U_{\xi}(t)$ are independent of ${\bm \lambda}$, and hence their gradient is zero. Further, it holds that $\phi^{-}(\bar{u}^{m})\bm \nabla\bar{u}^{m}=-\phi^{-}(\bar{u}^{m})\bm \nabla\phi^{-}(\bar{u}^{m})$. Hence, plugging  \eqref{eq:HJBeqn-j1-jN-1-re-variation-approximation} into~\eqref{eq:HJBeqn-j1-jN-1-re-variation}, we obtain
\begin{eqnarray}\label{eq:HJBeqn-j1-jN-1-re-variation-approximation0}
&&-\frac{1}{2}\int_{\Delta_{K-1}} \left(\phi^{-}(\bar{u}^{m}(T,{\bm \lambda}))\right)^2\D{\bm \lambda}- \frac{1}{2}\int_{Q_T}\left[(\bm \nabla\phi^{-}(\bar{u}^{m}))\bar{\bm \sigma}\bar{\bm \sigma}^{\top}
(\bm \nabla\phi^{-}(\bar{u}^{m}))^{\top}\right](t,{\bm \lambda})\D{\bm \lambda}\D t\nonumber\\
&&\qquad= \int_{Q_T}\left(-(\bm \nabla{\phi}^{-}(\bar{u}^{m})(t,{\bm \lambda}))\bar{\bm \theta}_{j_1,\ldots,j_n}(t,{\bm \lambda})+\bar{\xi}_{j_1,\ldots,j_n}(t,{\bm \lambda},\bar{u}^{m}(t,{\bm \lambda}))
-L_{0,\xi}\right){\phi}^{-}(\bar{u}^{m}(t,{\bm \lambda}))\D{\bm \lambda}\D t\nonumber\\
&&\qquad\quad+\frac{1}{2}\int_{Q_T}{\rm \bf div} (\bar{\bm \sigma}\bar{\bm \sigma}^{\top})(t,{\bm \lambda})\left(\bm \nabla{\phi}^{-}(\bar{u}^{m})(t,{\bm \lambda})\right)^{\top}
{\phi}^{-}(\bar{u}^{m}(t,{\bm \lambda}))\D{\bm \lambda}\D t,
\end{eqnarray}
which yields
\begin{eqnarray}\label{eq:HJBeqn-j1-jN-1-re-variation-approximationa1}
&&\frac{1}{2}\int_{\Delta_{K-1}} \left(\phi^{-}(\bar{u}^{m}(T,{\bm \lambda}))\right)^2\D{\bm \lambda}+ \frac{1}{2}\int_{Q_T}\left[(\bm \nabla\phi^{-}(\bar{u}^{m}))\bar{\bm \sigma}\bar{\bm \sigma}^{\top}
(\bm \nabla\phi^{-}(\bar{u}^{m}))^{\top}\right](t,{\bm \lambda})\D{\bm \lambda}\D t\nonumber\\
&&\qquad= \int_{Q_T}\left[(\bm \nabla{\phi}^{-}(\bar{u}^{m}))(t,{\bm \lambda})\bar{\bm \theta}_{j_1,\ldots,j_n}(t,{\bm \lambda})-\left(\bar{\xi}_{j_1,\ldots,j_n}(t,{\bm \lambda},\bar{u}^{m}(t,{\bm \lambda}))
-L_{0,\xi}\right)\right]{\phi}^{-}(\bar{u}^{m}(t,{\bm \lambda}))\D{\bm \lambda}\D t\nonumber\\
&&\qquad\quad-\frac{1}{2}\int_{Q_T}{\rm \bf div} (\bar{\bm \sigma}\bar{\bm \sigma}^{\top})(t,{\bm \lambda})\left(\bm \nabla{\phi}^{-}(\bar{u}^{m})(t,{\bm \lambda})\right)^{\top}
{\phi}^{-}(\bar{u}^{m}(t,{\bm \lambda}))\D{\bm \lambda}\D t\nonumber\\
&&\qquad \leq \int_{Q_T}(\bm \nabla{\phi}^{-}(\bar{u}^{m}))(t,{\bm \lambda})\bar{\bm \theta}_{j_1,\ldots,j_n}(t,{\bm \lambda}){\phi}^{-}(\bar{u}^{m}(t,{\bm \lambda}))\D{\bm \lambda}\D t\nonumber\\
&&\qquad\qquad-\frac{1}{2}\int_{Q_T}{\rm \bf div} (\bar{\bm \sigma}\bar{\bm \sigma}^{\top})(t,{\bm \lambda})\left(\bm \nabla{\phi}^{-}(\bar{u}^{m})(t,{\bm \lambda})\right)^{\top}
{\phi}^{-}(\bar{u}^{m}(t,{\bm \lambda}))\D{\bm \lambda}\D t.
\end{eqnarray}
Above, we have used the fact that, for all $(t,{\bm \lambda},v)\in Q_T\times\R$, $\bar{\xi}_{j_1,\ldots,j_n}(t,{\bm \lambda},v)-L_{0,\xi}\geq0$. This follows directly from the definition~\eqref{eq:min-beta0}. Moreover, using Young's inequality, 
for arbitrary $\delta>0$ there exists $C_{\delta,\theta}>0$ such that
\begin{eqnarray}
&&\int_{Q_T}\left|\left(\bm \nabla{\phi}^{-}(\bar{u}^{m})(t,{\bm \lambda})\right){\bar {\bm \theta}}_{j_1,\ldots,j_n}(t,{\bm \lambda})\right|{\phi}^{-}(\bar{u}^{m}(t,{\bm \lambda}))
\D{\bm \lambda}\D t\nonumber\\
&&\qquad\leq \frac{\delta}{4}\int_{Q_T}\left|\bm \nabla{\phi}^{-}(\bar{u}^{m})(t,{\bm \lambda})\right|^2\D{\bm \lambda}\D t+C_{\delta,\theta}\int_{Q_T}\left({\phi}^{-}(\bar{u}^{m}(t,{\bm \lambda}))\right)^2\D{\bm \lambda}\D t.
\label{eq:secondineq}
\end{eqnarray}
To deduce the above inequality, we have also used Assumption ({\bf A1}), which implies that $\bar {\bm \theta}_{j_1,\ldots,j_n}(t,{\bm \lambda})$ is bounded for all $(t,{\bm \lambda})\in Q_T$ by \eqref{eq:notation-theta-rho}. Similarly, there also exists a constant $C_{\delta,\sigma}>0$ so that
\begin{eqnarray}
&&\int_{Q_T}\left|{\rm \bf div} (\bar{\bm \sigma}\bar{\bm \sigma}^{\top})(t,{\bm \lambda})(\bm \nabla{\phi}^{-}(\bar{u}^{m})(t,{\bm \lambda}))^{\top}\right|{\phi}^{-}(\bar{u}^{m}(t,{\bm \lambda}))\D{\bm \lambda}\D t\nonumber\\
&&\qquad \leq \frac{\delta}{2}\int_{Q_T}\left|\bm \nabla{\phi}^{-}(\bar{u}^{m})(t,{\bm \lambda})\right|^2\D{\bm \lambda}\D t
+C_{\delta,\sigma}\int_{Q_T}\left({\phi}^{-}(\bar{u}^{m}(t,{\bm \lambda}))\right)^2\D{\bm \lambda}\D t,
\label{eq:thirdineq}
\end{eqnarray}
where we have used Assumption ({\bf A1}) which guarantees that both functions $b_i(\cdot,{\bm e}_k,{\bf z})$ and $h_i(\cdot,{\bm e}_k,{\bf z})$ are $C^1(\R_+)$. This implies that $\bm{\sigma}$ is bounded, which in turns implies that ${\rm \bf div} (\bar{\bm \sigma}\bar{\bm \sigma}^{\top})$ is bounded because $Q_T$ is a bounded domain.
On the other hand, it holds that
\begin{eqnarray}
\frac{1}{2}\int_{Q_T}\left[(\bm \nabla\phi^{-}(\bar{u}^{m}))\bar{\bm \sigma}\bar{\bm \sigma}^{\top}(\bm \nabla\phi^{-}(\bar{u}^{m}))^{\top}\right](t,{\bm \lambda})\D{\bm \lambda}\D t \geq \frac{1}{2}K_{\bm \sigma}\int_{Q_T}\left|\bm \nabla{\phi}^{-}(\bar{u}^{m})(t,{\bm \lambda})\right|^2\D{\bm \lambda}\D t,
\label{eq:lastineq}
\end{eqnarray}
where $K_{\bm \sigma}$ has been defined in Eq.~\eqref{eq:Ksigma}, and is finite by ({\bf A1}).
Combining the inequalities (\ref{eq:HJBeqn-j1-jN-1-re-variation-approximationa1})-(\ref{eq:lastineq}), we obtain
\begin{eqnarray*}
&&\frac{1}{2}\int_{\Delta_{K-1}} \left(\phi^{-}(\bar{u}^{m}(T,{\bm \lambda}))\right)^2\D{\bm \lambda}+\frac{1}{2}(K_{\bm \sigma}-\delta)\int_{Q_T}\left|\bm \nabla{\phi}^{-}(\bar{u}^{m})(t,{\bm \lambda})\right|^2
\D{\bm \lambda}\D t\nonumber\\
&&\qquad\leq
\left(C_{\delta,\theta}+ \frac{C_{\delta,\sigma}}{2} \right)\int_{Q_T}\left({\phi}^{-}(\bar{u}^{m}(t,{\bm \lambda}))\right)^2\D{\bm \lambda}\D t.
\end{eqnarray*}
Since $\delta$ can be arbitrarily chosen, we can take $\delta>0$ small enough so that $K_{\bm \sigma}-\delta>0$. We then obtain
\begin{eqnarray}\label{eq:GronwallLemma}
A(T) \leq 2\left(C_{\delta,\theta}+ \frac{C_{\delta,\sigma}}{2}\right)\int_0^T A(t)\D t,
\end{eqnarray}
where we define $A(t):=\int_{\Delta_{K-1}} \left(\phi^{-}(\bar{u}^{m}(t,{\bm \lambda}))\right)^2\D{\bm \lambda}$. By Gronwall's Lemma, for all $t\in[0,T]$, we have that $A(t)=0$ (and hence $\phi^{-}(\bar{u}^{m}(t))=0$, $\Delta_{K-1}$-a.s.). Using the definition of test function $\phi^{-}(\bar{u}^{m})$, we have that $\bar{u}^{m}(t)\geq L_{\xi}(t)$, $\Delta_{K-1}$-a.s..

\noindent {\it Proof of $\bar{u}^{m}(t)\leq U_{\xi}(t)$, $\Delta_{K-1}$-a.s..} In this case, we choose $\phi(t,{\bm \lambda})=\phi^{+}(\bar{u}^m(t,{\bm \lambda})):= \max (\bar{u}^{m}(t,{\bm \lambda})-U_{\xi}(t),0 )$ 
as the test function. As above, we plug such test function into the solution corresponding
to~\eqref{eq:HJBeqn-j1-jN-1-re-variation-approximation}. First
\begin{eqnarray*}
&&\int_0^T \left<\partial_t\bar{u}^{m}(t),{\phi}^{+}(\bar{u}^{m}(t))\right>\D t = \int_0^T \left<\partial_t(\bar{u}^{m}(t)-U_{\xi}(t)),{\phi}^{+}(\bar{u}^{m}(t))\right>\D t
+  \int_0^T \left<U_{0,\xi},{\phi}^{+}(\bar{u}^{m}(t))\right>\D t\nonumber\\
&&\quad=\frac{1}{2}\int_{\Delta_{K-1}} \left[\left(\phi^{+}(\bar{u}^{m}(T,{\bm \lambda}))\right)^2-\left(\phi^{+}(\bar{u}^{m}(0,{\bm \lambda}))\right)^2\right]\D{\bm \lambda}
+\int_0^T\left<U_{0,\xi},{\phi}^{+}(\bar{u}^{m}(t))\right>\D t\nonumber\\
&&\quad=\frac{1}{2}\int_{\Delta_{K-1}} \left(\phi^{+}(\bar{u}^{m}(T,{\bm \lambda}))\right)^2\D{\bm \lambda}+\int_0^T\left<U_{0,\xi},{\phi}^{+}(\bar{u}^{m}(t))\right>\D t,
\end{eqnarray*}
where the first equality is obtained using that $U'_{\xi}(t)=U_{0,\xi}$. The second equality follows from integration by parts, while for the third equality we used the initial condition $\bar{u}^{m}(0,{\bm \lambda})=0$ for all ${\bm \lambda}\in\Delta_{K-1}$, and $\phi^{+}(0)=0$, since $U_{\xi}(t)\geq0$. Using a similar computation as in \eqref{eq:nablacomp}, we deduce $\bm \nabla \bar{u}_{L,U}^{m}\phi^{+}(\bar{u}^{m})=0$. Moreover, $\phi^{+}(\bar{u}^{m})\bm \nabla\bar{u}^{m}=\phi^{+}(\bar{u}^{m})\bm \nabla\phi^{+}(\bar{u}^{m})$. Then, combining
\eqref{eq:HJBeqn-j1-jN-1-re-variation} and \eqref{eq:HJBeqn-j1-jN-1-re-variation-approximation}, it holds that
\begin{eqnarray}\label{eq:HJBeqn-j1-jN-1-re-variation-approximation2}
&&\frac{1}{2}\int_{\Delta_{K-1}} \left(\phi^{+}(\bar{u}^{m}(T,{\bm \lambda}))\right)^2\D{\bm \lambda}+ \frac{1}{2}\int_{Q_T}\left[(\bm \nabla\phi^{+}(\bar{u}^{m}))\bar{\bm \sigma}\bar{\bm \sigma}^{\top}(\bm \nabla\phi^{+}(\bar{u}^{m}))^{\top}\right](t,{\bm \lambda})\D{\bm \lambda}\D t\nonumber\\
&&\qquad= \int_{Q_T}\left[(\bm \nabla{\phi}^{+}(\bar{u}^{m}))(t,{\bm \lambda})\bar{\bm \theta}_{j_1,\ldots,j_n}(t,{\bm \lambda})
+\bar{\xi}_{j_1,\ldots,j_n}(t,{\bm \lambda},\bar{u}^{m}(t,{\bm \lambda}))-U_{0,\xi}\right]{\phi}^{+}(\bar{u}^{m}(t,{\bm \lambda}))\D{\bm \lambda}\D t\nonumber\\
&&\qquad\quad-\frac{1}{2}\int_{Q_T}{\rm \bf div} (\bar{\bm \sigma}\bar{\bm \sigma}^{\top})(t,{\bm \lambda})(\bm \nabla{\phi}^{+}(\bar{u}^{m})(t,{\bm \lambda}))^{\top}{\phi}^{+}(\bar{u}^{m}(t,{\bm \lambda}))\D{\bm \lambda}\D t.
\end{eqnarray}
From the first part of the proof, we know that $\bar{u}^{m}(t)\geq L_{\xi}(t)\geq L_{\xi}(T)$ for all $t\in[0,T]$, where the last inequality follows directly from~\eqref{eq:min-beta0}. Then, it holds that
\begin{align}\label{eq:betalessM}
\bar{\xi}_{j_1,\ldots,j_n}(t,{\bm \lambda},\bar{u}^m(t,{\bm \lambda})) &\leq C_{n,N} e^{B_n-\bar{u}^m(t,{\bm \lambda})} + \rho_{j_1,\ldots,j_n}(T-t,{\bm \lambda})\nonumber\\
&\leq
C_{n,N} e^{B_n-L_{\xi}(T)}+\sup_{(t,{\bm \lambda})\in Q_T}\rho_{j_1,\ldots,j_n}(T-t,{\bm \lambda})=U_{\xi}\leq U_{0,\xi},
\end{align}
where we recall that $B_n$ and $C_{n,N}$ have been defined in Eq.~\eqref{eq:BCdef} and $U_{\xi}$ in Eq.~\eqref{eq:min-beta0}. Hence, we obtain that
$\bar{\xi}_{j_1,\ldots,j_n}(t,{\bm \lambda},\bar{u}^m(t,{\bm \lambda})) -  U_{0,\xi}\leq0$. This implies that
\begin{eqnarray}\label{eq:HJBeqn-j1-jN-1-re-variation-approximation4}
&&\frac{1}{2}\int_{\Delta_{K-1}} \left(\phi^{+}(\bar{u}^{m}(T,{\bm \lambda}))\right)^2\D{\bm \lambda}+ \frac{1}{2}\int_{Q_T}\left[(\bm \nabla\phi^{+}(\bar{u}^{m}))\bar{\bm \sigma}\bar{\bm \sigma}^{\top}(\bm \nabla\phi^{+}(\bar{u}^{}))^{\top}\right](t,{\bm \lambda})\D{\bm \lambda}\D t\nonumber\\
&&\quad \leq \int_{Q_T}\left|\left(\bm \nabla{\phi}^{+}(\bar{u}^{m})\right)(t,{\bm \lambda})\bar{\bm \theta}_{j_1,\ldots,j_n}(t,{\bm \lambda})\right|
{\phi}^{+}(\bar{u}^{m}(t,{\bm \lambda}))\D{\bm \lambda}\D t\nonumber\\
 &&\qquad +\frac{1}{2}\int_{Q_T}\left|{\rm \bf div} (\bar{\bm \sigma}\bar{\bm \sigma}^{\top})(t,{\bm \lambda})(\bm \nabla{\phi}^{+}(\bar{u}^{m})(t,{\bm \lambda}))^{\top}\right|{\phi}^{+}(\bar{u}^{m}(t,{\bm \lambda}))\D{\bm \lambda}\D t.
\end{eqnarray}
Using Young's inequality and Gronwall's Lemma as in the proof for the case $\bar{u}^{m}(t)\geq L_{\xi}(t)$, we obtain
${\phi}^{+}(\bar{u}^{m}(t))=0$ for all $t\in[0,T]$, $\Delta_{K-1}$-a.s.. This is equivalent to $\bar{u}^{m}(t)\leq U_{\xi}(t)$ for all $t\in[0,T]$, $\Delta_{K-1}$-a.s. \hfill$\Box$\\

\subsubsection{Convergence to the Solution}
In this section, we first prove that solutions of the approximating problem are uniformly bounded in a suitably chosen Sobolev space. We then use this result to show that the sequence of approximating solutions converges to a Sobolev solution of the HJB PDE. We start with the following

\begin{lemma}\label{lem:bounded-L2H1}
Let $\bar{u}^{m}$ be the solution of the approximating Eq.~\eqref{eq:HJBeqn-j1-jN-1-re-variation-approximation}. Then there exists a constant $C>0$ independent of $m\in\N$ such that $\|\bar{u}^{m}\|_{L^2([0,T];{\rm H}^{1}(\Delta_{K-1}))}\leq C$.
\end{lemma}

\noindent{\it Proof.}\quad It is enough to prove $\int_{Q_T}|\bm \nabla\bar{u}^{m}(t,{\bm \lambda})|^2\D{\bm \lambda}\D t \leq C$. Choose the test function  $\phi(\bar{u}^{m}(t,{\bm \lambda}))=\sinh(a\bar{u}^{m}(t,{\bm \lambda}))$ for some constant $a>0$ which will be specified later. (If we were prescribing a general nonzero boundary function, the following analysis would carry through upon choosing $\phi(\bar{u}^{m}(t,{\bm \lambda}))=\sinh(a\bar{u}^{m}(t,{\bm \lambda})) - \sinh(a\bar{u}_Q(t,{\bm \lambda}))$).
Plugging the test function into the solution of the approximating equation~\eqref{eq:HJBeqn-j1-jN-1-re-variation-approximation}, we obtain

\begin{eqnarray}\label{eq:HJBeqn-j1-jN-1-re-variation-approximation1}
&&\int_0^T \left<\partial_t\bar{u}^{m}(t),\sinh(a\bar{u}^{m}(t))\right>\D t + \frac{1}{2}\int_{Q_T}a\cosh(a\bar{u}^{m}(t,{\bm \lambda}))\left[{(\bm \nabla\bar{u}^{m}) \bar{\bm \sigma}\bar{\bm \sigma}^{\top}(\bm \nabla\bar{u}^{m})^{\top}}\right](t,{\bm \lambda})
\D{\bm \lambda}\D t\nonumber\\
&&\qquad= \frac{1}{2}\int_{Q_T}\frac{\left[(\bm \nabla\bar{u}^{m})\bar{\bm \sigma}\bar{\bm \sigma}^{\top}(\bm \nabla\bar{u}^{m})^{\top}\right](t,{\bm \lambda})}
{1+\frac{1}{m}\left[(\bm \nabla\bar{u}^{m})\bar{\bm \sigma}\bar{\bm \sigma}^{\top}(\bm \nabla\bar{u}^{m})^{\top}\right](t,{\bm \lambda})}\sinh(a\bar{u}^{m}(t,{\bm \lambda}))\D{\bm \lambda}\D t\nonumber\\
&&\qquad\quad+\frac{\gamma}{2(1-\gamma)}\int_{Q_T}\frac{\left[(\bm \nabla\bar{u}^{m})\bar{\bm \sigma}_{\bf z}\bar{\bm \sigma}_{\bf z}^{\top}(\bm \nabla\bar{u}^{m})^{\top}\right](t,{\bm \lambda})}
{1+\frac{1}{m}\left[(\bm \nabla\bar{u}^{m})\bar{\bm \sigma}_{\bf z}\bar{\bm \sigma}_{\bf z}^{\top}(\bm \nabla\bar{u}^{m})^{\top}\right](t,{\bm \lambda})}\sinh(a\bar{u}^{m}(t,{\bm \lambda}))\D{\bm \lambda}\D t\nonumber\\
&&\qquad\quad+\int_{Q_T}\left[(\bm \nabla\bar{u}^{m})(t,{\bm \lambda})\bar{\bm \theta}_{j_1,\ldots,j_n}(t,{\bm \lambda})+\bar{\xi}_{j_1,\ldots,j_n}(t,{\bm \lambda},\bar{u}^{m}(t,{\bm \lambda}))\right]\sinh(a\bar{u}^{m}(t,{\bm \lambda}))
\D{\bm \lambda}\D t\nonumber\\
&&\qquad\quad-\frac{1}{2}\int_{Q_T}{\rm \bf div} (\bar{\bm \sigma}\bar{\bm \sigma}^{\top})(t,{\bm \lambda})(\bm \nabla\bar{u}^{m}(t,{\bm \lambda}))^{\top}\sinh(a\bar{u}^{m}(t,{\bm \lambda}))\D{\bm \lambda}\D t.
\end{eqnarray}
We start dealing with the first term on the r.h.s. of \eqref{eq:HJBeqn-j1-jN-1-re-variation-approximation1}. More precisely, we have
\begin{eqnarray*}
\int_0^T \left<\partial_t\bar{u}^{m}(t),\sinh(a\bar{u}^{m}(t))\right>\D t = \int_{\Delta_{K-1}}\int_0^T \sinh(a\bar{u}^{m}(t,{\bm \lambda}))\D\bar{u}^{m}(t,{\bm \lambda})\D{\bm \lambda}
= \frac{1}{a}\int_{\Delta_{K-1}} \cosh(a\bar{u}^{m}(T,{\bm \lambda})) \D{\bm \lambda}  > 0,
\end{eqnarray*}
where we used the fact that $\bar{u}^{m}(0,{\bm \lambda})=0$ for all ${\bm \lambda}\in\Delta_{K-1}$.
We next provide upper bounds for the first two terms on the r.h.s of \eqref{eq:HJBeqn-j1-jN-1-re-variation-approximation1}. We have
\begin{eqnarray*}
&&\frac{1}{2}\int_{Q_T}\frac{\left[(\bm \nabla\bar{u}^{m})\bar{\bm \sigma}\bar{\bm \sigma}^{\top}(\bm \nabla\bar{u}^{m})^{\top}\right](t,{\bm \lambda})}
{1+\frac{1}{m}\left[(\bm \nabla\bar{u}^{m})\bar{\bm \sigma}\bar{\bm \sigma}^{\top}(\bm \nabla\bar{u}^{m})^{\top}\right](t,{\bm \lambda})}|\sinh(a\bar{u}^{m}(t,{\bm \lambda}))|
\D{\bm \lambda}\D t\nonumber\\
&&\qquad\leq\frac{1}{2}\int_{Q_T}\left|\bm \nabla\bar{u}^{m}(t,{\bm \lambda})\bar{\bm \sigma}(t,{\bm \lambda})\right|^2\cosh(a\bar{u}^{m}(t,{\bm \lambda}))\D{\bm \lambda}\D t.
\end{eqnarray*}
Similarly, it holds that
\begin{eqnarray*}
&&\frac{\gamma}{2(1-\gamma)}\int_{Q_T}\frac{\left[(\bm \nabla\bar{u}^{m})\bar{\bm \sigma}_{\bf z}\bar{\bm \sigma}_{\bf z}^{\top}(\bm \nabla\bar{u}^{m})^{\top}\right](t,{\bm \lambda})}
{1+\frac{1}{m}\left[(\bm \nabla\bar{u}^{m})\bar{\bm \sigma}_{\bf z}\bar{\bm \sigma}_{\bf z}^{\top}(\bm \nabla\bar{u}^{m})^{\top}\right](t,{\bm \lambda})}|\sinh(a\bar{u}^{m}(t,{\bm \lambda}))|
\D{\bm \lambda}\D t\nonumber\\
&&\qquad\leq\frac{\gamma}{2(1-\gamma)}\int_{Q_T}\left|\bm \nabla\bar{u}^{m}(t,{\bm \lambda})\bar{\bm \sigma}(t,{\bm \lambda})\right|^2\cosh(a\bar{u}^{m}(t,{\bm \lambda}))\D{\bm \lambda}\D t.
\end{eqnarray*}

For the third term on the r.h.s. of \eqref{eq:HJBeqn-j1-jN-1-re-variation-approximation1}, using Young's inequality, for any $\delta>0$ there exists a constant $C_{\delta,\theta}>0$ such that
\begin{eqnarray*}
&&\int_{Q_T}\left|(\bm \nabla\bar{u}^{m})(t,{\bm \lambda})\bar{\bm \theta}_{j_1,\ldots,j_n}(t,{\bm \lambda})+\bar{\xi}_{j_1,\ldots,j_n}(t,{\bm \lambda},\bar{u}^{m}(t,{\bm \lambda}))\right|
\left|\sinh(a\bar{u}^{m}(t,{\bm \lambda}))\right|\D{\bm \lambda}\D t\nonumber\\
&&\quad\leq\delta\int_{Q_T}|\bm \nabla\bar{u}^{m}(t,{\bm \lambda})|^2\D{\bm \lambda}\D t+C_{\delta,\theta}\int_{Q_T}\cosh^2(a\bar{u}^{m}(t,{\bm \lambda}))\D{\bm \lambda}\D t+C_T,
\end{eqnarray*}
where $C_T>0$, and we used that $\bar{\xi}_{j_1,\ldots,j_n}(t,{\bm \lambda},\bar{u}^{m}(t,{\bm \lambda}))$ is bounded a.s., since $\bar{u}^{m}\in L^{\infty}(Q_T)$ by Lemma \ref{lem:lower-solution}.
For the last term on the r.h.s. of \eqref{eq:HJBeqn-j1-jN-1-re-variation-approximation1}, it holds that for any $\delta>0$ there exists $C_{\delta,\sigma}>0$ such that
\begin{eqnarray*}
&&\frac{1}{2}\int_{Q_T}\left|{\rm \bf div}(\bar{\bm \sigma}\bar{\bm \sigma}^{\top})(t,{\bm \lambda})(\bm \nabla\bar{u}^{m}(t,{\bm \lambda}))^{\top}\sinh(a\bar{u}^{m}(t,{\bm \lambda}))\right|\D{\bm \lambda}\D t\nonumber\\
&&\qquad\leq \delta\int_{Q_T}|\bm \nabla\bar{u}^{m}(t,{\bm \lambda})|^2\D{\bm \lambda}\D t + C_{\delta,\sigma}\int_{Q_T}\cosh^2(a\bar{u}^{m}(t,{\bm \lambda}))\D{\bm \lambda}\D t,
\end{eqnarray*}
where we have used Assumption ({\bf A1}) which ensures the boundedness of ${\rm \bf div}(\bar{\bm \sigma}\bar{\bm \sigma}^{\top})(t,{\bm \lambda})$. Using the previously derived
inequalities along with Eq.~\eqref{eq:HJBeqn-j1-jN-1-re-variation-approximation1}, we obtain
\begin{eqnarray}\label{eq:HJBeqn-j1-jN-1-re-variation-approximation2}
&&\frac{1}{2}\int_{Q_T}a\cosh(a\bar{u}^{m}(t,{\bm \lambda}))\left[(\bm \nabla\bar{u}^{m}){\bar{\bm \sigma}\bar{\bm \sigma}^{\top}(\bm \nabla\bar{u}^{m})^{\top}}\right](t,{\bm \lambda})
\D{\bm \lambda}\D t\nonumber\\
&&\qquad\leq \frac{1}{2(1-\gamma)}\int_{Q_T}\left|\bm \nabla\bar{u}^{m}(t,{\bm \lambda})\bar{\bm \sigma}(t,{\bm \lambda})\right|^2\cosh(a\bar{u}^{m}(t,{\bm \lambda}))\D{\bm \lambda}\D t\nonumber\\
&&\qquad\quad+2\delta\int_{Q_T}|\bm \nabla\bar{u}^{m}(t,{\bm \lambda})|^2\D{\bm \lambda}\D t+(C_{\delta,\theta}+C_{\delta,\sigma})\int_{Q_T}\cosh^2(a\bar{u}^{m}(t,{\bm \lambda}))\D{\bm \lambda}\D t+C_T,
\end{eqnarray}
for some $C_T>0$. Recalling that $0<\gamma<1$ and using $\cosh(x)\geq1$ for all $x\in\R$, it follows that
\begin{eqnarray*}
\int_{Q_T}\left(\frac{a}{2}-\frac{1}{2(1-\gamma)}\right)\cosh(a\bar{u}^{m}(t,{\bm \lambda}))\left|(\bm \nabla\bar{u}^{m})(t,{\bm \lambda})\bar{\bm \sigma}(t,{\bm \lambda})\right|^2
\D{\bm \lambda}\D t\geq \left(\frac{a}{2}-\frac{1}{2(1-\gamma)}\right) K_{\bm \sigma}\int_{Q_T}|\bm \nabla\bar{u}^{m}(t,{\bm \lambda})|^2\D{\bm \lambda}\D t.
\end{eqnarray*}
Above, $K_{\bm \sigma}$ is defined in Eq.~\eqref{eq:Ksigma}, {and we have chosen $a>0$ large enough so that $\frac{a}{2}-\frac{1}{2(1-\gamma)}>0$}. Using the above upper bond along with the inequality~\eqref{eq:HJBeqn-j1-jN-1-re-variation-approximation2}, we obtain
\begin{eqnarray}\label{eq:HJBeqn-j1-jN-1-re-variation-approximation4}
\left[\left(\frac{a}{2}-\frac{1}{2(1-\gamma)}\right)K_{\bm \sigma}-2\delta\right]\int_{Q_T}|\bm \nabla\bar{u}^{m}(t,{\bm \lambda})|^2\D{\bm \lambda}\D t\leq (C_{\delta,\theta}+C_{\delta,\sigma})\int_{Q_T}\cosh^2(a\bar{u}^{m}(t,{\bm \lambda}))\D{\bm \lambda}\D t+C_T.
\end{eqnarray}
{By the above choice of $a>0$, we can take $\delta>0$ small enough to guarantee that} $(\frac{a}{2}-\frac{1}{2(1-\gamma)}) K_{\bm \sigma}-2\delta>0$. The desired result follows using that $(\bar{u}^m;\ n\in\N)$ is uniformly bounded, as established in Lemma \ref{lem:lower-solution}.\hfill$\Box$

We can then provide the convergence result.
\begin{theorem}\label{thm:existence-solution}
Eq.~\eqref{eq:hjb-default-jn-T-t} admits a solution $\bar{u}\in L^2([0,T];{\rm H}^{1}(\Delta_{K-1}))\cap L^{\infty}(Q_T)$.
\end{theorem}

\noindent{\it Proof.}\quad The proof relies on standard arguments in PDE analysis.
From Lemma \ref{lem:bounded-L2H1} we know that $\|\bar{u}^{m}\|_{L^2([0,T];{\rm H}^{1}(\Delta_{K-1}))}$ is uniformly bounded. This means that we can extract a subsequence ($m_k, k \in \N$) satisfying $\lim_{k \rightarrow \infty} m_k = \infty$ so that $\bar{u}^{m_k}\too \bar{u}$ in $L^2([0,T];L^2(\Delta_{K-1}))$. In order to take limits, as $k\too\infty$, on both sides of the approximating equation \eqref{eq:HJBeqn-j1-jN-1-re-variation-approximation} in the space $L^2([0,T];L^2(\Delta_{K-1}))$, we use a similar argument as that in the proof of Lemma \ref{lem:bounded-L2H1}. We also use the uniform boundedness of the approximating solution $\bar{u}^m$, and the fact that $\bar{u} \in L^{\infty}(Q_T)$ by Lemma \ref{lem:lower-solution}. We remark that both $L_{\xi}(t)$ and $U_{\xi}(t)$ are independent of $m$. Hence, there exists a sequence of positive constants $C_{m_k}$ satisfying $\lim_{k \rightarrow \infty} C_{m_k} = 0$, such that
\begin{equation}
||\bm \nabla(\bar{u}^{m_k}-\bar{u})||_{L^2([0,T];L^2(\Delta_{K-1}))} \leq C_{m_k}.
\label{eq:conveq}
\end{equation}
Taking limits on both sides of Eq.~\eqref{eq:HJBeqn-j1-jN-1-re-variation-approximation} in the space $L^2([0,T];L^2(\Delta_{K-1}))$ as $k\too\infty$, we have that the limit $\bar{u}$ is indeed a Sobolev solution to Eq.~\eqref{eq:hjb-default-jn-T-t} (see section~7.1 of Part II in \cite{Evans}).  \hfill$\Box$


\subsubsection{Uniqueness of the Solution}
\begin{theorem}\label{thm:unique-solution}
The Sobolev solution to Eq.~\eqref{eq:hjb-default-jn-T-t} is unique.
\end{theorem}

\noindent{\it Proof.}\quad Let $\bar{u}$ be a Sobolev solution to~\eqref{eq:hjb-default-jn-T-t}. We establish a smooth and increasing transformation $\nu(\cdot)$ of the solution as $\bar{u}(t,{\bm \lambda})=\nu(v(t,{\bm \lambda}))$, where $\nu(\cdot):\R\too\R$.\footnote{{The introduction of the transform $\nu$ is needed to overcome the difficulties arising from the simultaneous presence of quadratic gradient and nonlinearity in our HJB equation. Later in the proof we will choose a specific function $\nu$ given by the expression~\eqref{eq:T}. Such a choice allows us to obtain a negative coefficient for the quadratic term in the estimate~\eqref{eq:unique-DeB-ine4}. This in turn makes it possible to apply Gronwall's lemma to get comparison results. } }
This implies that $v(t,{\bm \lambda})=\nu^{-1}(u(t,{\bm \lambda}))$ is the transformed solution of Eq.~\eqref{eq:hjb-default-jn-T-t}, where $\nu^{-1}$ is the inverse of $\nu$. Hence, if we prove that the Sobolev solution of the transformed equation corresponding to $v$ is unique, we can conclude that $\bar{u}$ is the unique Sobolev solution to~\eqref{eq:hjb-default-jn-T-t}. Let $\varphi\in L^2([0,T];{\rm H}_0^1(\Delta_{K-1}))\cap L^{\infty}(Q_T)$. We choose the test function $\phi(t,{\bm \lambda})=\varphi(t,{\bm \lambda})/\nu'(v(t,{\bm \lambda}))$. Then, the variational form \eqref{eq:HJBeqn-j1-jN-1-re-variation} may be rewritten as:
\begin{eqnarray}\label{eq:HJBeqn-j1-jN-1-re-variation-trans}
&&\int_0^T \left<\partial_tv(t),\varphi(t)\right>\D t + \frac{1}{2}\int_{Q_T}(\bm \nabla\varphi)(t,{\bm \lambda})\left[\bar{\bm \sigma}\bar{\bm \sigma}^{\top}(\bm \nabla v)^{\top}\right](t,{\bm \lambda})\D{\bm \lambda}\D t\nonumber\\
&&\qquad=\frac{1}{2}\int_{Q_T}\frac{\nu''(v)}{\nu'(v)}(t,{\bm \lambda})\left[(\bm \nabla v)\bar{\bm \sigma}\bar{\bm \sigma}^{\top}(\bm \nabla v)^{\top}\right](t,{\bm \lambda}) \varphi(t,{\bm \lambda})\D{\bm \lambda}\D t\nonumber\\
&&\qquad\quad+\frac{1}{2}\int_{Q_T}\left[\nu'(v)(\bm \nabla v)\bar{\bm \sigma}\bar{\bm \sigma}^{\top}(\bm \nabla v)^{\top}\right](t,{\bm \lambda})\varphi(t,{\bm \lambda})\D{\bm \lambda}\D t\nonumber\\
&&\qquad\quad+\frac{\gamma}{2(1-\gamma)}\int_{Q_T}\left[\nu'(v)(\bm \nabla v)\bar{\bm \sigma}_{\bf z}\bar{\bm \sigma}_{\bf z}^{\top}(\bm \nabla v)^{\top}\right](t,{\bm \lambda})\varphi(t,{\bm \lambda})\D{\bm \lambda}\D t\nonumber\\
&&\qquad\quad+\int_{Q_T}\left[(\bm \nabla v)(t,{\bm \lambda})\bar{\bm \theta}_{j_1,\ldots,j_n}(t,{\bm \lambda})+\frac{\bar{\xi}_{j_1,\ldots,j_n}(t,{\bm \lambda},\nu(v(t,{\bm \lambda})))}{\nu'(v(t,{\bm \lambda}))}\right]
\varphi(t,{\bm \lambda})\D{\bm \lambda}\D t\nonumber\\
&&\qquad\quad-\frac{1}{2}\int_{Q_T}{\rm \bf div} (\bar{\bm \sigma}\bar{\bm \sigma}^{\top})(t,{\bm \lambda})(\bm \nabla v)(t,{\bm \lambda})^{\top}\varphi(t,{\bm \lambda})\D{\bm \lambda}\D t.
\end{eqnarray}
The above variational form corresponds to the following equation:
\begin{eqnarray}\label{eq:HJBeqn-j1-jN-1-re-pde-trans}
&&\frac{\partial v}{\partial t}(t,{\bm \lambda}) - \frac{1}{2}{\rm Tr}\left[\bar{\bm \sigma}\bar{\bm \sigma}^{\top}D^2v\right](t,{\bm \lambda})
-\frac{1}{2}\left[\frac{\nu''(v)}{\nu'(v)}(\bm \nabla v)\bar{\bm \sigma}\bar{\bm \sigma}^{\top}(\bm \nabla v)^{\top}\right](t,{\bm \lambda})\nonumber\\
&&\qquad=\frac{1}{2}\left[\nu'(v)(\bm \nabla v)\bar{\bm \sigma}\bar{\bm \sigma}^{\top}(\bm \nabla v)^{\top}\right](t,{\bm \lambda})
+\frac{\gamma}{2(1-\gamma)}\left[\nu'(v)(\bm \nabla v)\bar{\bm \sigma}_{\bf z}\bar{\bm \sigma}_{\bf z}^{\top}(\bm \nabla v)^{\top}\right](t,{\bm \lambda}) \nonumber\\
&&\qquad\quad+(\bm \nabla v)(t,{\bm \lambda})\bar{\bm \theta}_{j_1,\ldots,j_n}(t,{\bm \lambda})+\frac{\bar{\xi}(t,{\bm \lambda},\nu(v(t,{\bm \lambda})))}{\nu'(v(t,{\bm \lambda}))}.
\end{eqnarray}

Let $\bar{u}_1$ and $\bar{u}_2$ be two Sobolev solutions to Eq.~\eqref{eq:hjb-default-jn-T-t} with the same initial values and boundary conditions. The corresponding transformations are given by $\bar{u}_i=\nu(v_i)$ with $i=1,2$. Let $v_{1,2}(t,{\bm \lambda}):=v_1(t,{\bm \lambda})-v_2(t,{\bm \lambda})$ be the difference of the two transformed solutions. Then, $v_{1,2}\in L^2([0,T];{\rm H}_0^1(\Delta_{K-1}))\cap L^{\infty}(Q_T)$, since $v_i=\nu^{-1}(\bar{u}_i)$ for $i=1,2$, and $\nu^{-1}$ is smooth and increasing. Using the variational form \eqref{eq:HJBeqn-j1-jN-1-re-variation-trans}
and $\varphi$, we have
\begin{eqnarray}\label{eq:HJBeqn-j1-jN-1-re-variation-trans-diff}
&&\int_0^T \left<\partial_tv_{1,2}(t),\varphi(t)\right>\D t + \frac{1}{2}\int_{Q_T}(\bm \nabla\varphi)({\bm \lambda})\left[\bar{\bm \sigma}\bar{\bm \sigma}^{\top}(\bm \nabla v_{1,2})^{\top}\right](t,{\bm \lambda})\D{\bm \lambda}\D t\nonumber\\
&&\qquad=\int_{Q_T}\left[B(t,{\bm \lambda},v_1(t,{\bm \lambda}),(\bm \nabla v_1)(t,{\bm \lambda}))-B(t,{\bm \lambda},v_2(t,{\bm \lambda}),(\bm \nabla v_2)(t,{\bm \lambda}))\right] \varphi(t,{\bm \lambda})\D{\bm \lambda}\D t,
\end{eqnarray}
where the function $B(t,{\bm \lambda},v,\chi)$, for $(t,{\bm \lambda},v,\chi)\in Q_T\times[\nu^{-1}(L_{\xi}(T)),\nu^{-1}(U_{\xi}(T))]\times\R^{K-1}$, is given by
\begin{eqnarray}\label{eq:B}
B(t,{\bm \lambda},v,\chi)&:=& \frac{1}{2}\left(\frac{\nu''(v)}{\nu'(v)}+\nu'(v)\right)\chi[\bar{\bm \sigma}\bar{\bm \sigma}^{\top}](t,{\bm \lambda})\chi^{\top}
+\frac{\gamma}{2(1-\gamma)}\nu'(v)\chi[\bar{\bm \sigma}_{\bf z}\bar{\bm \sigma}_{\bf z}^{\top}](t,{\bm \lambda})\chi^{\top} \nonumber\\
&&+\chi\bar{\bm \theta}_{j_1,\ldots,j_n}(t,{\bm \lambda})+\frac{\bar{\xi}_{j_1,\ldots,j_n}(t,{\bm \lambda},\nu(v))}{\nu'(v)}-\frac{1}{2}{\rm \bf div} (\bar{\bm \sigma}\bar{\bm \sigma}^{\top})(t,{\bm \lambda})\chi^{\top}.
\end{eqnarray}

Next, we take ${\varphi(t,{\bm \lambda})}:=\left[(v^+_{1,2}(t,{\bm \lambda}))\right]^n$, $n\in\N$. Here, $v^+_{1,2}$ denotes the positive part of $v_{1,2}$. {Since $v_{1,2}\in L^2([0,T];{\rm H}_0^1(\Delta_{K-1}))\cap L^{\infty}(Q_T)$, and the mapping $u\to u^+$ from ${\rm H}_0^1$ to ${\rm H}_0^1$ is bounded, the positive part of $v_{1,2}$ also belong to $L^2([0,T];{\rm H}_0^1(\Delta_{K-1}))\cap L^{\infty}(Q_T)$. Hence, $v_{1,2}^+$ satisfies the requirement to be a test function in our variational form \eqref{eq:HJBeqn-j1-jN-1-re-variation-trans-diff}.}
Then
\begin{eqnarray*}
\int_0^T \left<\partial_t{v_{1,2}}(t),{\varphi(t)}\right>\D t = \int_0^T \left<\partial_t{v^+_{1,2}}(t),(v^+_{1,2}(t))^n\right>\D t=\frac{1}{n+1}\int_{\Delta_{K-1}}\left({v^+_{1,2}}(T,{\bm \lambda})\right)^{n+1}\D{\bm \lambda}.
\end{eqnarray*}
Above, we have used that ${v_{1,2}}(0,{\bm \lambda})=0$ for all ${\bm \lambda}\in\Delta_{K-1}$. Further, we also have
\begin{eqnarray*}
&&\frac{1}{2}\int_{Q_T}\left[(\bm \nabla\varphi)\bar{\bm \sigma}\bar{\bm \sigma}^{\top}(\bm \nabla {v_{1,2} })^{\top}\right](t,{\bm \lambda})\D{\bm \lambda}\D t = \frac{n}{2}\int_{Q_T}\left[{(v_{1,2}^+)^{n-1}}(\bm \nabla {v^+_{1,2}})\bar{\bm \sigma}\bar{\bm \sigma}^{\top}(\bm \nabla {v^+_{1,2}})^{\top}\right](t,{\bm \lambda})\D{\bm \lambda}\D t\nonumber\\
&&\qquad\quad \geq \frac{{K}_{\bm \sigma}}{2}\int_{Q_T}n\left[({v^+_{1,2}})^{n-1}\left|\bm \nabla {v^+_{1,2}}\right|^2\right](t,{\bm \lambda})\D{\bm \lambda}\D t,
\end{eqnarray*}
where we recall the finiteness of the quantity $K_{\bm \sigma}$ defined by~\eqref{eq:Ksigma}. Using the above derived equalities and inequalities, we obtain from Eq.~\eqref{eq:HJBeqn-j1-jN-1-re-variation-trans-diff} that
\begin{eqnarray}\label{eq:uniqueness-ine1}
&&\frac{1}{n+1}\int_{\Delta_{K-1}}\left(v(T,{\bm \lambda})^+\right)^{n+1}\D{\bm \lambda}+\frac{{K}_{\bm \sigma}}{2}\int_{Q_T}n\left[(v^+)^{n-1}\left|\bm \nabla v^+\right|^2\right](t,{\bm \lambda})\D{\bm \lambda}\D t\nonumber\\
&&\qquad\leq\int_{Q_T}\left[B(t,{\bm \lambda},v_1,\bm \nabla v_1)-B(t,{\bm \lambda},v_2,\bm \nabla v_2)\right](t,{\bm \lambda}) (v^{+}(t,{\bm \lambda}))^{n}\D{\bm \lambda}\D t.
\end{eqnarray}
For any $l\in[0,1]$, define $v_l:=lv_1+(1-l)v_2$. Using the mean-value theorem, it follows that
\begin{eqnarray*}
B(t,{\bm \lambda},v_1,\bm \nabla v_1)-B(t,{\bm \lambda},v_2,\bm \nabla v_2) =
\int_0^1 \left[\frac{\partial B}{\partial {v}}(t,{\bm \lambda},v_l,\bm \nabla v_l){v_{1,2}} + \frac{\partial B}{\partial\chi}(t,{\bm \lambda},v_l,\bm \nabla v_l)(\bm \nabla {v_{1,2}})^{\top}\right]\D l,
\end{eqnarray*}
where we can compute
\begin{align}\label{eq:unique-DeB}
\frac{\partial B}{\partial v}(t,{\bm \lambda},v,\chi)&= \frac{1}{2}\left[\left(\frac{\nu''(v)}{\nu'(v)}\right)'+\nu''(v)\right]\chi[\bar{\bm \sigma}\bar{\bm \sigma}^{\top}](t,{\bm \lambda})\chi^{\top}
+\frac{\gamma}{2(1-\gamma)}\nu''(v)\chi[\bar{\bm \sigma}_{\bf z}\bar{\bm \sigma}_{\bf z}^{\top}](t,{\bm \lambda})\chi^{\top}\nonumber\\
&\quad+\frac{\partial_v\bar{\xi}_{{j_1,\ldots,j_n}}(t,{\bm \lambda},\nu(v))}{\nu'(v)}-\bar{\xi}_{{j_1,\ldots,j_n}}(t,{\bm \lambda},\nu(v))
\frac{\nu''(v)}{(\nu'(v))^2},\\
\frac{\partial B}{\partial\chi}(t,{\bm \lambda},v,\chi)&= \left(\frac{\nu''(v)}{\nu'(v)}+\nu'(v)\right)\chi[\bar{\bm \sigma}\bar{\bm \sigma}^{\top}](t,{\bm \lambda})+\frac{\gamma}{1-\gamma}\nu'(v)\chi[\bar{\bm \sigma}_{\bf z}
\bar{\bm \sigma}_{\bf z}^{\top}](t,{\bm \lambda})+\bar{\bm \theta}_{{j_1,\ldots,j_n}}(t,{\bm \lambda})^{\top}\nonumber\\
&\quad-\frac{1}{2}{\rm \bf div} (\bar{\bm \sigma}\bar{\bm \sigma}^{\top})(t,{\bm \lambda}).\nonumber
\end{align}
Using the Young's inequality, for any $\delta>0$ there exists a constant $C_{\delta}>0$ such that
\begin{eqnarray*}\label{eq:uniqueness-ine3}
&&\left|\frac{\partial B}{\partial\chi}(t,{\bm \lambda},v_l,\bm \nabla v_l)(\bm \nabla {v^+_{1,2}})^{\top}\right|({v_{1,2}^{+}})^{n}
=\left|\frac{{v^{+}_{1,2}}}{n}\frac{\partial B}{\partial\chi}(t,{\bm \lambda},v_l,\bm \nabla v_l)(\bm \nabla {v^+_{1,2}})^{\top}\right|n({ v^+_{1,2}})^{n-1}\nonumber\\
&&\qquad\leq \left[C_{\delta}\left|\frac{{v^{+}_{1,2}}}{n}\frac{\partial B}{\partial\chi}(t,{\bm \lambda},v_l,\bm \nabla v_l)\right|^2
+\delta\left|\bm \nabla { v^+_{1,2}}\right|^2\right]n({v^+_{1,2}})^{n-1}\nonumber\\
&&\qquad=\frac{C_{\delta}}{n}\left|\frac{\partial B}{\partial\chi}(t,{\bm \lambda},v_l,\bm \nabla v_l)\right|^2({v^+_{1,2}})^{n+1}
+\delta\left|\bm \nabla { v^+_{1,2}}\right|^2n({ v^+_{1,2}})^{n-1}.
\end{eqnarray*}
Hence, the inequality \eqref{eq:uniqueness-ine1} yields
\begin{eqnarray}\label{eq:uniqueness-ine4}
&&\frac{1}{n+1}\int_{\Delta_{K-1}}\left({v^+_{1,2}}(T,{\bm \lambda})\right)^{n+1}\D{\bm \lambda}+ \left(\frac{\hat{K}_{\bm \sigma}}{2}-\delta \right)n\int_{Q_T}\left[({v^+_{1,2}})^{n-1}\left|\bm \nabla {v^+_{1,2}}\right|^2\right](t,{\bm \lambda})\D{\bm \lambda}\D t\nonumber\\
&&\qquad\leq\int_0^1\int_{Q_T}\left[\frac{\partial B}{\partial v_l}(t,{\bm \lambda},v_l,\bm \nabla v_l)+\frac{C_{\delta}}{n}\left|\frac{\partial B}{\partial\chi}(t,{\bm \lambda},v_l,\bm \nabla v_l)\right|^2\right](t,{\bm \lambda})({v^+_{1,2}}(t,{\bm \lambda}))^{n+1}\D{\bm \lambda}\D t\D l.
\end{eqnarray}

Next, we specify the transformation $\nu$ as follows. For $a,k>0$,
\begin{eqnarray}\label{eq:T}
\nu(v) = -\frac{1}{a}\log\left(k^{-1}+e^{-ak v}\right).
\end{eqnarray}
Notice that the above transformation $\nu$ is smooth and increasing. Moreover, it holds that
\begin{eqnarray*}
\nu'(v) = k \frac{e^{-ak v}}{k^{-1}+e^{-ak v}},\ \ \ \
\nu''(v) = -ak\frac{e^{-akv}}{(k^{-1}+e^{-akv})^2}.
\end{eqnarray*}
Then, we get
\begin{align*}
\frac{\nu''(v)}{\nu'(v)}&=-\frac{a}{k^{-1}+e^{-akv}},\ \ \ \
\left(\frac{\nu''(v)}{\nu'(v)}\right)'=-a^2k\frac{e^{-akv}}{(k^{-1}+e^{-akv})^2},\nonumber\\
\frac{\partial_{w}\bar{\xi}_{j_1,\ldots,j_n}(t,{\bm \lambda},\nu(v))}{\nu'(v)} &=-\frac{k^{-1}+e^{-akv}}{ke^{-akv}}\sum_{i\in\{j_{n+1},\ldots,j_N\}}\tilde{h}_{i;j_1,\ldots,j_{n}}
(t,{\bm \lambda})e^{w_{j_1,\ldots,j_n,i}\left(T-t,\frac{{\bm \lambda}\cdot {\bm h}_{i;j_1,\ldots,j_{n}}^{\bot}(t)}{\tilde{h}_{i;j_1,\ldots,j_{n}}(t,{\bm \lambda})}\right)-\nu(v)}\leq0,\nonumber\\
-{\bar{\xi}}_{j_1,\ldots,j_n}(t,{\bm \lambda},\nu(v))\frac{\nu''(v)}{(\nu'(v))^2}&=\frac{a}{k}\bar{\rho}_{j_1,\ldots,j_n}(t,{\bm \lambda})e^{akv}\nonumber\\
&\quad+\frac{a}{k}\sum_{i\in\{j_{n+1},\ldots,j_N\}}\tilde{h}_{i;j_1,\ldots,j_{n}}
(t,{\bm \lambda})e^{w_{j_1,\ldots,j_n,i}\left(T-t,\frac{{\bm \lambda}\cdot {\bm h}_{i;j_1,\ldots,j_{n}}^{\bot}(t)}{\tilde{h}_{i;j_1,\ldots,j_{n}}(t,{\bm \lambda})}\right)-\nu(v)+akv}.
\end{align*}
We recall that ${\bar{\xi}}_{j_1,\ldots,j_n}(t,{\bm \lambda},{v})$ has been defined in Eq.~\eqref{eq:beta-Jn}. The above inequality follows from the fact that $\tilde{h}_{i}(t,{\bm \lambda},{\bf z})>0$ for all $i\in\{1,\ldots,N\}$ and $({\bm \lambda},{\bf z})\in\Delta_{K-1}\times{\cal S}$.
Hence, using \eqref{eq:unique-DeB} along with the previously derived expressions, we have, for $(t,{\bm \lambda},v,\chi)\in Q_T\times[\nu^{-1}(L_{\xi}(T)),\nu^{-1}(U_{\xi}(T))]\times\R^{K-1}$,
\begin{align}\label{eq:unique-Deb-ine}
\frac{\partial B}{\partial v}(t,{\bm \lambda},v,\chi)&\leq-ak(a+1)\frac{e^{-akv}}{(k^{-1}+e^{-akv})^2}\frac{1}{2}\chi[\bar{\bm \sigma}\bar{\bm \sigma}^{\top}](t,{\bm \lambda})\chi^{\top}-ak\frac{e^{-akv}}{(k^{-1}+e^{-akv})^2}\frac{\gamma}{2(1-\gamma)}\chi[\bar{\bm \sigma}_{\bf z}\bar{\bm \sigma}_{\bf z}^{\top}](t,{\bm \lambda})\chi^{\top}\nonumber\\
&\quad+\frac{a}{k}\bar{\rho}_{j_1,\ldots,j_n}(t,{\bm \lambda})e^{akv}+\frac{a}{k}\sum_{i\in\{j_{n+1},\ldots,j_N\}}\tilde{h}_{i;j_1,\ldots,j_{n}}
(t,{\bm \lambda})e^{w_{j_1,\ldots,j_n,i}\left(T-t,\frac{{\bm \lambda}\cdot h_{i;j_1,\ldots,j_{n}}^{\bot}(t)}{\tilde{h}_{i;j_1,\ldots,j_{n}}(t,{\bm \lambda})}\right)-\nu(v)+akv}\nonumber\\
&\leq-ak(a+1)\frac{e^{-akv}}{(k^{-1}+e^{-akv})^2}\frac{1}{2}K_{\bm \sigma}\left|\chi\right|^2
-ak\frac{e^{-akv}}{(k^{-1}+e^{-akv})^2}\frac{\gamma}{2(1-\gamma)}K_{{\bm \sigma}_{\bf z}}\left|\chi\right|^2\\
&\quad+\frac{a}{k}\left[\sup_{(t,{\bm \lambda})\in Q_T}\left|\rho_{j_1,\ldots,j_n}(t,{\bm \lambda})\right|\right]e^{akv}+\frac{a}{k}\max_{i\in\{j_{n+1},\ldots,j_N\}}\left[\sup_{(t,{\bm \lambda})\in Q_T}\tilde{h}_{i;j_1,\ldots,j_{n}}(t,{\bm \lambda})\right]e^{B_n-\nu(v)+akv},\nonumber
\end{align}
where we recall that $K_{{\bm \sigma}}$ has been defined in Eq.~\eqref{eq:Ksigma}, while $B_n$ has been defined in Eq.~\eqref{eq:BCdef}.
Using Assumption ({\bf A1}), and recalling that $v\in[\nu^{-1}(L_{\xi}(T)),\nu^{-1}(U_{\xi}(T))]$, and $\nu(v)\in[L_{\xi}(T),U_{\xi}(T)]$, we can conclude that there exist constants $\nu_1>0$ and $\nu_2>0$ such that for all $(t,{\bm \lambda})\in Q_T$,
\begin{eqnarray}\label{eq:unique-DeB-ine2}
\frac{\partial B}{\partial v}(t,{\bm \lambda},v,\bm \nabla v) \leq -\nu_1\left|\bm \nabla v \right|^2+\nu_2.
\end{eqnarray}

Next, we estimate $\left|\frac{\partial B}{\partial\chi}(t,{\bm \lambda},v,\bm \nabla v)\right|^2$. From \eqref{eq:unique-DeB}, we obtain
\begin{align*}
\left|\frac{\partial B}{\partial\chi}(t,{\bm \lambda},v,\bm \nabla v)\right|^2&\leq 4\Bigg[\left|\left(\frac{\nu''(v)}{\nu'(v)}+\nu'(v)\right)(\bm \nabla v)[\bar{\bm \sigma}\bar{\bm \sigma}^{\top}](t,{\bm \lambda})\right|^2+\frac{\gamma^2}{(1-\gamma)^2}\left|\nu'(v)(\bm \nabla v)[\bar{\bm \sigma}_{\bf z}\bar{\bm \sigma}_{\bf z}^{\top}](t,{\bm \lambda})\right|^2\nonumber\\
&\quad+\left|\bar{\bm \theta}_{j_1,\ldots,j_n}(t,{\bm \lambda})^{\top}-\frac{1}{2}{\rm \bf div}(\bar{\bm \sigma}\bar{\bm \sigma}^{\top})(t,{\bm \lambda})\right|^2\Bigg]\nonumber\\
&\leq4K^{\bm \sigma}\sup_{v\in[\nu^{-1}(L_{\xi}(T)),\nu^{-1}(U_{\xi}(T))]}\left[\left|\frac{\nu''(v)}{\nu'(v)}+\nu'(v)\right|^2+\frac{\gamma^2}{(1-\gamma)^2}\left|\nu'(v)\right|^2\right]
\left|\bm \nabla v\right|^2\nonumber\\
&\quad+4\sup_{(t,{\bm \lambda})\in Q_T}\left|\bar{\bm \theta}_{j_1,\ldots,j_n}(t,{\bm \lambda})^{\top}-\frac{1}{2}{\rm \bf div}(\bar{\bm \sigma}\bar{\bm \sigma}^{\top})(t,{\bm \lambda})\right|^2,
\end{align*}
where $K^{\bm \sigma}$ has been defined in \eqref{eq:Ksigma}. This implies the existence of positive constants $\eta_1$ and $\eta_2$ such that, for all $(t,{\bm \lambda})\in Q_T$,
\begin{eqnarray}\label{eq:unique-DeB-ine3}
\left|\frac{\partial B}{\partial\chi}(t,{\bm \lambda},v,\bm \nabla v)\right|^2\leq \eta_1\left|\bm \nabla v \right|^2+\eta_2.
\end{eqnarray}
Using~\eqref{eq:unique-DeB-ine2} and~\eqref{eq:unique-DeB-ine3}, we can bound the first integrand term in Eq.~\eqref{eq:uniqueness-ine4} as
\begin{eqnarray}\label{eq:unique-DeB-ine4}
\frac{\partial B}{\partial v_l}(t,{\bm \lambda},v_l,\bm \nabla v_l)+\frac{C_{\delta}}{n}\left|\frac{\partial B}{\partial\chi}(t,{\bm \lambda},v_l,\bm \nabla v_l)\right|^2\leq \left(\frac{\eta_1C_{\delta}}{n}-\nu_1\right)|\bm \nabla {v_{1,2}}|^2 + \frac{\eta_2C_{\delta}}{n}+\nu_2.
\end{eqnarray}
Choosing $n\in\N$ sufficiently large so that $\frac{\eta_1C_{\delta}}{n}-\nu_1<0$, and $\delta>0$ small enough so that $\frac{K_{\bm \sigma}}{2}-\delta>0$, we deduce the following bound from Eq.~\eqref{eq:uniqueness-ine4}:
\begin{eqnarray}\label{eq:uniqueness-ine5}
\int_{\Delta_{K-1}}\left({v^+_{1,2}}(T,{\bm \lambda})\right)^{n+1}\D{\bm \lambda}
\leq(n+1)\left(\frac{\eta_2C_{\delta}}{n}+\nu_2\right)\int_{Q_T}({v^+_{1,2}}(t,{\bm \lambda}))^{n+1}\D{\bm \lambda}\D t.
\end{eqnarray}
Using Gronwall's Lemma, it follows that ${v^+_{1,2}}(t)=0$, $\Delta_{K-1}$-a.s. for all $t\in[0,T]$. It means that $v_1(t)\leq v_2(t)$, $\Delta_{K-1}$-a.s.. Using a similar procedure, we can show that $v_1(t)\geq v_2(t)$, $\Delta_{K-1}$-a.s., hence concluding that $v_1(t)=v_2(t)$, $\Delta_{K-1}$-a.s.. Since $\nu$ is smooth and increasing, we have $\nu(v_1(t))=\nu(v_2(t))$, for all $t\in[0,T]$, $\Delta_{K-1}$-a.s.., i.e., $\bar{u}_1(t)=\bar{u}_2(t)$, $\Delta_{K-1}$-a.s. for all $t\in[0,T]$. This completes the proof of uniqueness. \hfill$\Box$\\


\section{Verification Theorem}\label{sec:verif}
This section proves a verification result based on the analysis done in the previous section. This establishes a rigorous connection between the original control problem, given by \eqref{eq:partialobs}, and the fully observed control problem specified by \eqref{eq:value-fcn}. Eq.~\eqref{eq:connect} indicates that the optimal expected utility achieved by the investor when solving his partially observed control problem can be directly computed by taking the exponential of the value function of the risk-sensitive control problem. The verification theorem~\ref{thm:verification-thm} shows that the value function is the unique Sobolev solution of a recursive system of second-order semilinear parabolic PDEs with quadratic gradient growth.

\begin{theorem}\label{thm:verification-thm}
Recall Assumption ({\bf A1}). For $(t,{\bm \lambda})\in Q_T$, and ${\bf z}={\bf 0}^{j_1,\ldots,j_n} \in {\cal S}$ with $n=1,\ldots,N$, let $w(t,{\bm \lambda},{\bf z})$ be the unique Sobolev solution to Eq.~\eqref{eq:HJBeqn1fix}, where $w(t,{\bm \lambda},{\bf z})=\bar{u}(T-t,{\bm \lambda})$ with $\bar{u}$ being the unique Sobolev solution to Eq.~\eqref{eq:hjb-default-jn-T-t}.
Then $w(t,{\bm \lambda},{\bf z})$ coincides with the value function defined by \eqref{eq:value-fcn}. {\Red Moreover, for each ${\bf z}\in{\cal S}$, define
\begin{eqnarray}\label{eq:optimal-strategy-Jn}
{{\bm \pi}_{\bf z}^*(t,{\bm\lambda})} {:=} \frac{1}{1-\gamma}\left[({\bm \Sigma}^{\top}{\bm \Sigma})^{-1}\left({\bm \Sigma}{\bm \sigma}(t,{\bm \lambda},{\bf z})^{\top}(\bm \nabla w)^{\top}(t,{\bm \lambda},{\bf z})-{\bm \Gamma}(t,{\bm \lambda},{\bf z})\right)\right]_{\bf z}.
\end{eqnarray}
Then ${{\bm \pi}_{\bf z}^*(t,{\bm\lambda})}\in\argmax_{{\bm\pi}}\Phi({\bm \nabla} w(t,{\bm\lambda},{\bf z});t,{{\bm \lambda}},{\bf z},{{\bm \pi}})$, $Q_T$-a.s., is the optimal feedback function and $\hat{\bm\pi}(t):={\bm\pi}^*_{{\bm H}(t)}(t,\tilde{\bm p}^*(t))$
is the corresponding optimal strategy. The process $\tilde{\bm p}^*(t)\in\Delta_{K-1}$ satisfies the SDE~\eqref{eq:sde-tildep} with ${\bm\pi}(t)$ replaced by the optimal strategy $\hat{\bm\pi}(t)$.
The function $\Phi(\cdot)$ is given by the expression in~\eqref{eq:Phi}.}
\end{theorem}

\noindent{\it Proof.}\quad Using the definition of the function $\eta(t,{\bm e}_k,{\bf z},{{\bm \pi}})$ given in \eqref{eq:eta} and \eqref{eq:tildeg}, we deduce that
\begin{eqnarray}\label{eq:tilde-eta2}
\tilde{\eta}(t,{\bm \lambda},{\bf z},{{\bm \pi}})=-r + \sum_{i=1}^N \pi_i(t)(r-\tilde{b}_i(t,{\bm \lambda},{\bf z}) - \tilde{h}_i(t,{\bm \lambda},{\bf z}) )+\frac{1-\gamma}{2}{{\bm \pi}}^{\top}{\bm \Sigma}{\bm \Sigma}^{\top}{{\bm \pi}}.
\end{eqnarray}
Notice that $\tilde{\eta}(t,{\bm \lambda},{\bf z},{{\bm \pi}})$ is convex in ${{\bm \pi}}$. By minimizing it over ${{\bm \pi}}$, it follows that $-\tilde{\eta}(t,{\bm \lambda},{\bf z},{{\bm \pi}})$ is dominated by a constant $C$, given that $Q_T$ is a bounded domain and ${\cal S}$ is a finite set.

Similarly to the approximating equation \eqref{eq:HJBeqn-j1-jN-1-re-variation-approximation}, for each $m\in\N$ and ${\bf z} \in{\cal S}$ we consider
\begin{align}\label{eq:wn2}
\frac{\partial w^{{m}}}{\partial t}(t,{\bm \lambda},{\bf z})
&=-\frac{1}{2}{\rm Tr}\left[{\bm \sigma}{\bm \sigma}^{\top}D^2w^{m}\right](t,{\bm \lambda},{\bf z})-\frac{1}{2}\frac{\left[(\bm \nabla w^{{m}}{\bm \sigma}{\bm \sigma}^{\top}(\bm \nabla w_{L,U}^{m})^{\top}\right](t,{\bm \lambda},{\bf z})}{1+\frac{1}{m}\left[(\bm \nabla w^{m}){\bm \sigma}{\bm \sigma}^{\top}(\bm \nabla w^{m})^{\top}\right](t,{\bm \lambda},{\bf z})}\nonumber\\
&\quad-\frac{\gamma}{2(1-\gamma)}\frac{\left[(\bm \nabla w^{{m}}){\bm \sigma}_{\bf z}{\bm \sigma}_{\bf z}^{\top}(\bm \nabla w^{m}_{L,U})^{\top}\right](t,{\bm \lambda},{\bf z})}{1+\frac{1}{m}\left[(\bm \nabla w^{{m}}){\bm \sigma}_{\bf z}{\bm \sigma}_{\bf z}^{\top}(\bm \nabla w^{m})^{\top}\right](t,{\bm \lambda},{\bf z})}\nonumber\\
&\quad-(\bm \nabla w^{{m}})(t,{\bm \lambda},{\bf z}){\bm \theta}(t,{\bm \lambda},{\bf z})-\xi(t,{\bm \lambda},{\bf z},w^{{m}}(t,{\bm \lambda},{\bf z})),
\end{align}
with terminal condition $w^{m}(T,{\bm \lambda},{\bf z})=0$ for all $({\bm \lambda},{\bf z})\in\Delta_{K-1}\times{\cal S}$. Above, the coefficient
\[
\xi(t,{\bm \lambda},{\bf z},v)=\sum_{i=1}^N(1-z_i)\tilde{h}_i(t,{\bm \lambda},{\bf z})e^{w\left(t, \frac{{\bm \lambda}\cdot{\bm h}_{i}^{\bot}(t,{\bf z})}{\tilde{h}_i(t,{\bm \lambda},{\bf z})}, {\bf z}^{i}\right)-v}+\rho(t,{\bm \lambda},{\bf z}).
\]
By Lemma~\ref{lem:lower-solution}, $w^{{m}}(\cdot,{\bf z})\in[L_{\xi}(T),U_{\xi}(T)]$. This implies $w^{{m}}_{L,U} = w^{{m}}$, so that we may rewrite Eq.~\eqref{eq:wn2} as
\begin{align}\label{eq:wn3}
\frac{\partial w^{m}}{\partial t}(t,{\bm \lambda},{\bf z})
&=-\frac{1}{2}{\rm Tr}\left[{\bm \sigma}{\bm \sigma}^{\top}D^2w^{m}\right](t,{\bm \lambda},{\bf z})-\frac{1}{2}\frac{\left[(\bm \nabla w^{m}){\bm \sigma}{\bm \sigma}^{\top}(\bm \nabla w^{m})^{\top}\right](t,{\bm \lambda},{\bf z})}{1+\frac{1}{m}\left[(\bm \nabla w^{m}){\bm \sigma}{\bm \sigma}^{\top}(\bm \nabla w^{m})^{\top}\right](t,{\bm \lambda},{\bf z})}\nonumber\\
&\quad-\frac{\gamma}{2(1-\gamma)}\frac{\left[(\bm \nabla w^{m}){\bm \sigma}_{\bf z}{\bm \sigma}_{\bf z}^{\top}(\bm \nabla w^{m})^{\top}\right](t,{\bm \lambda},{\bf z})}{1+\frac{1}{m}\left[(\bm \nabla w^{m}){\bm \sigma}_{\bf z}{\bm \sigma}_{\bf z}^{\top}(\bm \nabla w^{m})^{\top}\right](t,{\bm \lambda},{\bf z})}\nonumber\\
&\quad-(\bm \nabla w^{m})(t,{\bm \lambda},{\bf z}){\bm \theta}(t,{\bm \lambda},{\bf z})-\xi(t,{\bm \lambda},{\bf z},w^{m}(t,{\bm \lambda},{\bf z})).
\end{align}
Notice that $w^{m}$ is a weak solution to the approximating problem. This prevents a direct application of the classical It\^{o}'s formula. Next, we apply the generalized It\^{o}'s formula as given in \cite{Krylov}, pag. 122, requiring that $w^{m} \in {\rm W}_p^{2}(Q_T)$ for any $p \geq 2$. Despite we only know that the solution $w^{m}\in L^2([0,T];{\rm H}^2(\Delta_{K-1}))\cap {\rm H}^{1}([0,T];L^2(\Delta_{K-1}))$, we next show that $w^{m} \in {\rm W}_p^{2}(Q_T)$ for any $p \geq 2$.
For $p\geq 2$, ${\bf z}\in{\cal S}$, and a given $v(\cdot,{\bf z})\in {\rm W}_p^{0,1}(Q_T)$ with $v(\cdot,{\bf z})\in[L_{\xi}(T),U_{\xi}(T)]$, consider the linear parabolic equation:
\begin{eqnarray}\label{eq:wn4}
\frac{\partial w^{m}}{\partial t}(t,{\bm \lambda},{\bf z})+\frac{1}{2}{\rm Tr}\left[{\bm \sigma}{\bm \sigma}^{\top}D^2w^{m}\right](t,{\bm \lambda},{\bf z})+(\bm \nabla w^{m})(t,{\bm \lambda},{\bf z}){\bm \theta}(t,{\bm \lambda},{\bf z})
=g(v)(t,{\bm \lambda},{\bf z}),
\end{eqnarray}
where $w^{m}(T,{\bm \lambda},{\bf z})=0$ for all $({\bm \lambda},{\bf z})\in\Delta_{K-1}\times{\cal S}$. For $(t,{\bm \lambda})\in Q_T$, and $v(\cdot,{\bf z})\in {\rm W}_p^{0,1}(Q_T)$ where $v(\cdot,{\bf z})\in[L_{\xi}(T),U_{\xi}(T)]$ for fixed ${\bf z} \in{\cal S}$, we are setting
\begin{align*}
g(v)(t,{\bm \lambda},{\bf z})&=-\frac{1}{2}\frac{\left[(\bm \nabla v){\bm \sigma}{\bm \sigma}^{\top}(\bm \nabla v)^{\top}\right](t,{\bm \lambda},{\bf z})}{1+\frac{1}{m}\left[(\bm \nabla v){\bm \sigma}{\bm \sigma}^{\top}(\bm \nabla v)^{\top}\right](t,{\bm \lambda},{\bf z})}-\frac{\gamma}{2(1-\gamma)}\frac{\left[(\bm \nabla v){\bm \sigma}_{\bf z}{\bm \sigma}_{\bf z}^{\top}(\bm \nabla v)^{\top}\right](t,{\bm \lambda},{\bf z})}{1+\frac{1}{m}\left[(\bm \nabla v){\bm \sigma}_{\bf z}{\bm \sigma}_{\bf z}^{\top}(\bm \nabla v)^{\top}\right](t,{\bm \lambda},{\bf z})}\nonumber\\
&\quad-\xi(t,{\bm \lambda},{\bf z},v(t,{\bm \lambda},{\bf z})).
\end{align*}
The above defined function $g(v)$ is bounded, satisfying $\max_{{\bf z}\in{\cal S}}\sup_{(t,{\bm\lambda})\in Q_T}|g(v)(t,{\bm\lambda},{\bf z})|\leq C_m$, where $C_m$ is a positive constant independent of $v$. Then $g(v)(\cdot,{\bf z})\in L^p(Q_T)$ for any $p\geq2$ and ${\bf z}\in{\cal S}$, and further $\|g(\cdot,{\bf z})\|_{L^p(Q_T)}\leq C_m$. Applying Theorem 7.32 in \cite{Lieberman}, pag. 182, whose assumptions are satisfied in light of Lemma \ref{lem:sing2-b}, we deduce that for each ${\bf z}\in{\cal S}$ and  $m\in\N$, there exists a unique solution $w^{m}(\cdot,{\bf z})\in {\rm W}_p^{1,2}(Q_T)$ such that
\begin{eqnarray}\label{eq:normWpwn}
\left\|w^{m}(\cdot,{\bf z})\right\|_{{\rm W}_p^{1,2}(Q_T)}\leq C_m<+\infty,
\end{eqnarray}
where $C_m$ is a finite positive constant depending only on $m\in\N$.

Next, for $m\in\N$, define the process
\begin{eqnarray}\label{eq:Mpin}
M^{{{\bm \pi}},m}(s):={{\cal B}}(s){\cal Y}^m(s):= e^{-\gamma\int_t^s\tilde{\eta}(u,\tilde{\bm p}(u),{\bm H}(u),{{\bm \pi}}(u))\D u}e^{w^{m}(s,\tilde{\bm p}(s),{\bm H}(s))},\ \ \ s\geq t.
\end{eqnarray}
{Notice that the state process $\tilde{\bm p}(s):=\tilde{\bm p}^{{\bm\pi}}(s)$ depends on the control ${\bm\pi}\in{\cal U}_t$. Moreover, recalling Eq.~\eqref{eq:sde-tildep},
we have that $\tilde{\bm p}^{{\bm\pi}}(s)$, $s\geq t$, satisfies the following SDE under $\tilde{\Px}$: $\tilde{\bm p}^{{\bm\pi}}(t) = {\bm\lambda}$, ${\bm H}(t)={\bf z}$, and for $s>t$,
\begin{eqnarray*}
\D\tilde{\bm p}^{{\bm\pi}}(s) = {\bm \beta}_{\gamma}(s,\tilde{\bm p}^{{\bm\pi}}(s),{\bm H}(s),{\bm \pi}(s))\D s + {\bm \sigma}(s,\tilde{\bm p}^{{\bm\pi}}(s),{\bm H}(s))\D\tilde{\bm W}(s)+
\sum_{i=1}^N{\bf J}_i(s,\tilde{\bm p}^{{\bm\pi}}(s-),{\bm H}(s-))\D\tilde{\Xi}_i(s).
\end{eqnarray*}
} Letting the generator of $(\tilde{\bm p}(t),{\bm H}(t);\ t\geq0)$, given by \eqref{eq:generator}, operate on $f(u,{\bm \lambda},{\bf z})=e^{w^{m}(u,{\bm \lambda},{\bf z})}$ we obtain
\begin{align*}
\A^{\bm\pi}f(u,\tilde{{\bm p}}(u),{\bm H}(u)) &=  {\cal Y}_u^m\Bigg\{\frac{\partial w^{m}}{\partial u}(u,\tilde{{\bm p}}(u),{\bm H}(u)) + (\bm \nabla w^{m})(u,\tilde{{\bm p}}(u),{\bm H}(u)){{\beta}_\gamma(u,\tilde{{\bm p}}(u),{\bm H}(u),{\bm \pi}(u))}\nonumber\\
&\quad+\frac{1}{2}{\rm Tr}\left[{\bm \sigma}{\bm \sigma}^{\top}D^2w^{m}\right](u,\tilde{{\bm p}}(u),{\bm H}(u))\nonumber\\
&\quad+\sum_{i=1}^N\left[e^{w^{m}\left(u, \frac{\tilde{\bm p}(u)\cdot{\bm h}_i^{\bot}(u,{\bm H}(u))}{\tilde{h}_i(u,\tilde{\bm p}(u),{\bm H}(u))},{\bm H}^{i}(u)\right)-w^{m}(u,\tilde{{\bm p}}(u),{\bm H}(u))}-1\right](1-z_i)\tilde{h}_i(u,\tilde{\bm p}(u),{\bm H}(u))\Bigg\}.
\end{align*}
Using the generalized It\^{o}'s formula (see Theorem 2.10.1 in \cite{Krylov}, pag.122), it follows that
\begin{align}\label{eq:veri-thm-ito}
M^{{{\bm \pi}},m}(s) 
&= M^{{{\bm \pi}},m}(t) + \int_t^s M_{u}^{{{\bm \pi}},m}R^m(u,\tilde{\bm p}(u),{{\bm \pi}}(u),{\bm H}(u))\D u + {\cal M}^{{\bm \pi},m}(s),
\end{align}
where the function
\begin{align}\label{eq:R}
R^m(u,{\bm \lambda},{{\bm \pi}},{\bf z})&:=\frac{\partial w^{m}}{\partial u}(u,{\bm \lambda},{\bf z}) + (\bm \nabla w^{m})(u,{\bm \lambda},{\bf z}){{{\bm \beta}}_\gamma(u,{\bm \lambda},{\bf z},{{\bm \pi}})}+\frac{1}{2}{\rm Tr}\left[{\bm \sigma}{\bm \sigma}^{\top}D^2w^{m}\right](u,{\bm \lambda},{\bf z})\nonumber\\
&\quad-\gamma{\tilde{\eta}(u,{\bm \lambda},{\bf z},{{\bm \pi}})}
+\sum_{i=1}^N\left[e^{w^{m}\left(u,\frac{{\bm \lambda}\cdot{\bm h}_{i}^{\bot}(u,{\bf z})}{\tilde{h}_i(u,{\bm \lambda},{\bf z})},{\bf z}^{i}\right)-w^{m}(u,{\bm \lambda},{\bf z})}-1\right](1-z_i)\tilde{h}_i(u,{\bm \lambda},{\bf z}).
\end{align}
Above, we recall that ${\bm \beta}_\gamma(t,{\bm \lambda},{\bf z},{{\bm \pi}}) = {{\bm \beta}}_{\varpi}(t,{\bm \lambda}) + \gamma{\bm \sigma}(t,{\bm \lambda},{\bf z}){\bm \Sigma}^{\top}{{\bm \pi}}$ defined in Eq.~\eqref{eq:coeffdef} is linear in ${\bm \pi}$. Moreover, the $\tilde{\Px}$-(local) martingale is given by
\begin{align}\label{eq:M}
{\cal M}^{{\bm \pi},m}(s) &:= {\cal M}^{{{\bm \pi},n,c}}(s)+{\cal M}^{{{\bm \pi},n,d}}(s),
\end{align}
where ${\cal M}^{{{\bm \pi},m,c}}(s)$ and ${\cal M}^{{{\bm \pi},m,d}}(s)$ are the local martingale components associated with Brownian motion and default events, respectively given by
\begin{align}
{\cal M}^{{{\bm \pi},m,c}}(s) &= \int_t^s M^{{{\bm \pi}},m}(u)(\bm \nabla w^{m})(u,\tilde{\bm p}(u),{\bm H}(u)){\bm \sigma}(u,\tilde{\bm p}(u),{\bm H}(u))\D\tilde{\bm W}_u,\\\notag
{\cal M}^{{{\bm \pi},m,d}}(s) &= \sum_{i=1}^N\int_t^s M^{{{\bm \pi}},m}(u-)\left[e^{w^{m}\left(u, \frac{\tilde{\bm p}(u-)\cdot{\bm h}_{i}^{\bot}(u-,{\bm H}(u-))}{\tilde{h}_i(u,\tilde{\bm p}(u-),{\bm H}(u-))},{\bm H}^{i}(u-)\right)-w^{m}(u,\tilde{\bm p}(u-),{\bm H}(u-))}-1\right]\D\tilde{\Xi}_i(u).
\end{align}

Notice that $R^m(u,{\bm \lambda},{{\bm \pi}},{\bf z})$ is concave in ${\bm \pi}\in\R^N$ since $-\gamma\tilde{\eta}(u,{\bm \lambda},{\bf z},{{\bm \pi}})$ is concave in ${\bm \pi}$. For $m\in\N$, consider the sequence of maximizers of $R^m(u,{\bm \lambda},{{\bm \pi}},{\bf z})$, obtained using the first-order condition and given by
\begin{eqnarray}\label{eq:approx-strategy}
{{\bm \pi}}^{m,*}_{\bf z}(t,{\bm\lambda}) := \frac{1}{1-\gamma}\left[({\bm \Sigma}^{\top}{\bm \Sigma})^{-1}\left({\bm \Sigma}{\bm \sigma}(t,{\bm \lambda},{\bf z})^{\top}(\bm \nabla w^{m})(t,{\bm \lambda},{\bf z})^{\top}-{\bm \Gamma}(t,{\bm \lambda},{\bf z})\right)\right]_{\bf z}.
\end{eqnarray}
First, we have $R^m(u,{\bm \lambda},{{\bm \pi}},{\bf z})\leq R^m(u,{\bm \lambda},{\bf z}):=R^m(u,{\bm \lambda},{{\bm \pi}}_{\bf z}^{m,*}(t,{\bm\lambda}),{\bf z})$,
where
\begin{align}\label{eq:Rnstar}
R^m(u,{\bm \lambda},{\bf z})&=\frac{\partial w^{m}}{\partial u}(u,{\bm \lambda},{\bf z}) + \frac{1}{2}{\rm Tr}\left[{\bm \sigma}{\bm \sigma}^{\top}D^2 w^{m}\right](u,{\bm \lambda},{\bf z})+ \frac{1}{2}\left[(\bm \nabla w^{m}){\bm \sigma}{\bm \sigma}^{\top}(\bm \nabla w^{m})^{\top}\right](u,{\bm \lambda},{\bf z})\nonumber\\
 &\quad+ \frac{\gamma}{2(1-\gamma)}\left[(\bm \nabla w^{m}){\bm \sigma}_{\bf z}{\bm \sigma}_{\bf z}^{\top}(\bm \nabla w^{m})^{\top}\right](u,{\bm \lambda},{\bf z})+(\bm \nabla w^{m})(u,{\bm \lambda},{\bf z}){\bm \theta}(u,{\bm \lambda},{\bf z})\nonumber\\
 &\quad+ {\xi}(u,{\bm \lambda},{\bf z},w^{m}(u,{\bm \lambda},{\bf z})).
\end{align}
Using Eq.~\eqref{eq:wn3}, we can further simplify \eqref{eq:Rnstar} to
\begin{align}\label{eq:Rnstar2}
R^m(u,{\bm \lambda},{\bf z})&=\frac{1}{2}\left|(\bm \nabla w^{m})(u,{\bm \lambda},{\bf z}){\bm \sigma}^{\top}(u,{\bm \lambda},{\bf z})\right|^2\left[1-\frac{1}{1+\frac{1}{m}\left|(\bm \nabla w^{m})(u,{\bm \lambda},{\bf z}){\bm \sigma}^{\top}(u,{\bm \lambda},{\bf z})\right|^2}\right]\\
&\quad+\frac{\gamma}{2(1-\gamma)}\left|(\bm \nabla w^{m})(u,{\bm \lambda},{\bf z}){\bm \sigma}_{\bf z}^{\top}(u,{\bm \lambda})\right|^2\left[1-\frac{1}{1+\frac{1}{m}\left|(\bm \nabla w^{m})(u,{\bm \lambda},{\bf z}){\bm \sigma}_{\bf z}^{\top}(u,{\bm \lambda})\right|^2}\right].\nonumber
\end{align}
For $a\in(0,1)$, define the stopping time
\[
\tau_a:=\inf\left\{u\geq t;\ \sum_{k=1}^{K-1}\tilde{p}_k(u)>1-a,\ {\rm or}\ \min_{1\leq k\leq K-1}\{\tilde{p}_k(u)\}<a\right\}.
\]
It then follows from \eqref{eq:veri-thm-ito} that
\begin{eqnarray}\label{eq:inewn}
{M}^{{{\bm \pi}},m}(s\wedge\tau_a) \leq M^{{{\bm \pi}},m}(t) + \int_t^{s\wedge\tau_a}{M}^{{{\bm \pi}},m}(u) R^m(u,\tilde{\bm p}(u),{\bm H}(u))\D u+{\cal M}^{{\bm \pi},m}(s\wedge\tau_a),
\end{eqnarray}
with the inequality becoming an equality when ${{\bm \pi}}={{\bm \pi}}_{\bf z}^*$, given in~\eqref{eq:optimal-strategy-Jn}.
We know that for each ${\bf z}\in{\cal S}$, and $m\in\N$, the solution
$w^{{m}}(\cdot,{\bf z})\in L^2([0,T];{\rm H}^2(\Delta_{K-1}))\cap {\rm H}^{1}([0,T];L^2(\Delta_{K-1}))$, due to the previous analysis of Eq.~\eqref{eq:HJBeqn-j1-jN-1-re-variation-approximation}. 
Moreover, from Lemma \ref{lem:bounded-L2H1}, we obtain that $\int_{Q_T}|\bm \nabla w^{m}(t,{\bm \lambda},{\bf z})|^2\D{\bm \lambda}\D t\leq C$ for each ${\bf z}\in{\cal S}$, where $C>0$ is a constant independent of $m$. Further, we have that $w^{m} \in[L_{\xi}(T),U_{\xi}(T)]$. Therefore, we can extract a subsequence  ($m_k, k \in \N$) satisfying $\lim_{k \rightarrow \infty} m_k = \infty$ so that
$R^{m_k}(t,{\bm\lambda},{\bf z})\too0$, $\Delta_{K-1}$-a.s. for fixed $(t,{\bf z})\in[0,T]\times{\cal S}$.
Notice that we can bound $e^{w^{m}(t,{\bm \lambda},{\bf z})}$ by a positive constant which is independent of $m$. Then, it follows that $e^{w^{m}(t,{\bm \lambda},{\bf z})}R^m(t,{\bm \lambda},{\bf z})\too0$, $\Delta_{K-1}$-a.s. as $m\too\infty$.

Next, we show that the local martingale in~\eqref{eq:M} is a true martingale. Since $w^{(m)}$ is uniformly bounded by Lemma \ref{lem:lower-solution}, the integrand process in ${\cal M}^{{{\bm \pi},m,d}}(t)$ is bounded hence ${\cal M}^{{{\bm \pi},m,d}}(t)$ is a bounded martingale, which immediately yields that
$\Ex^{\tilde{\Px}}_t\left[{\cal M}^{{\bm \pi},m}(s\wedge\tau_a)\right]=\Ex^{\tilde{\Px}}_t\left[{\cal M}^{{\bm \pi},m,c}(s\wedge\tau_a)\right]$. Then for each $m\in\N$,
\begin{align*}
\Ex_t^{\tilde{\Px}}\left[\left|{\cal M}^{{\bm \pi},m,c}(s\wedge\tau_a)\right|^2\right]&\leq \Ex_t^{\tilde{\Px}}\left[\int_t^T\left|M^{{{\bm \pi}},m}(u)\right|^2\left|\bm \nabla w^{m}(u,\tilde{\bm p}(u),{\bm H}(u)){\bm \sigma}^{\top}(u,\tilde{\bm p}(u),{\bm H}(u))\right|^2\D u\right]\nonumber\\
&\leq C\max_{{\bf z} \in{\cal S}}\left\|\left|\bm \nabla w^{m}(\cdot,{\bf z})\right|^2\right\|_{{\rm W}_K^{1,2}(Q_T)}<+\infty,
\end{align*}
where the first inequality comes from the Burkh\"older-Davis-Gundy inequality {(see, e.g., Theorem 23.12 in \cite{Kallenberg})}. The second inequality is obtained using Krylov{'}s estimate for semimartingales (see \cite{Melnikov}), whereas for the last inequality we used that the process $\left|M^{{{\bm \pi}},m}(u)\right|$ is bounded, $\left|{\bm \sigma}(t,{\bm \lambda})\right|$ is bounded as a consequence of Assumption ({\bf A1}), and $w^{m}(\cdot,{\bf z})\in {\rm W}_{p}^{1,2}(Q_T)$ (see Eq~\eqref{eq:normWpwn} above) for $p=2 K$ and ${\bf z} \in{\cal S}$. This implies that the local martingale ${\cal M}^{{\bm \pi},n,c}(t)$ is a true martingale, and hence $\Ex^{\tilde{\Px}}_t\left[{\cal M}^{{\bm \pi},m}(s\wedge\tau_a)\right]=\Ex^{\tilde{\Px}}_t\left[{\cal M}^{{\bm \pi},m,c}(s\wedge\tau_a)\right]=0$.
It remains to show that ${{\bm \pi}}^{*}_{\bf z}(t,{\bm\lambda})$ given in \eqref{eq:optimal-strategy-Jn} is the optimal feedback strategy. For this, it is enough to show the existence of a subsequence
($m_k, k \in \N$) satisfying $\lim_{k \rightarrow \infty} m_k = \infty$, so that ${{\bm \pi}}^{m_k,*}_{\bf z}(\cdot)$ converges to the optimal strategy given by Eq.~\eqref{eq:optimal-strategy-Jn} in $L^2([0,T];L^2(\Delta_{K-1}))$, as $k\too\infty$.

By \eqref{eq:conveq}, for each ${\bf z}\in{\cal S}$ {$w^{m_k}(\cdot,{\bf z})\too w(\cdot,{\bf z})$ in $L^2([0,T];{\rm H}^{1}(\Delta_{K-1}))$ as $k\too\infty$, hence we deduce immediately that ${{\bm \pi}}^{m_k,*}_{\bf z}\too{{\bm \pi}}^{*}_{\bf z}$ in $L^2([0,T];L^2(\Delta_{K-1}))$, as $m\too\infty$, given that ${\bm \sigma}$ is bounded as a consequence of Assumption ({\bf A1}).} Hence, we can extract a subsequence $m_{j_k}$ of $m_k$, $k \in \N$, such that ${{\bm \pi}}^{m_{j_k},*}(t,{\bm \lambda},{\bf z})\too{{\bm \pi}}^{*}(t,{\bm \lambda},{\bf z})$, $\Delta_{K-1}$-a.s. for each $t\in[0,T]$ and ${\bf z}\in{\cal S}$. Taking the limit, as $k\too\infty$, in \eqref{eq:inewn}, and setting $s=T$ we conclude that
\begin{eqnarray}\label{eq:inewn2}
\Ex_t^{\tilde{\Px}}\left[{M}^{{{\bm \pi}}}(T\wedge\tau_a)\right]\leq M^{{{\bm \pi}}}(t) = e^{w(t,\tilde{\bm p}(t),{\bm H}(t))}
\end{eqnarray}
with the inequality becoming an equality when ${{\bm \pi}}={{\bm \pi}}^{*}_{\bf z}$.
Recall the functions $\Phi$ and $\Phi^*$ given by \eqref{eq:Phi} and \eqref{eq:Phistar} respectively. For $(t,{\bm\lambda})\in Q_T$ and $R>0$, consider a multifunction $F_{{\bf z},R}$ defined on $Q_T$ as
\[
F_{{\bf z},R}(t,{\bm\lambda}):=\big\{{\bm\pi}\in B_R;\ \Phi({\bm\nabla}w(t,{\bm\lambda,{\bf z}});t,{\bm\lambda},{\bf z},{\bm\pi})=\Phi^*({\bm\nabla}w(t,{\bm\lambda,{\bf z}});t,{\bm\lambda},{\bf z})\big\}.
\]
Above, $B_R\subset\R^N$ denotes the closed ball of radius $R$ in $\R^N$. {\Red Using a measurable selection argument similar to \cite{Ahmed} (see pages 512-513 therein), there exists
a measurable selection ${\bm\pi}^*_{{\bf z},R}(t,{\bm\lambda})\in F_{{\bf z},R}(t,{\bm\lambda})$ for $(t,{\bm\lambda})\in Q_T$. {Using the explicit forms of $\Phi$ and $\Phi^*$, we obtain that} ${\bm\pi}^*_{{\bf z}}(t,{\bm\lambda}):=\lim_{R\too\infty}{\bm\pi}^*_{{\bf z},R}(t,{\bm\lambda})$ is also measurable since the pointwise limit of measurable selections ${\bm\pi}^*_{{\bf z},R}(t,{\bm\lambda})$ is also measurable.} Thus $\hat{\bm\pi}(t):={\bm\pi}^*_{{\bm H}(t)}(t,\tilde{\bm p}^*(t))$ defines a time-$t$ optimal strategy.
It remains to show that such a strategy is admissible, i.e. that the condition~\eqref{cond:admiss} is satisfied. 
Define the following increasing sequence of stopping times indexed by $l\in\N$,
\[
\varsigma_l := \inf\{t>0;\ |\bm \nabla w(t,\tilde{\bm p}^*(t),{\bm H}(t))|\geq l\}\wedge T.
\]
Notice that by virtue of Theorem~\ref{thm:existence-solution}, $w \in L^2([0,T];{\rm H}^{1}(\Delta_{K-1}))$, hence
 yielding that the sequence of stopping times $\varsigma_l$'s is well defined.
Using the expression of our optimal feedback control given by~\eqref{eq:optimal-strategy-Jn} and recalling the Assumption ({\bf A1}), we obtain
\begin{eqnarray*}
\Ex^{\hat{\Px}}\left[e^{\int_{\varsigma_l}^{\varsigma_{l+1}}\left|{\bm \pi}^{*}_{{\bm H}(s)}\right|^2\D s} \right] \leq C_T \left\{1 + \Ex^{\hat{\Px}} \left[ e^{\int_0^{\varsigma_{l+1}}\left|(\bm \nabla w)^{\top}(s,\tilde{\bm p}^*(s),{\bm H}(s))\right|^2 \D s} \right] \right\} <+\infty,
\end{eqnarray*}
where $C_T>0$ is a positive constant depending on $T$. The boundedness of the above expectation directly follows from the definition of the stopping times $\varsigma_l$'s. This ends the proof of the verification theorem. \hfill$\Box$

{\Red The above theorem stresses the important link between strategies and value functions. The stock investment strategy depends on the current distress state through the function ${\bm \Gamma}(t,{\bm \lambda},{\bf z})$, and on all distress states reachable from it through the gradient of the value function $w(t,{\bm \lambda},{\bf z})$. Hence, the investor's strategy is forward-looking and depending on all possible
states of financial distress reached when other stocks in the portfolio enter into a state of distress.
}

\section{Numerical Analysis}\label{sec:numerical}
We provide a numerical study to assess the impact of contagious distress on the optimal investment strategies. The drift and default intensity coefficients of the stock price dynamics in Eq.~\eqref{eq:price-dynamics-i-stock1} are chosen to be time homogenous, i.e. 
\begin{align*}
\frac{\D P_i(t)}{P_{i}(t)} = \left( b_i({\bm X}(t)) + h_i({\bm X}(t)) \right) \D t + \vartheta_i \D W_i(t),\ \ \ \ \ P_i(0) = P^{\circ,i} , \ \  \ \ i=1,2.
\end{align*}
The generator ${\bm A}= [\varpi_{i,j}]_{i,j=1,2}$ of the Markov chain is also chosen to be time homogeneous. Section \ref{sec:compmethod} describes the setup used for the numerical examples.
Section \ref{sec:analysis} analyzes the optimal strategies.

\subsection{Setup}\label{sec:compmethod}
We describe the setup and  numerical method used to solve the system of HJB PDEs. 
The investor's portfolio consists of two stocks whose price processes are modulated by a two-states hidden Markov chain, i.e., $N=K=2$. This yields
\begin{align*}
{\bm\sigma}(\lambda_1)&=\lambda_1\left[\vartheta_1^{-1}(\mu_1({\bm e}_1)-\tilde{\mu}_1(x)),\vartheta_2^{-1}(\mu_2({\bm e}_1)-\tilde{\mu}_2(\lambda_1))\right]\\\notag
&=\lambda_1(1-\lambda_1)\left[\vartheta_1^{-1}\left(\mu_1({\bm e}_1)-\mu_1({\bm e}_2)\right),\vartheta_2^{-1}\left(\mu_2({\bm e}_1)-\mu_2({\bm e}_2)\right)\right],\\\notag
{\bm\sigma}_{\bf z}(\lambda_1)&=\lambda_1\left[\vartheta_1^{-1}(1-z_1)(\mu_1({\bm e}_1)-\tilde{\mu}_1(\lambda_1)),
\vartheta_2^{-1}(1-z_2)(\mu_2({\bm e}_1)-\tilde{\mu}_2(\lambda_1))\right]\\\notag
&=\lambda_1(1-\lambda_1)\left[\vartheta_1^{-1}(1-z_1)\left(\mu_1({\bm e}_1)-\mu_1({\bm e}_2)\right),\vartheta_2^{-1}(1-z_2)\left(\mu_2({\bm e}_1)-\mu_2({\bm e}_2)\right)\right], \notag
\end{align*}
and the column vector
\begin{align*}
{\bm\Gamma}(\lambda_1)&=\left[\Gamma_1(\lambda_1),\Gamma_2(\lambda_1)\right]^{\top}=\left[r-\tilde{b}_1(\lambda_1)-\tilde{h}_1(\lambda_1),r-\tilde{b}_2(\lambda_1)
-\tilde{h}_2(\lambda_1)\right]^{\top}\\\notag
&=\left[
    \begin{array}{c}
      r-b_1({\bm e}_2)-h_1({\bm e_2})-\lambda_1\big(b_1({\bm e}_1)-b_1({\bm e}_2)+h_1({\bm e_1})-h_1({\bm e_2})\big) \\
      r-b_2({\bm e}_2)-h_2({\bm e_2})-\lambda_1\big(b_2({\bm e}_1)-b_2({\bm e}_2)+h_2({\bm e_1})-h_2({\bm e_2})\big)  \\
    \end{array}
  \right].
\end{align*}
Hence, the feedback function characterizing the optimal trading strategy is given by
\begin{align}\label{eq:pi2}
{\bm\pi}^*_{\bf z}(\lambda_1,y)&=\frac{1}{1-\gamma}\left[
                                            \begin{array}{cc}
                                              \vartheta_1^{-2} & 0 \\
                                              0 & \vartheta_2^{-2} \\
                                            \end{array}
                                          \right]\left[
                                                   \begin{array}{c}
                                                     \lambda_1(1-\lambda_1)y\big(\mu_1({\bm e}_1)-\mu_1({\bm e}_2)\big)-\Gamma_1(\lambda_1) \\
                                                    \lambda_1(1-\lambda_1)y\big(\mu_2({\bm e}_1)-\mu_2({\bm e}_2)\big)-\Gamma_2(\lambda_1) \\
                                                   \end{array}
                                                 \right]_{\bf z}\nonumber\\
                                          &=\frac{1}{1-\gamma}\left[
                                               \begin{array}{c}
                                                  \vartheta_1^{-2}(1-z_1)\left\{\lambda_1(1-\lambda_1)y\big(\mu_1({\bm e}_1)-\mu_1({\bm e}_2)\big)-\Gamma_1(\lambda_1)\right\}\\
                                                  \vartheta_2^{-2}(1-z_2)\left\{\lambda_1(1-\lambda_1)y\big(\mu_2({\bm e}_1)-\mu_2({\bm e}_2)\big)-\Gamma_2(\lambda_1)\right\}\\
                                               \end{array}
                                             \right].
\end{align}
Depending on the distress state, this takes the following forms:
\begin{equation}\notag\label{eq:strategy}
\begin{cases}
             \pi_{(0,1),1}^*(t,\lambda_1)=\vartheta_1^{-2}\left[\lambda_1(1-\lambda_1)\frac{\partial w(t,\lambda_1,(0,1))}{\partial \lambda_1}\big(\mu_1({\bm e}_1)-\mu_1({\bm e}_2)\big)-\Gamma_1(\lambda_1)\right],   & {\bf z}=(z_1,z_2)=(0,1),\\
             \pi_{(0,1),2}^*(t,\lambda_1)=0,
\\ \\
              \pi_{(1,0),1}^*(t,\lambda_1)=0,  & {\bf z}=(z_1,z_2)=(1,0),\\
              \pi_{(1,0),2}^*(t,\lambda_1)=\vartheta_2^{-2}\left[\lambda_1(1-\lambda_1)\frac{\partial w(t,\lambda_1,(1,0))}{\partial \lambda_1}\big(\mu_2({\bm e}_1)-\mu_2({\bm e}_2)\big)-\Gamma_2(\lambda_1)\right],
\\ \\
              \pi_{(0,0),1}^*(t,\lambda_1)=\vartheta_1^{-2}\left[\lambda_1(1-\lambda_1)\frac{\partial w(t,\lambda_1,(0,0))}{\partial \lambda_1}\big(\mu_1({\bm e}_1)-\mu_1({\bm e}_2)\big)-\Gamma_1(\lambda_1)\right],   & {\bf z}=(z_1,z_2)=(0,0),\\
              \pi_{(0,0),2}^*(t,\lambda_1)=\vartheta_2^{-2}\left[\lambda_1(1-\lambda_1)\frac{\partial w(t,\lambda_1,(0,0))}{\partial \lambda_1}\big(\mu_2({\bm e}_1)-\mu_2({\bm e}_2)\big)-\Gamma_2(\lambda_1)\right].
              \end{cases}
\end{equation}
In the above expressions, for $(t,x)\in[0,T]\times\Delta\times{\cal S}$, $w(t,\lambda_1,(0,1))$, $w(t,\lambda_1,(1,0))$, and $w(t,\lambda_1,(0,0))$ are the unique Sobolev solutions of the following recursive HJB system of semilinear PDEs: on $(t,\lambda_1)\in[0,T)\times\Delta$,
\begin{eqnarray}\label{eq:hjb2}
\frac{\partial w(t,\lambda_1,{\bf z})}{\partial t} +a(\lambda_1)\frac{\partial^2 w(t,\lambda_1,{\bf z})}{\partial \lambda_1^2}+F_{\bf z}\left(t,\lambda_1,w(t,\lambda_1,{\bf z}),\frac{\partial w(t,\lambda_1,{\bf z})}{\partial \lambda_1}\right)=0,
\end{eqnarray}
and $w(T,\lambda_1,{\bf z})=0$. The function
\begin{align*}
F_{\bf z}(t,\lambda_1,y,v) &:=b_{\bf z}(\lambda_1)v^2+d(\lambda_1)v+\sum_{i=1}^2(1-z_i)\big[h_i({\bm e}_2)+(h_i({\bm e}_1)-h_i({\bm e}_2))\lambda_1\big]w_{\bf z}^{i}(t,\lambda_1)e^{-y}+e_{\bf z}(\lambda_1),
\end{align*}
and the coefficients
\begin{align*}
a(\lambda_1) &:=\frac{\lambda_1^2}{2}(1-\lambda_1)^2\sum_{i=1}^2\vartheta_i^{-2}\big(\mu_i({\bm e}_1)-\mu_i({\bm e}_2)\big)^2,\\\notag
b_{\bf z}(\lambda_1) &:= \frac{\lambda_1^2}{2}(1-\lambda_1)^2\sum_{i=1}^2\vartheta_i^{-2}\left[1+\frac{\gamma(1-z_i)}{(1-\gamma)}\right]\big(\mu_i({\bm e}_1)-\mu_i({\bm e}_2)\big)^2,\\\notag
d(\lambda_1)&:=\varpi_{2,1}+(\varpi_{1,1}-\varpi_{2,1})\lambda_1-\frac{\gamma}{1-\gamma}\lambda_1(1-\lambda_1)\sum_{i=1}^2(\mu_i({\bm e}_1)-\mu_i({\bm e}_2))\Gamma_i(\lambda_1),\\\notag
e_{\bf z}(\lambda_1)&:=\gamma r-\sum_{i=1}^2(1-z_i)\big[h_i({\bm e}_2)+(h_i({\bm e}_1)-h_i({\bm e}_2))\lambda_1\big]+\frac{\gamma}{2(1-\gamma)}\sum_{i=1}^2\vartheta_i^{-2}(1-z_i)\Gamma_i^2(\lambda_1),\\\notag
w_{\bf z}^{i}(t,\lambda_1)&:=\exp\left[w\left(t,\frac{h_i({\bm e}_1)\lambda_1}{h_i({\bm e}_2)+(h_i({\bm e}_1)-h_i({\bm e}_2))\lambda_1},{\bf z}^i\right)\right].
\end{align*}
Notice that Eq.~\eqref{eq:hjb2} identifies a recursive system of parabolic PDEs. We solve Eq.~\eqref{eq:hjb2} using the approach described next: {\sf(I)} we solve Eq.~\eqref{eq:hjb2} numerically for the distress states ${\bf z}=(z_1,z_2)=(0,1)$ and ${\bf z}=(1,0)$ respectively using the fact that $w_{(0,1)}^{1}(t,\lambda_1)=w_{(1,0)}^{2}(t,\lambda_1)=e^{\gamma r(T-t)}$, and {\sf(II)} we substitute the numerical solutions obtained in {\sf(I)} for the distress states ${\bf z}=(0,1)$ and ${\bf z}=(1,0)$ into Eq.~\eqref{eq:hjb2} and set ${\bf z}=(0,0)$. Hence, we need to solve three semilinear PDEs. When ${\bf z}=(0,1)$, Eq.~\eqref{eq:hjb2} reduces to
\begin{eqnarray}\label{eq:hjb3}
\frac{\partial u}{\partial t} +a(\lambda_1)\frac{\partial^2u}{\partial \lambda_1^2}+F_{(0,1)}\left(t,\lambda_1,u,\frac{\partial u}{\partial \lambda_1}\right)=0,
\end{eqnarray}
where $u(t,\lambda_1):=w(t,\lambda_1,(0,1))$, and the function
\begin{align*}
F_{(0,1)}(t,\lambda_1,y,v) &=b_{(0,1)}(\lambda_1)v^2+d(\lambda_1)v+\big[h_1({\bm e}_2)+(h_1({\bm e}_1)-h_1({\bm e}_2))\lambda_1\big]e^{\gamma r(T-t)}e^{-y}
+e_{(0,1)}(\lambda_1).
\end{align*}
Notice that Eq.~\eqref{eq:hjb3} is a semilinear parabolic PDE, which admits a unique Sobolev solution by the analysis conducted in Section~\ref{sec:default-contagion}. In order to numerically compute it, we first apply a semi-discretization in time to Eq.~\eqref{eq:hjb3}, and then transform it into an elliptic problem to which we apply the finite element method (FEM). We use the {\sf General Form PDE} interface with {\sf Time-Dependent Study} built-in {\sf COMSOL Multiphysics} to compute the solution. In the symmetric case ${\bf z}=(1,0)$, Eq.~\eqref{eq:hjb2} reduces to
\begin{eqnarray}\label{eq:hjb4}
\frac{\partial u}{\partial t} +a(\lambda_1)\frac{\partial^2u}{\partial \lambda_1^2}+F_{(1,0)}\left(t,\lambda_1,u,\frac{\partial u}{\partial \lambda_1}\right)=0,
\end{eqnarray}
where $u(t,\lambda_1):=w(t,\lambda_1,(1,0))$, and the function
\begin{align*}
F_{(1,0)}(t,\lambda_1,y,v) &=b_{(1,0)}(\lambda_1)v^2+d(\lambda_1)v+\big[h_2({\bm e}_2)+(h_2({\bm e}_1)-h_2({\bm e}_2))\lambda_1\big]e^{\gamma r(T-t)}e^{-y}
+e_{(1,0)}(\lambda_1).
\end{align*}
We can now plug the above solutions $w(t,\lambda_1,(0,1))$ and $w(t,\lambda_1,(1,0))$ into Eq.~\eqref{eq:hjb2} and obtain the solution associated with the distress state ${\bf z}=(0,0)$. Then Eq.~\eqref{eq:hjb2} may be rewritten as:
\begin{eqnarray}\label{eq:hjb5}
\frac{\partial u}{\partial t} +a(\lambda_1)\frac{\partial^2 u}{\partial \lambda_1^2}+F_{(0,0)}\left(t,\lambda_1,u,\frac{\partial u}{\partial \lambda_1}\right)=0,
\end{eqnarray}
where $u(t,\lambda_1):=w(t,\lambda_1,(0,0))$, and the function
\begin{align*}
F_{(0,0)}(t,\lambda_1,y,v) &:=b_{(0,0)}(\lambda_1)v^2+d(\lambda_1)v+e_{(0,0)}(\lambda_1)\\\notag
&\quad+\big[h_1({\bm e}_2)+(h_1({\bm e}_1)-h_1({\bm e}_2))\lambda_1\big]\exp\left[w\left(t,\frac{h_1({\bm e}_1)\lambda_1}{h_1({\bm e}_2)+(h_1({\bm e}_1)-h_1({\bm e}_2))\lambda_1},(1,0)\right)\right]e^{-y}\\\notag
&\quad+\big[h_2({\bm e}_2)+(h_2({\bm e}_1)-h_2({\bm e}_2))\lambda_1\big]\exp\left[w\left(t,\frac{h_2({\bm e}_1)\lambda_1}{h_2({\bm e}_2)+(h_2({\bm e}_1)-h_2({\bm e}_2))\lambda_1},(0,1)\right)\right]e^{-y}.
\end{align*}
We use the same software to solve Eq.~\eqref{eq:hjb5}.

\subsection{Analysis of Strategies}\label{sec:analysis}
We analyze the dependence of the optimal strategies on contagious distress, volatilities and risk aversion. We choose the following benchmark parameters setup. We set the drifts $b_1({\bm e}_1)=1$, $b_1({\bm e}_2)=0.5$, $b_2({\bm e}_1)=1.2$, $b_2({\bm e}_2)=0.4$, the default intensities
$h_1({\bm e}_1)=1$, $h_1({\bm e}_2)=0.1$, $h_2({\bm e}_1)=1$, $h_2({\bm e}_2)=0.1$, the volatilities $\vartheta_1=0.4$, $\vartheta_2=0.6$, the risk aversion parameter $\gamma=0.3$,
the interest rate $r=0$, and the maturity $T=3$. The generator of the Markov chain is specified by $\varpi_{1,1}=0.5$ and $\varpi_{2,2}=0.4$
(thus $\varpi_{1,2}=-\varpi_{1,1}$ and $\varpi_{2,1}=-\varpi_{2,2}$). In all plots, we report the dependence of the strategies on the parameter $\lambda_1$ tracking the filter probability of being in the first regime. We set the investment time $t=0$. {\Red Note that in each figure, the sub-figures reported in the different panels may have different scales.}

\begin{figure}
\centering
\begin{tabular}{cc}
\epsfig{file={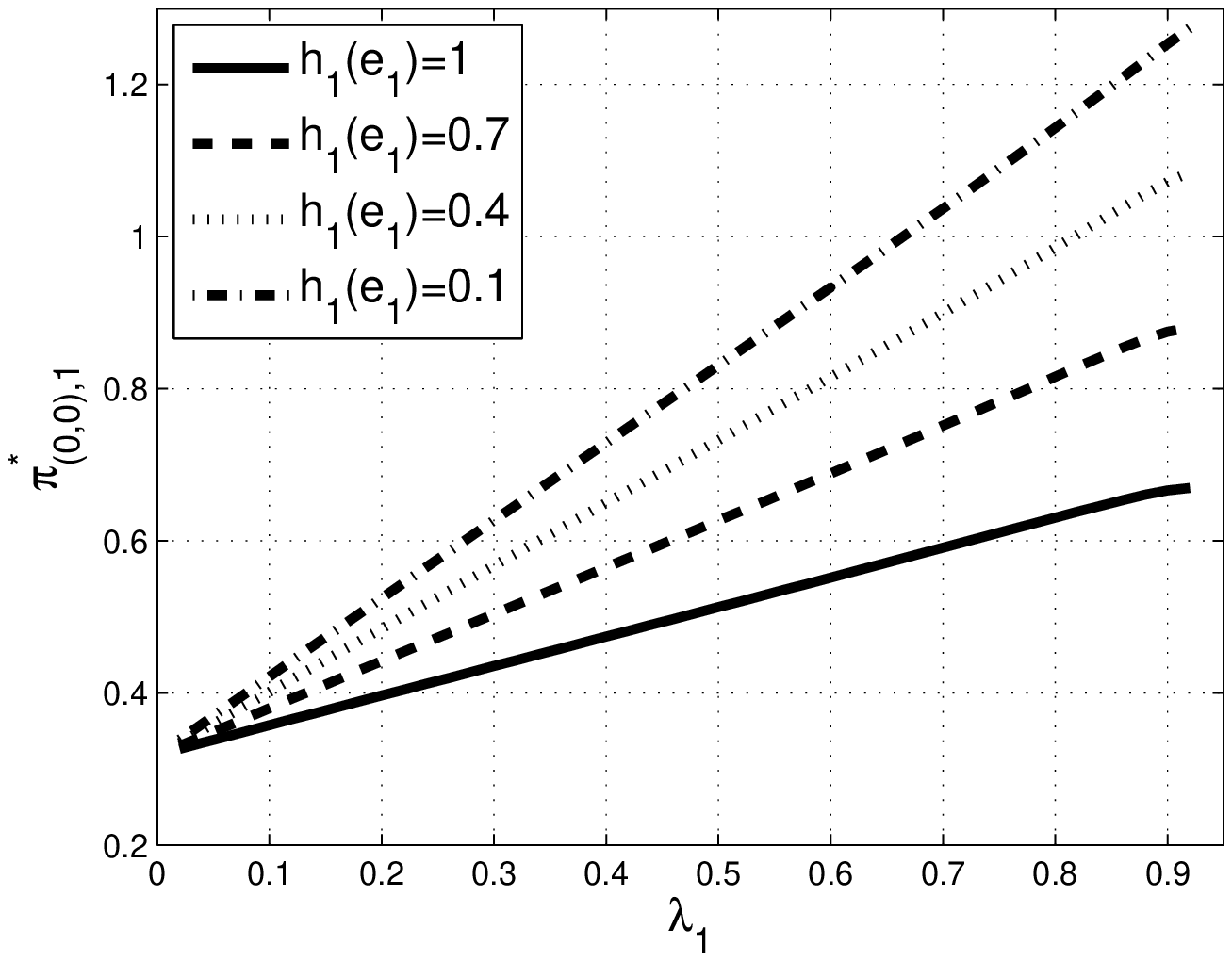},width=0.3\linewidth,clip=}
\epsfig{file={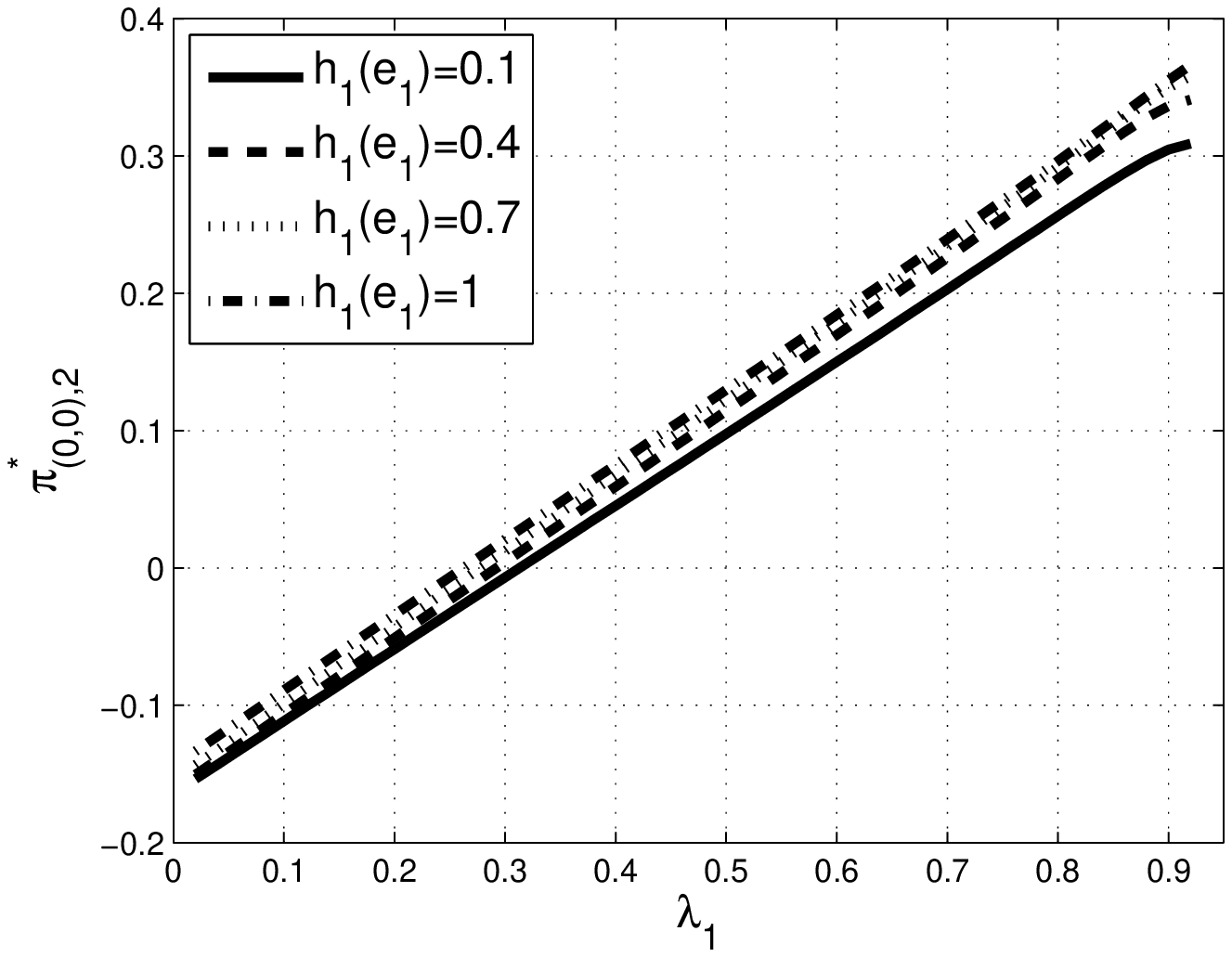},width=0.3\linewidth,clip=} \\
\epsfig{file={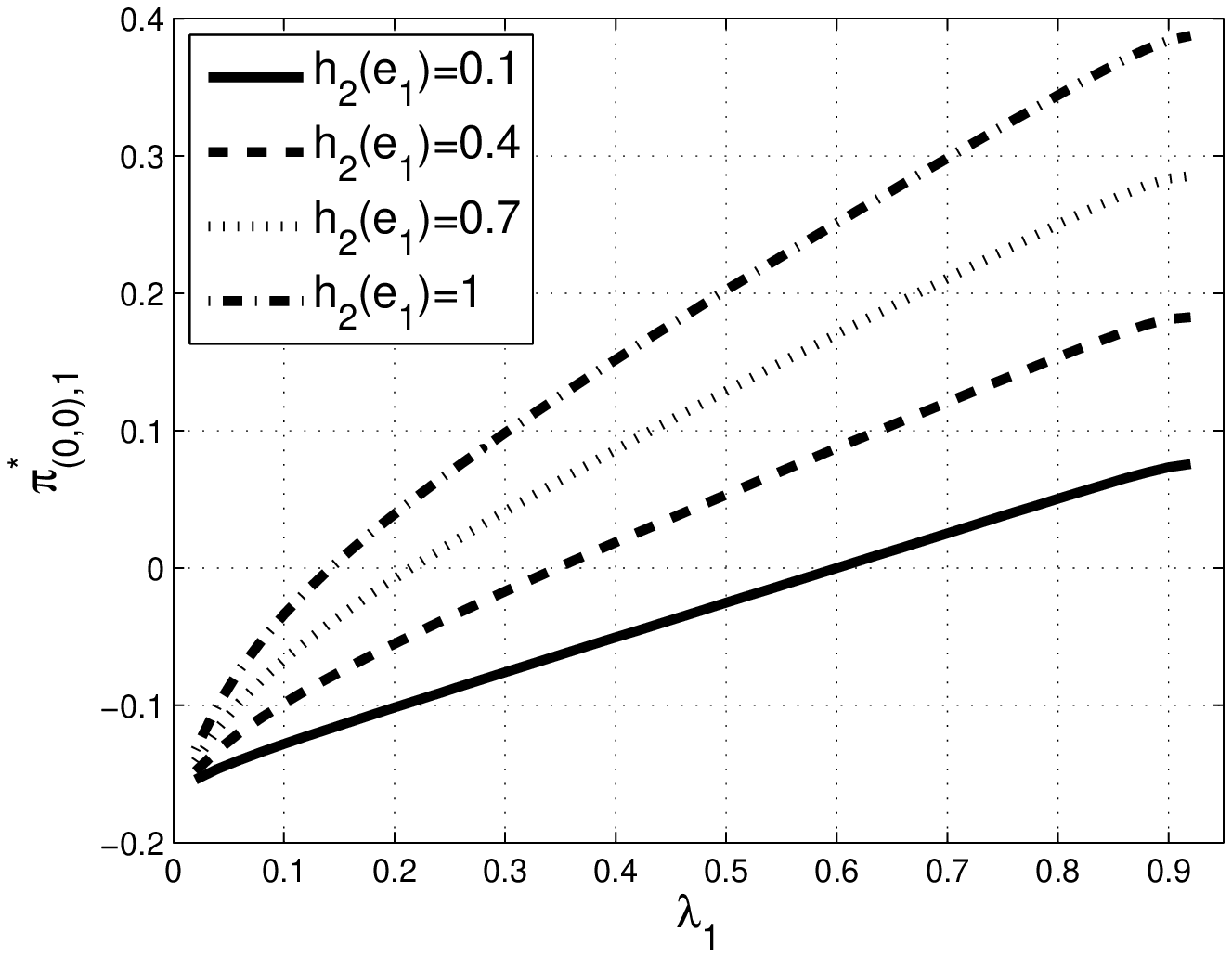},width=0.3\linewidth,clip=}
\epsfig{file={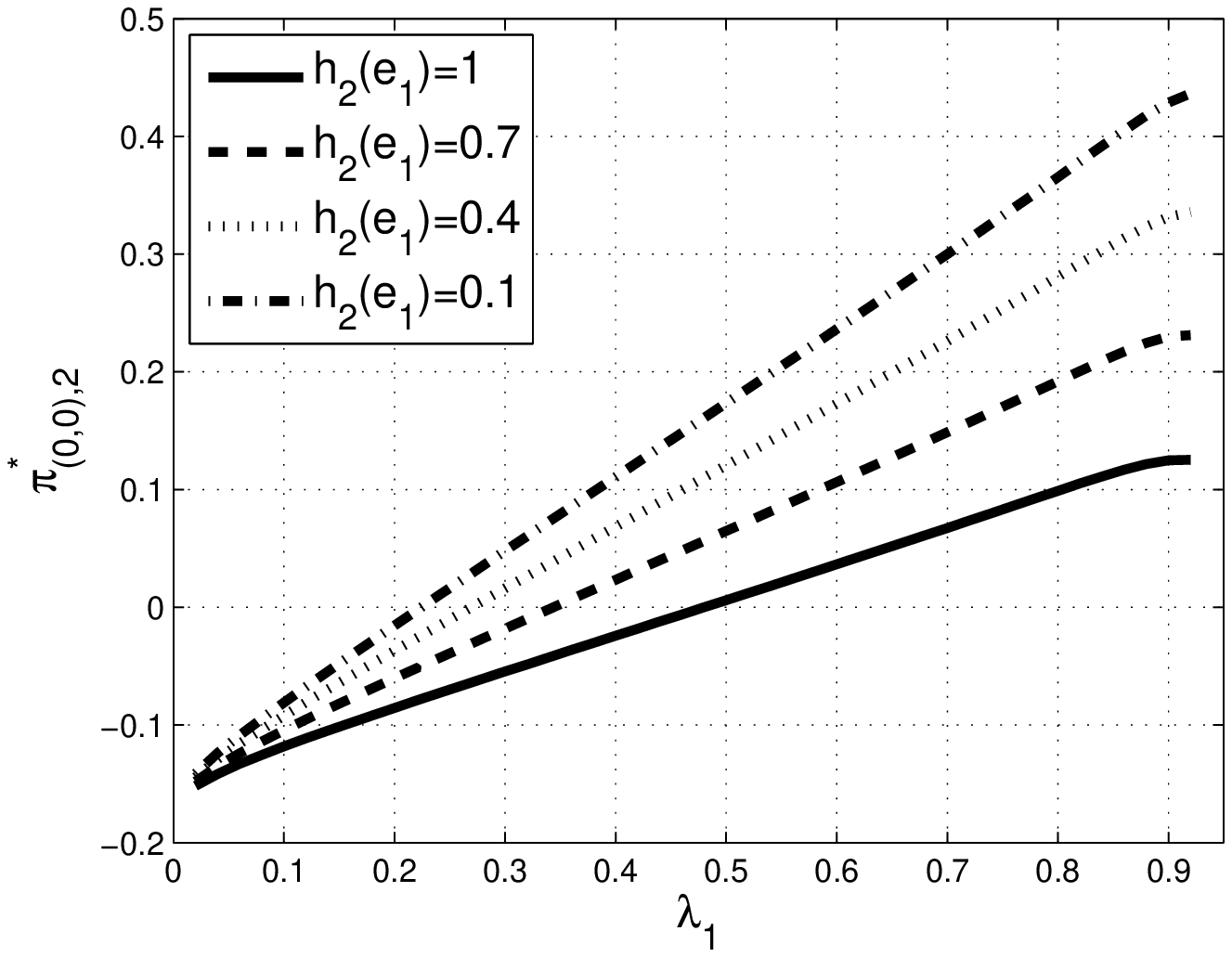},width=0.3\linewidth,clip=} \\
\end{tabular}
\caption{The top panels report the fraction of wealth invested in stock 1 and 2 for different levels of distress intensity of stock ``1'' in regime ``1''. The bottom panels give the corresponding fractions when the distress intensity of stock ``2'' in regime ``1'' is varied.}
\label{fig:prestrategy}
\end{figure}

Figure \ref{fig:prestrategy} suggests that the fraction of wealth invested in stocks increases as the filter probability of the hidden chain being in the first regime gets higher. This happens because, ceteris paribus, the growth rate of both stocks is the highest in regime ``1''. 
When the probability of being in the first regime is small ($\lambda_1 \approx 0$), the stock investment strategies are only mildly sensitive to the distress intensity in regime ``1'' given that the stocks dynamics are essentially modulated by the parameters of regime ``2''. {\Red As the probability of being in regime ``1'' increases, the investor's holdings become more sensitive to the distress risk associated with this regime. 
Indeed, when $h_1(e_1)$ increases, he decreases his holdings in stock ``1'' and purchases additional shares of the safer stock ``2''. He behaves similarly if, ceteris paribus, the distress intensity of stock ``2'' increases (see the bottom panels of figure \ref{fig:prestrategy}).}


\begin{figure}
\centering
\begin{tabular}{cc}
\epsfig{file={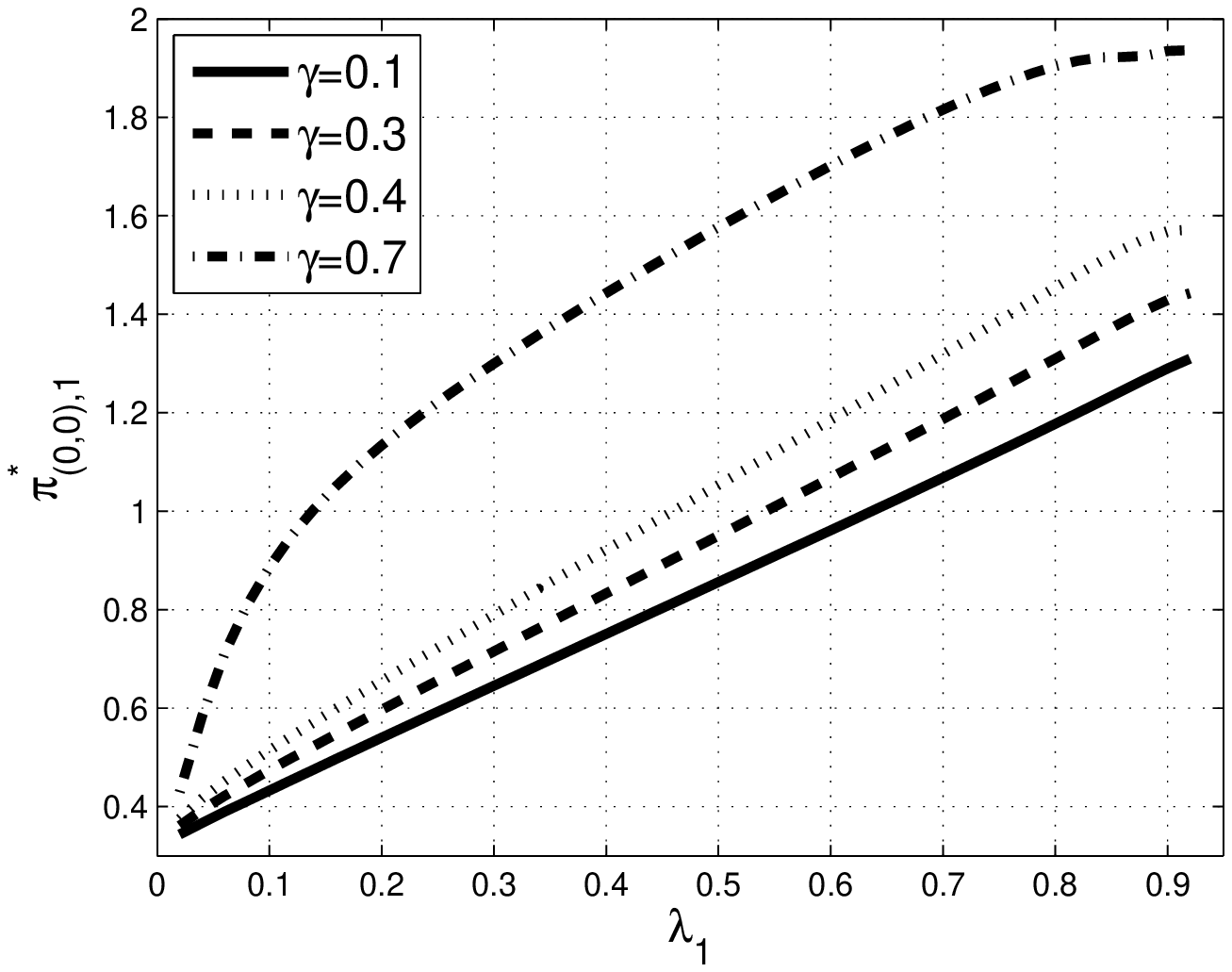},width=0.32\linewidth,clip=}
\epsfig{file={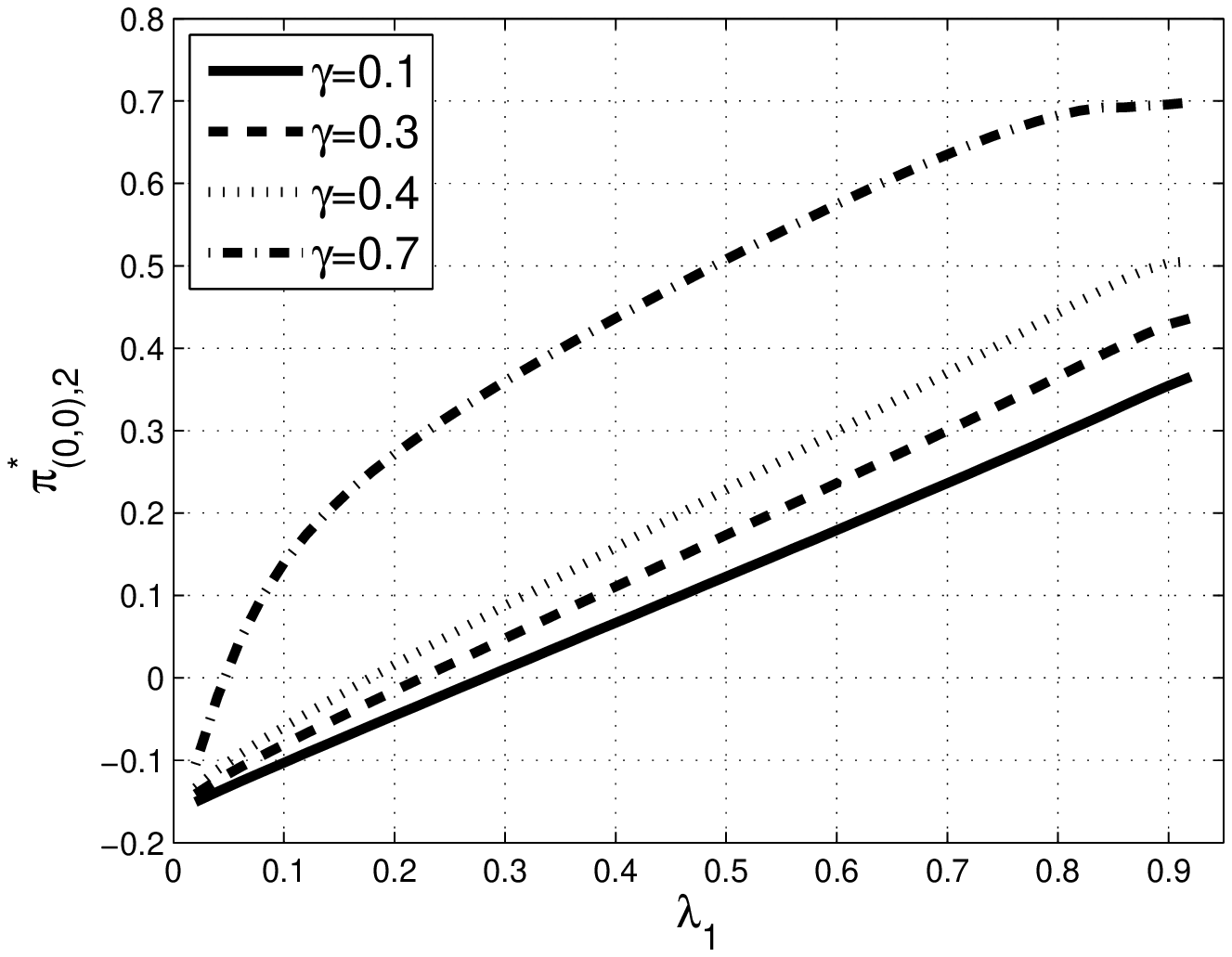},width=0.32\linewidth,clip=}\\
\epsfig{file={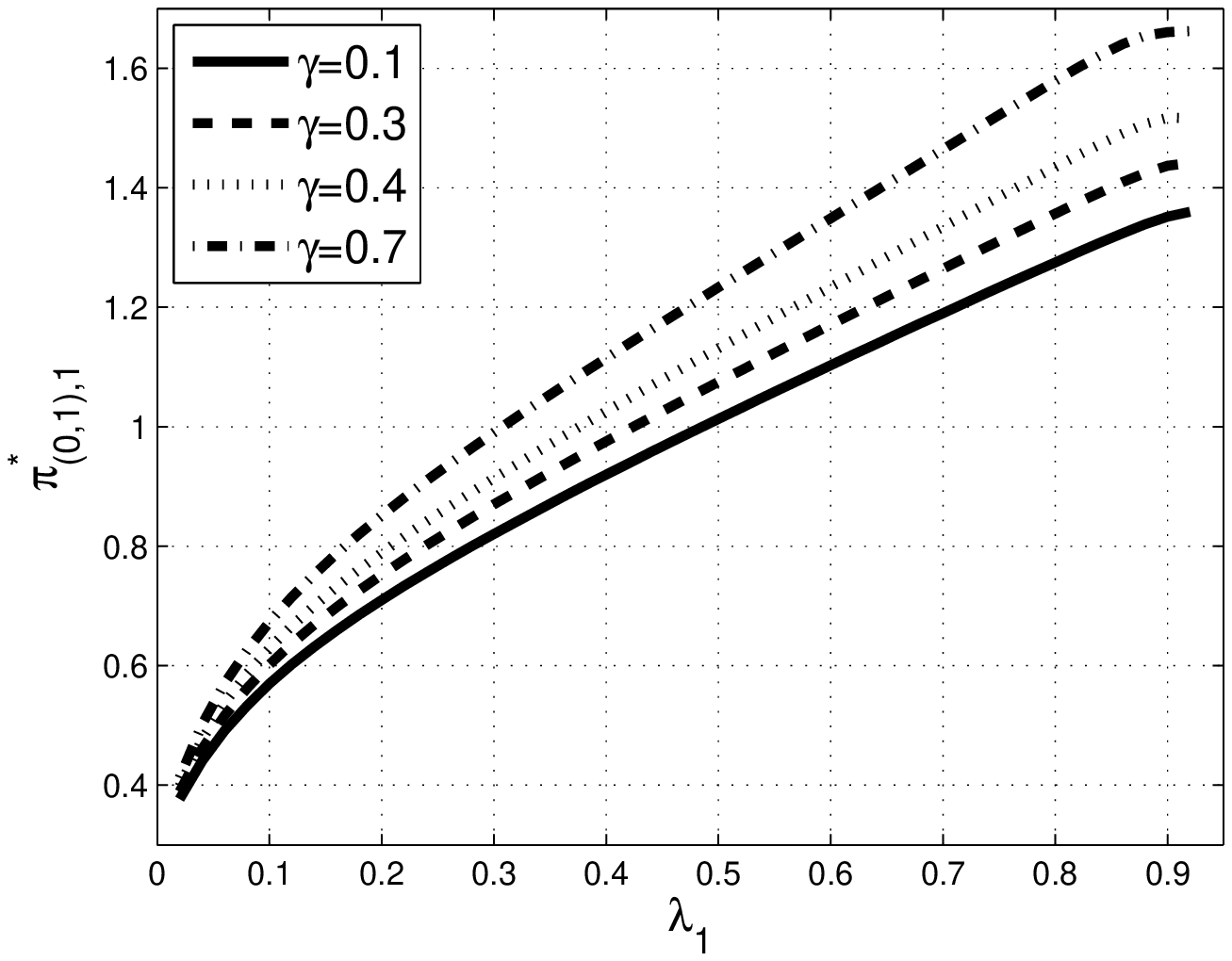},width=0.32\linewidth,clip=}
\epsfig{file={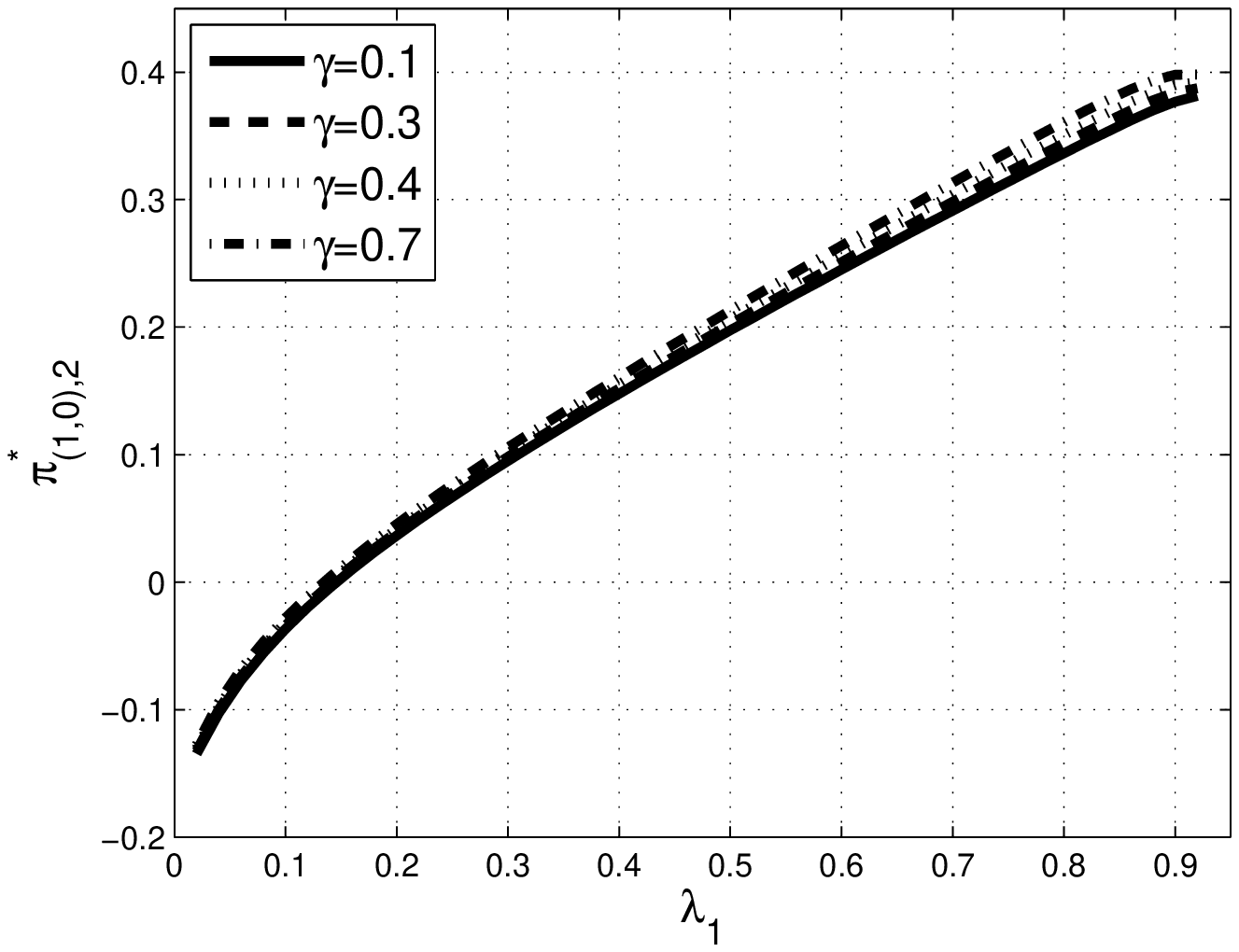},width=0.32\linewidth,clip=} \\
\end{tabular}
\caption{The top panels report the fraction of wealth invested in stock 1 and 2 when both are alive. The bottom panels give the fraction of wealth invested in a stock when the other enters into distress.}
\label{fig:gammatrategy}
\end{figure}

\begin{figure}
\centering
\begin{tabular}{cc}
\epsfig{file={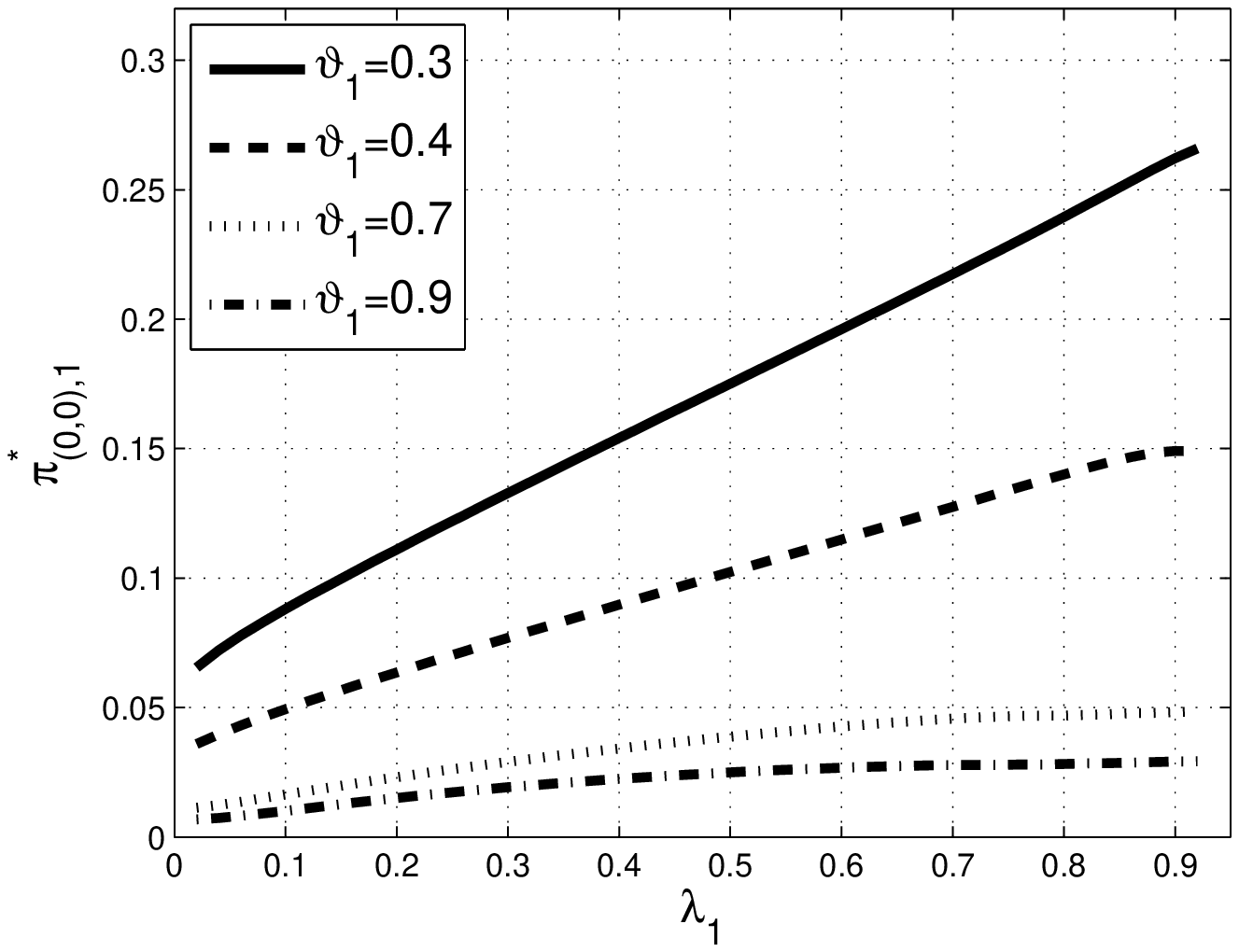},width=0.3\linewidth,clip=}
\epsfig{file={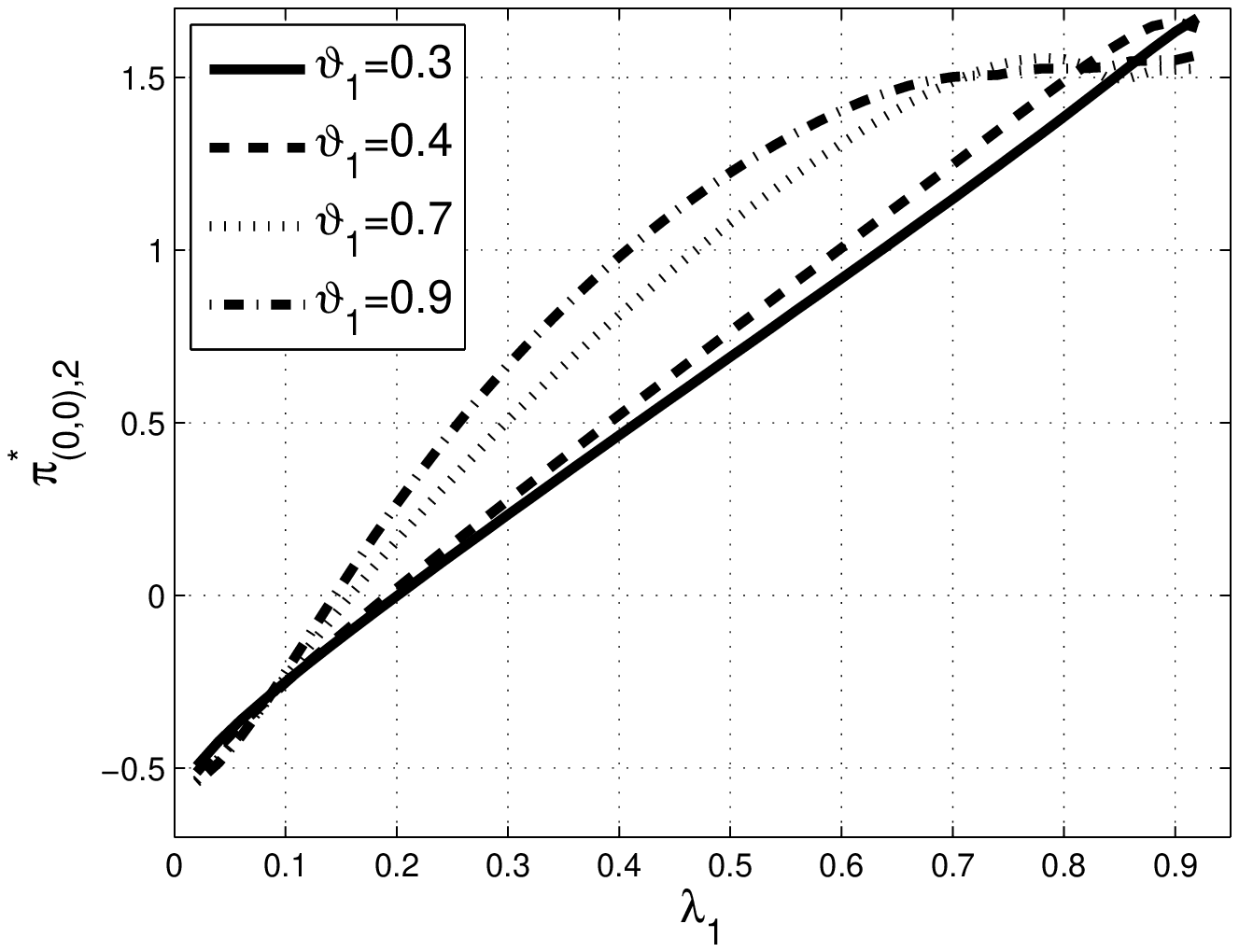},width=0.3\linewidth,clip=}\\
\epsfig{file={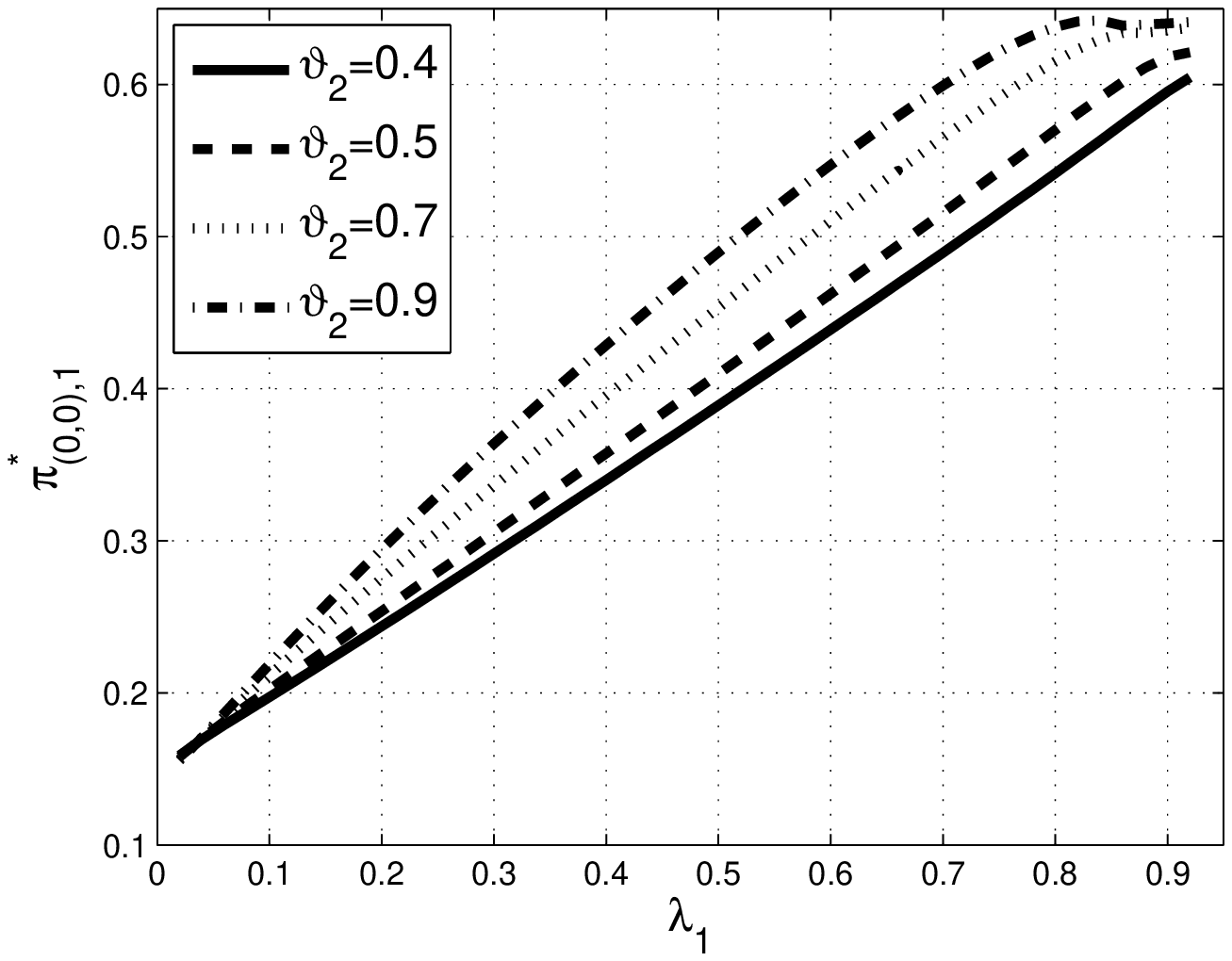},width=0.3\linewidth,clip=}
\epsfig{file={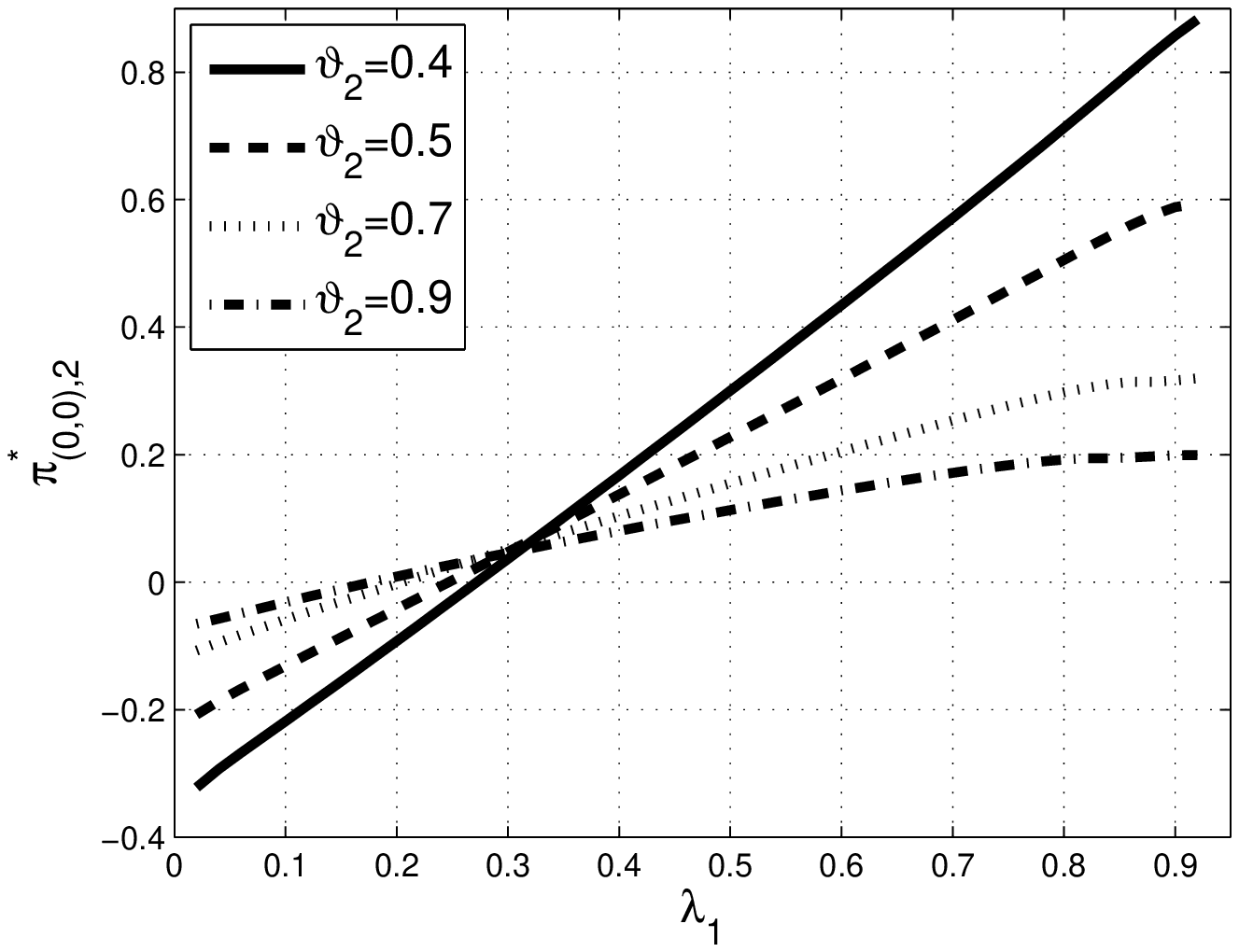},width=0.3\linewidth,clip=} \\
\end{tabular}
\caption{The top panels report the fraction of wealth invested in stock 1 and 2, for different levels of stock ``1'' volatility. The bottom panels report the same dependence, but for different values of stock ``2'' volatility.}
\label{fig:voltrategy}
\end{figure}

Figure \ref{fig:gammatrategy} indicates that the investor increases his holdings in the risky stocks as he becomes more risk-seeking ($\gamma \uparrow$). Since stock ``1'' has the smallest volatility ($\vartheta_1 < \vartheta_2$), he allocates a higher fraction of wealth to stock ``1'' relative to ``2''. A direct comparison of top and bottom panels indicate that the distress of a stock has a contagious effect on the non-distressed stock and leads the investor to reduce the fraction of wealth allocated to it. Moreover, after the stock entered into distress, the strategy of the investor becomes less sensitive to his risk aversion level. For example, after ``1'' enters into distress, an investor with power parameter $\gamma = 0.7$ behaves quite similarly to an investor with parameter $\gamma = 0.1$, i.e. both have the same allocation strategy in the non-distressed stock ``2''. If stock ``2'' enters into distress, his investment decisions are more sensitive to the risk aversion parameter, relative to the case when stock ``1'' enters into distress (compare bottom panels of figure \ref{fig:gammatrategy}). 
Figure \ref{fig:voltrategy} indicates that volatility induces both a substitution and a transmission effect on the strategies. The bottom panels suggest an ``investment substitution''
effect. As the volatility of stock ``2'' increases, the investor decreases his long/short position in the risky stock ``2'' and increases his position in stock ``1''. {\Red Being the investor risk-averse, ceteris paribus, he prefers to allocate a higher fraction of his wealth to the stock with the highest instantaneous expected return, rather than bearing the increased market risk coming from his position in stock ``2''.}
The top panels, instead, suggest the presence of a transmission effect. If the volatility of stock ``1'' is high, the investor reduces his holdings in stock ``1'' and his strategy is not sensitive to the filter probability. However, it appears from the top right panel of figure \ref{fig:voltrategy} that he only uses the savings from his reduced position in stock ``1'' to purchase shares of stock ``2'' when $\lambda_1$ is sufficiently low. If the probability of being in regime ``1'' exceeds a certain threshold, the investor would reduce his long position in both stocks if the volatility parameter is large enough
(compare $\vartheta_1 = 0.9$ with $\vartheta_1 = 0.3$ in the top panel). This indicates that when the chain is in regime ``1'' (where distress and market risk are both high), the investor reduces his holdings in {both stocks} and invests more in the risk-free money market account.

\appendix

\section{Proofs of Technical Lemmas}\label{app:utilmax}

\noindent{\textbf{Proof of Lemma \ref{lem:sing2-b}.}}

We only consider the case $K=2$. The proof for $K>2$ is similar. For $K=2$, notice that
${\bm \lambda}_1=\lambda_1 \in(0,1)$ and ${\bm \lambda}_2=\lambda_2 \in(0,1)$. Using the above definition of ${\bm \sigma}(t,{\bm \lambda},{\bf z})$, we obtain
\begin{align*}
&{\bm \sigma} {\bm \sigma}^{\top}(t,\lambda_1,{\bf z})=(\lambda_1)^2\sum_{i=1}^N \frac{1}{\vartheta_i^2({\bf z})} \left({b_i(t,{\bm e}_1,{\bf z}) + h_i(t,{\bm e}_1,{\bf z})} -\vartheta_i^2({\bf z})/2-\tilde{\mu}_i(t,\lambda_1,{\bf z})\right)^2\nonumber\\
&\quad=(\lambda_1)^2 \sum_{i=1}^N\frac{1}{\vartheta_i^2({\bf z})}\left[{b_i(t,{\bm e}_1,{\bf z}) + h_i(t,{\bm e}_1,{\bf z})} -\vartheta_i^2({\bf z})/2-\mu_i(t,{\bm e}_2,{\bf z})-(\mu_i(t,{\bm e}_1,{\bf z})-\mu_i(t,{\bm e}_2,{\bf z}))\lambda_1\right]^2.
\end{align*}
Using the equality $\mu_i(t,{\bm e}_j,{\bf z})=b_i(t,{\bm e}_j,{\bf z}) + h_i(t,{\bm e}_j,{\bf z}) -\vartheta_i^2({\bf z})/2$, it follows that
\begin{align*}
{\bm \sigma} {\bm \sigma}^{\top}(t,\lambda_1,{\bf z})&=(\lambda_1)^2\sum_{i=1}^N \frac{1}{\vartheta_i^2({\bf z})} \big[b_i(t,{\bm e}_1,{\bf z}) +  h_i(t,{\bm e}_1,{\bf z})-b_i(t,{\bm e}_2,{\bf z}) - h_i(t,{\bm e}_2,{\bf z})\nonumber\\
&\quad-(b_i(t,{\bm e}_1,{\bf z}) + h_i(t,{\bm e}_1,{\bf z}) -b_i(t,{\bm e}_2,{\bf z}) - h_i(t,{\bm e}_2,{\bf z}))\lambda_1\big]^2.
\end{align*}
Since $(t,\lambda_1)\in Q_T$, which is a bounded domain, by the Assumption ({\bf A1}) we obtain that the gradient of ${\bm \sigma}{\bm \sigma}^{\top}(t,\lambda_1,{\bf z})$ satisfies
\[
\sup_{(t,\lambda_1,{\bf z})\in Q_T\times{\cal S}}\left|\left(\bm \nabla_{(t,\lambda_1)}{\bm \sigma}{\bm \sigma}^{\top}\right)(t,\lambda_1,{\bf z})\right|\leq C,
\]
where $C>0$ is a positive constant. Choosing the linear increasing function $\varrho(x)=Cx$, $x>0$, we obtain the inequality \eqref{eq:sing2-b} from the mean-value theorem. Moreover, $\varrho(0)=0$. This concludes the proof. \hfill$\Box$\\

\noindent{\textbf{Proof of Lemma \ref{eq:generatorp-H}.}}

Consider the bivariate process $(\tilde{\bm p}(t),{\bm H}(t))$. Using It\^o's formula, for any $s>t$,
\begin{align*}
&f(s,\tilde{{\bm p}}(s),{\bm H}(s))=f(t,\tilde{\bm p}(t),{\bm H}(t)) + \int_t^s \frac{\partial f}{\partial u}(u,\tilde{\bm p}(u),{\bm H}(u))\D u + \int_t^s \bm \nabla f(u,\tilde{\bm p}(u),{\bm H}(u)){{\bm \beta}}_\gamma(u,\tilde{\bm p}(u),{\bm H}(u),{{\bm \pi}}(u))\D u\nonumber\\
&\quad+\frac{1}{2}\int_t^s{\rm Tr}\left[{\bm \sigma}{\bm \sigma}^{\top}(u,\tilde{\bm p}(u),{\bm H}(u))D^2f(u,\tilde{{\bm p}}(u),{\bm H}(u))\right]\D u + \int_t^s \bm \nabla f(u,\tilde{\bm p}(u),{\bm H}(u)){\bm \sigma}(u,\tilde{\bm p}(u),{\bm H}(u))\D\tilde{\bm W}(u)\nonumber\\
&\quad+\sum_{t<u\leq s}\left(f(u,\tilde{\bm p}(u),{\bm H}(u))-f(u,\tilde{\bm p}(u-),{\bm H}(u-))\right).
\end{align*}
Since our model excludes events in which two or more stocks simultaneously enter into the distress state, we have
\begin{eqnarray*}
&\sum_{t<u\leq s}\left(f(u,\tilde{\bm p}(u),{\bm H}(u))-f(u,\tilde{\bm p}(u-),{\bm H}(u-))\right) \\
&=\sum_{i=1}^N\int_t^s\left(f(u,\tilde{{\bm p}}(u-)+{\bf J}_i(u,\tilde{\bm p}(u-),{\bm H}(u-)), {\bm H}^{i}({u-}))
-f(u,\tilde{\bm p}(u-),{\bm H}(u-))\right)\D H_i(u),
\end{eqnarray*}
{where we recall from Section \ref{sec:model} that ${\bm H}^{i}(t)$ is obtained from ${\bm H}(t)$ as defined in Eq.~\eqref{eq:flip} and ${\bf J}_i(t,{\bm\lambda},{\bf z})$ is defined in \eqref{eq:coeffdef}. Notice that for each $i=1,\ldots,N$,
\begin{align*}
{{\bm \lambda}}+{\bf J}_i(t,{\bm \lambda},{\bf z}) = {\bm \lambda} + diag({\bm \lambda})\frac{1}{\tilde{h}_i(t,{\bm \lambda},{\bf z})}\left[{\bm h}_{i}^{\bot}(t,{\bf z})-{\bf1}_{K-1}\tilde{h}_i(t,{\bm \lambda},{\bf z})\right] = \frac{1}{\tilde{h}_i(t,{\bm \lambda},{\bf z})}\left[{\bm \lambda}\cdot{\bm h}^{\bot}_{i}(t,{\bf z})\right].
\end{align*}
Hence, it holds that
\begin{align*}
& f(s,\tilde{{\bm p}}(s),{\bm H}(s))=f(t,\widetilde{\bm p}(t),{\bm H}(t)) + \int_t^s \frac{\partial f}{\partial u}(u,\tilde{\bm p}(u),{\bm H}(u))\D u + \int_t^s \bm \nabla f(u,\tilde{\bm p}(u),{\bm H}(u)){{\bm \beta}}_\gamma(u,\tilde{\bm p}(u),{\bm H}(u),{{\bm \pi}}(u))\D u\nonumber\\
&\quad+\frac{1}{2}\int_t^s{\rm Tr}\left[{\bm \sigma}{\bm \sigma}^{\top}(u,\tilde{\bm p}(u),{\bm H}(u))D^2f(u,\tilde{{\bm p}}(u),{\bm H}(u))\right]\D u + \int_t^s \bm \nabla f(u,\tilde{\bm p}(u),{\bm H}(u)){\bm \sigma}(u,\tilde{\bm p}(u),{\bm H}(u))\D\tilde{\bm W}(u)\nonumber\\
&\quad+\sum_{i=1}^N\int_t^s\left[f\left(u, \frac{1}{\tilde{h}_i(u,\tilde{\bm p}(u-),{\bm H}(u-))}\left[\tilde{\bm p}(u-)\cdot{\bm h}_{i}^{\bot}(u,{\bm H}(u-))\right],{\bm H}^{i}(u-)\right)
-f(u,\tilde{\bm p}(u-),{\bm H}(u-))\right]\D \tilde{\Xi}_i(u)\nonumber\\
&\quad+\sum_{i=1}^N\int_t^s\left[f\left(u, \frac{1}{\tilde{h}_i(u,\tilde{\bm p}(u),{\bm H}(u))}\left[\tilde{\bm p}(u)\cdot{\bm h}_{i}^{\bot}(u,{\bm H}(u))\right],{\bm H}^{i}(u)\right)
-f(u,\tilde{\bm p}(u),{\bm H}(u))\right] \\
& \; \; \; \; \; \; \; \times(1-H_i(u))\tilde{h}_i(u,\tilde{\bm p}(u),{\bm H}(u))\D u,
\end{align*}
which yields the generator \eqref{eq:generator}. \hfill$\Box$

{

\end{document}